\documentclass[12pt, a4paper]{article}
\pdfoutput=1
\usepackage{jheppub}
\usepackage{latexsym, graphicx}
\usepackage{amsmath, amssymb}
\usepackage{color, hyperref}

\title{Relativistic single-electron wavepacket in quantum electromagnetic fields: Quantum coherence, correlations, and the Unruh effect}
\author[a]{Shih-Yuin Lin}
\author[b]{and Bei-Lok Hu}

\affiliation[a]{Department of Physics, National Changhua University of Education, Changhua 500207, Taiwan}
\affiliation[b]{Maryland Center for Fundamental Physics and Joint Quantum Institute \\
University of Maryland, College Park, Maryland 20742-4111, USA}

\emailAdd{sylin@cc.ncue.edu.tw} 
\emailAdd{blhu@umd.edu}
\date{26 December 2023}

\abstract{Conventional formulation of QED since the 50s works very well for stationary states and for scattering problems, but with newly arisen challenges from the 80s on, where real time evolution of particles in a nonequilibrium setting are required, and quantum features such as coherence, dissipation, correlation and entanglement in a system interacting with its quantum field environment are sought after, new ways to formulate QED suitable for these purposes beckon. In this paper we present a linearized effective theory using a Gaussian wavepacket description of a charged relativistic particle coupled to quantum electromagnetic fields to study the interplay between single electrons and quantum fields in free space,  at a scale well below the Schwinger limit. The proper values of the regulators in our effective theory are determined from the data of individual experiments, and will be time-dependent in the laboratory frame if the single electrons are accelerated. Using this new theoretical tool, we address the issues of decoherence of flying electrons in free space and the impact of Unruh effect on the electrons. Our result suggests that vacuum fluctuations may be a major source of blurring the interference pattern in electron microscopes. For a single electron accelerated in a uniform electric field, we identify the Unruh effect in the two-point correlators of the deviations from the electron's classical trajectory. From our calculations we also bring out some subtleties, involving the bosonic versus fermionic spectral functions.}


\begin{document}

\maketitle

\section{Introduction}
\label{intro}

Quantum electrodynamics (QED) established in the early 50s is admittedly one of the most successful theories in modern physics.  Even in the simplest case of an electron interacting with the vacuum state of an electromagnetic (EM) field, QED predicts with high accuracy a wide range of phenomena from the Lamb shift \cite{La47, Be47, We48} to pair production \cite{Sa31, HE36, Sc51}.  
Conventional formulation of QED works very well for stationary states and for scattering problems \cite{IZ80}. In those applications it is convenient to calculate the statistics in momentum space to compare with the experimental or observational data. Dealing with a stationary system at finite temperatures one can use imaginary time formulation. For scattering problems one can calculate the transition amplitude from an in state to an out asymptotic state.  

\subsection{New Challenges and new issues requiring new formulations}

New challenges and developments begun in the 80s and 90s called for and empowered new formulations for more contemporary problems, such as those in relativistic transport, quantum radiation, and  quantum information. If one wishes to follow the evolution of a nonstationary system in real time, such as in quantum transport problems \cite{CH88, CHR00} (e.g., in relativistic heavy ion collision experiments \cite{RHIC,qgp}),  the asymptotic out-state cannot be stipulated at will, because there are no asymptotic out regions at all moments in the evolution that such `out' states can be well-defined. The quantum state at any moment of time $t > t_0$  is determined by solving the dynamics.
Forcing an in-out treatment on an evolutionary problem will result in equations of motion which are not real nor causal, see, e.g, \cite{Jordan,CalHu87}.
For such problems, one needs to use the in-in (Schwinger-Keldysh or closed-time-path) formalism \cite{Sch61,Kel64,Chou,CalHu88}.   
The in-in formalism is also a natural framework for treating nonequilibrium quantum processes \cite{Berges,CalHu08}, where dissipative mechanisms acquire their real-time physical meanings in a truly statistical mechanical sense (versus identifying the imaginary parts in the transition amplitudes in the in-out formulation).  We mention  radiation reaction,  such as in the derivation of the Lorentz-Abraham-Dirac or the Mino-Sasaki-Tanaka-Quinn-Wald  equations \cite{MST,QuinnWald,JH02,GalHu05,GHL06} and quantum processes in the early universe using the in-in formulation \cite{HuVer20} such as cosmological particle creation \cite{CalHu87,CamVer94}, noise and fluctuations \cite{CalHu94,HuSin95,CamVer96} and entropy generation \cite{Parentani,LCHent,HHent} as the representative main-stream problems.   

In the 90s with the increasing awareness of the role quantum information concepts and methodology may play in helping to address the fundamental issues  of theoretical physics,  two kinds of quantum processes are of special interest in the study of particle-field interaction dynamics, namely, decoherence and entanglement.  Environment-induced decoherence \cite{EID} relies on a better understanding of the effects of noise, and in the case when the environment is a quantum field which a quantum system (like the electron) interacts with, a better understanding of quantum noise.  Quantum entanglement is intimately related to quantum correlations. These provided a strong motive for understanding traditional field theory and particle physics in an open quantum system perspective or for reformulating them in this context. We mention this development because correlation, dissipation and decoherence are indeed the main themes of this work. For a discussion of how these processes, viewed from the nonequilibrium quantum field theory and open quantum system perspective, are rooted in and related to each other, see, e.g., \cite{cddn,CalHu99}.

\subsection{Particle wavepackets for the treatment of decoherence}

Having outlined the different formulations of, and described the new issues in, particle-field interactions, let us remark briefly on how to construct a theory for a relativistic particle in motion which can highlight the features of quantum correlations, dissipation (in this case in the form of radiation reaction) and decoherence with clarity and economy 
\footnote{In the radiation reaction literature (see e.g., \cite{QuinnWald} and references in \cite{JH02,GalHu05,GHL06}) there are considerable discussions of extended objects, with point mass or charge as a limit, and the application of effective field theory \cite{EFT, GH09} in its treatment. Our preference is for conceptual and structural simplicity to facilitate direct comparisons with experiments, such as the decoherence of single electrons in an electron microscope \cite{TE89}.}. 
Here, the charged particles are `point-like' in the spirit of classical electrodynamics \cite{Ro65, Sp99, MST, QuinnWald, JH02, JH05, GalHu05, GHL06}, and the size of the `point' charge corresponds to the minimal width of the quantum mechanical wavepacket of the particle allowed in our effective theory, which would be of the order of the electron Compton wavelength \cite{HHL23}.

Indeed, our effective theory is designed to be valid well below the Schwinger limit (above which particle pairs are produced).  So far the strongest background  EM fields in laboratory are those produced by intense lasers. To our knowledge,  the most powerful lasers currently in the world can at most achieve an intensity of $10^{23}$ ${\rm W}/{\rm cm}^2$ (e.g. \cite{ELINP, SULF}), which is well below the Schwinger limit $~ 10^{29}$ ${\rm W}/{\rm cm}^2$ \cite{Sa31, HE36, Sc51}. This means that the single electrons produced in a contemporary laboratory, even though in relativistic motion, can be treated as a quantum mechanical particle whose wavefunction can have single-particle interpretation safely, and the linear approximation of interaction would be good enough. Note that we are {\it not} describing the collective behavior of an electron beam consisting of a bunch of electrons interacting with each other. Generalization of our present work to a congruence of particles could be useful in addressing similar issues in quantum kinetic theory \cite{CH88}. 

The effective theory in this paper would be equivalent to those describing point particles moving in quantum fields in terms of the worldline influence functional formalism from the semiclassal level to the stochastic limit [e.g. Ref. \cite{JH02, JH05} with the quantum state of the particle's position (the `worldline') initially Gaussian].
We want to see an electron with both particle features -- a localized object, and wave features -- albeit not represented as a plane wave, as is often assumed in textbooks
\footnote{The wavefunction of our quantum mechanical particle is normalizable, unlike the conventional plane waves in scattering problems. The probability is assumed to be conserved in the range of validity of our effective theory.}. 
We can distinguish the EM field-induced {\it decoherence} on the particle, and the intrinsic quantum {\it dispersion} of the wavepacket present even without the field as its environment.

\subsection{Moving electrons and the Unruh effect}

An accelerating single electron interacting with an EM field can be used to address important properties of the Unruh effect \cite{Unr76} in the sense explained below.  In 1976, Unruh showed that a particle or atom undergoing uniform  acceleration in the Minkowski vacuum of a quantum field  will experience vacuum fluctuations with a thermal spectrum. While this effect can be related to the Sokolov-Ternov effect in QED about the spin depolarization in storage rings \cite{BL83, AS07}, further proposals about the motional degrees of freedom of single electrons under high acceleration (e.g. \cite{CT99}) have not offered detailed, complete calculations in QED yet \cite{JH05, SSH08, OYZ16}.

\subsection{Some special features}

The structure of our effective theory bears some resemblance with the Unruh-DeWitt (UD) harmonic-oscillator (HO) detector theory \cite{Unr76, DeW79, LH06, LH07}, with the internal degrees of freedom of a UD HO detector replaced by the {\it deviation} of the charged particle from its classical trajectory determined by the classical Euler-Lagrange equations (rather than put in by hand,  as in many UD detector theories.)
To discern how our effective theory differs from the conventional formulation of QED, we mention a few special features below:

\paragraph{a. Linear and Gaussian approximations}
Since we are working well below the Schwinger limit, the nonlinear QED effect should not be significant. This allows us to consider a linearized quantum theory (although the classical background can be highly nonlinear) with Gaussian wavepackets of the charged particle centered at the classical trajectory, while the vacuum state of EM fields is also Gaussian \cite{Ha92}. To justify our use of the Gaussian approximation  the initial width of the particle wavepacket should be no less than its Compton wavelength divided by the Lorentz factor of its classical motion  \cite{HHL23}. 
By virtue of the linearity of our effective theory  the quantum state of the combined system if initially Gaussian  will always be Gaussian.
Also by virtue of the linearity of our effective theory, our operators after the coupling is switched on can be expanded as linear combinations of the initial operators, each initial operator is associated with a time-dependent amplitude of the mode function. Then we are able to write down the evolution equations for the mode functions, which are independent of quantum states. 

\paragraph{b. Fluctuations, dissipation, and mass renormalization}
In their equations of motion, the particle mode functions are driven by the field mode functions, which can be interpreted as vacuum fluctuations if the quantum field is initially in the Minkowski vacuum state. Then the particle mode function will produce something similar to the self field and radiation reaction in classical electrodynamics. In the point-particle limit, we have to introduce a mass renormalization on the mode functions of particle to absorb the divergence from that self field. Then the equations of motion for the mode functions of particle looks similar to the Lorentz-Abraham-Dirac equation in classical electrodynamics \cite{Lo92a, Lo92b, Ab05, Di38}.

\paragraph{c. Quantum state renormalization}
If we assume the particle state and field state are pure before the particle-field coupling is switched on, then the mass renormalization due to the particle-field coupling would violate the uncertainty relation at very early times. So we have to renormalized the quantum state from free theory in order to keep the uncertainty relation whenever the coupling is on. 

\paragraph{d. Time-dependent regulators}
To deal with the divergences arising in the mode sum of the particle-deviation correlators, we introduce two kinds of the regulators. The first kind of the regulator works like a UV cutoff suppressing the contribution from the field modes of wavelength shorter than the Compton wavelength with length contraction \cite{HHL23}, which will apparently depend on the coordinate time in the laboratory frame if the speed of the particle is varying in time.  
The second kind of the regulators is the cutoff of the coincidence limit corresponding to the resolution of the experiment in terms of the particle's proper time, which can also be time-dependent in the laboratory frame. 
To derive the Unruh effect for a uniformly accelerated charged particle, the values of the second kind regulators must be much greater than the first kind regulator \cite{LH10}.
The values of these regulators are explicitly present in our final result, just like the case of the Lamb shift in QED \cite{Be47}. We set the values of these regulators according to the setting of electron interference experiments in electron microscopes \cite{TE89}. With these values our result looks reasonable.

\subsection{Key findings and organization}

Extracting from the many new results we mention two main findings. One is a direct comparison with observations from established electron microscope experiments, identifying the decoherence of an electron by the EM field it interacts with. The other is a hitherto unknown novel feature in the theoretical structure of acceleration radiation and the Unruh effect.

\paragraph{a. Evidence of decoherence in electron interference experiments}

Comparing our theoretical results with well-established experiments we found that the purity of a single electron in the electron interference experiment described in Ref. \cite{TE89} could be close to 1/2 in the transverse direction when arriving the screen, while the purity was not seriously decreased during the acceleration stage in the early history of the flying electron. Our result suggests that vacuum fluctuations may play a major role
in blurring the interference pattern in Ref. \cite{TE89}.

\paragraph{b. Unruh effect on the motional degrees of freedom} 

We demonstrated that the Unruh effect can naturally be identified in the two-point correlators of the particle's deviation from its classical trajectory in each direction. The calculated purity of the particle-motion deviation in the direction of linear acceleration with the consideration of the Unruh effect decays significantly faster than the results for the particle {\it at rest} with the flying time and regulators replaced by those with time dilation (which corresponds to the condition that the Unruh effect is artificially removed.) If a photoelectron is born with a highly relativistic motion and then stopped by a negative voltage, then in the period of deceleration the electron may behave differently from those with the Unruh effect. Moreover, we found that some terms in the two-point correlators have a Planck factor corresponding to a fermionic bath at the Unruh temperature, rather than a bosonic bath that the other terms  correspond  to. Thus, one cannot trivially apply the Unruh effect to a system by simply introducing a bosonic or fermionic environment.

This paper is organized as follows: In Section \ref{SecElecDyn} we present our effective theory for a combined system of a spinless relativistic charged particle and electromagnetic fields, where we quantize the theory and obtain the counterpart of the LAD equation for the mode functions in our quantum theory. 
In Section \ref{SecRestParticle}, we apply our theory to the case of a single electron at rest. We solve the mode functions and calculate the regularized symmetric two-point correlators of the particle, with which we obtain the purity of the particle's reduced state and compare our result with the electron-interference experiment in Ref. \cite{TE89}.
Then we apply our theory to the case of a charged particle accelerated in uniform electric fields in Section \ref{SecUACinUE}. We show how the Unruh effect emerges in the correlators, and compare our result on quantum decoherence of single electrons with those in the acceleration stage in Ref. \cite{TE89}. Finally, we summarize our findings in Section \ref{SecSumOut}.

\section{Relativistic particle wavepacket interacting with quantum fields}
\label{SecElecDyn}

Consider a relativistic particle of mass $m$, charge $q$, moving along a worldline $z^\mu$, interacting with EM fields $A^\mu$. The dynamics of the combined system is described by the action $S=S_z + S_I + S_{F}$ with \cite{Ro65}
\begin{eqnarray}
	S_z&=& -m c \int d\tau\, \sqrt{-\frac{dz_\mu}{d\tau}\frac{dz^\mu}{d\tau}} 
	    =  -m c^2 \int dt \sqrt{1 -\frac{1}{c^2}\frac{dz_i}{dt}\frac{dz^i}{dt}} \label{SwithN}\\
	S_I&=&  q\int d^4 x 
	        \int d\tau \frac{dz^\mu}{d\tau} \delta^4\left[x-z(\tau)\right] A_\mu(x) \nonumber\\
	&=&\int dt\left[ q c A_0\big(t, {\bf z}(t)\big) + q\frac{dz^i}{dt}A_i\big(t,{\bf z}(t)\big) \right], \label{Sintt} \\
  S_{F} &=& \int \frac{dt d^3 x}{\mu^{}_0} \left[-\frac{1}{4}F_{\mu\nu}F^{\mu\nu}-\frac{\bar{\alpha}}{2}\left( \partial_\mu A^\mu\right)^2\right].
	  \label{SField}
\end{eqnarray}
in the SI unit \cite{Gr17}. Here, $F_{\mu\nu} = \partial_\mu A_\nu - \partial_\nu A_\mu$ with $A^\mu=(\Phi/c, {\bf A})$ 
where $\Phi$ and ${\bf A}$ are the `scalar' and vector potentials, respectively, in the non-relativistic expressions of electromagnetism,
 $\bar{\alpha}$ is an arbitrary constant, $\mu^{}_0 = 4\pi \times 10^{-7} \, {\rm N}\cdot{\rm A}^{-2}$ is the vacuum permeability, and we are working in Minkowski spacetime with metric
$\eta_{\mu\nu}={\rm diag}\{-1,1,1,1\}$.
$S_F$ is the action of free EM fields   constrained by the Lorenz gauge $\partial_\mu A^\mu=0$. 
$S_z+S_I$ is form-invariant under a reparametrization of the proper time $\tau$  of the particle, $\tau \to \tau' = \tau'(\tau)$, implying that the number of the degrees of freedom of $z^\mu(\tau)$ is $4-1=3$. To fix this reparametrization freedom, we choose the Minkowski-time gauge $z^0=ct$ for the particle and keep $ {\bf z} = z^{i(=1,2,3)}$ as dynamical variables.

Let $z^i(t) = \bar{z}^i(t) + \tilde{z}^i(t)$ and $A^\mu(t,{\bf x}) = \bar{A}^\mu(t,{\bf x}) + \tilde{A}^\mu(t,{\bf x})$, where $\bar{z}^i(t)$ and $\bar{A}^\mu(t, {\bf x})$ are the classical solutions satisfying the Euler-Lagrange equations $\delta S/\delta z^i(t) = 0$ and $\delta S/\delta A^\mu_{\bf x}(t) = 0$, where $A^\mu_{\bf x}(t) \equiv A^\mu(t,{\bf x})$.
Later we will only quantize the deviations from classical solutions, namely, $\tilde{z}^i$ and $\tilde{A}^\mu$. 
Recall that the vacuum state of EM fields is Gaussian, and in Ref. \cite{HHL23}, 
one can see that the charge density of a Klein-Gordon wavepacket initially Gaussian with the width greater than $O(\lambda^{}_C/\bar{\gamma})$,
where $\lambda^{}_C$ is its Compton wavelength and $\bar{\gamma}$ is the Lorentz factor of its classical motion,
will be approximately a Gaussian function centered around its classical trajectory for a long time. Thus, it is justified to assume here that in the regime of our interest the quantum states of $\tilde{z}^i$ and $\tilde{A}^\mu$ are Gaussian, and the evolution of the combined system can be approximately described by the action expanded about the classical solutions up to the quadratic order,
\begin{equation}
S  \approx S[\bar{z},\bar{A}] + \int dt\, L_2 \label{Stot}
\end{equation}
where
\begin{eqnarray}
\int dt\, L_2 &\equiv& 
\frac{1}{2} \int dt\, dt'\, \left\{ \sum_{i,j} 
   \tilde{z}^i(t) \left.\frac{\delta^2 S}{\delta z^i(t) \delta z^j(t')} 
  \right|_{\bar{z},\bar{A}}\tilde{z}^j(t') + \right.\nonumber\\
	&& \left. 2\sum_{i,\mu,{\bf x}} \tilde{z}^i(t) \left. \frac{\delta^2 S}{\delta z^i(t) \delta A^\mu_{\bf x}(t')}
	\right|_{\bar{z},\bar{A}} \tilde{A}^\mu_{\bf x}(t') +
	\sum_{\mu,{\bf x},\nu,{\bf y}} \tilde{A}^\nu_{\bf y}(t) \left. \frac{\delta^2 S}{\delta A^\nu_{\bf y}(t) \delta A^\mu_{\bf x}(t')}
	\right|_{\bar{z},\bar{A}} \tilde{A}^\mu_{\bf x}(t')\right\} \nonumber\\
 &=&\int dt \frac{m}{2}\bar{\gamma}(t) \bar{M}_{ij}(t)\dot{\tilde{z}}^i(t) \dot{\tilde{z}}^j(t) + \int \frac{dt}{\mu^{}_0} \sum_{\bf x}
  \left[ -\frac{1}{4}\tilde{F}_{\mu\nu}\tilde{F}^{\mu\nu}(x) - \frac{\bar{\alpha}}{2}\left( \partial_\mu \tilde{A}^\mu(x)\right)^2\right]+ 
	\nonumber\\
&& q\int dt\sum_{\bf x}\left[ 
    \bar{v}^\mu\tilde{z}^i\partial_i \tilde{A}_\mu^{\bf x}  
  + \dot{\tilde{z}}^i \tilde{A}_i^{\bf x}
	+ \dot{\tilde{z}}^j\tilde{z}^i \partial_i \bar{A}_j^{\bf x} 
	+ \frac{\bar{v}^\mu}{2} 
	  \tilde{z}^i \tilde{z}^j \partial_i\partial_j \bar{A}_\mu^{\bf x}
\right]\delta^3({\bf x}-\bar{\bf z}) 
\label{L2}
\end{eqnarray} 
with the linear terms of $\tilde{z}$ and $\tilde{A}^\mu$ vanishing after introducing the Euler-Lagrangian equations. Here we have chosen $z^0=ct$ and so $\bar{v}^\mu \equiv (c, \bar{v}^j) = (c, \partial_t \bar{z}^j)$, $\dot{\tilde{z}}^i\equiv \partial_t \tilde{z}^i$, $\tilde{F}_{\mu\nu}\equiv \partial_\mu \tilde{A}_\nu - \partial_\nu \tilde{A}_\mu$, $\sum_{\bf x} \equiv \int d^3 x$, $\bar{\gamma} \equiv \left(1-\frac{\bar{v}_i \bar{v}^i}{c^2}\right)^{-\frac{1}{2}}$, and
\begin{equation}
\bar{M}_{ij}(t) \equiv \eta_{ij} + \frac{\bar{\gamma}^2}{c^2} \bar{v}_i \bar{v}_j. \label{Mijdef}
\end{equation}
The structure of (\ref{L2}) is similar to the Unruh-DeWitt detector theory \cite{Unr76, DeW79}, with the internal degree of freedom of the detector replaced by the particle-motion deviations $\tilde{z}^i(t)$ or their derivatives here.

\subsection{Radiation reaction at the classical level}
\label{SecCED}

The Euler-Lagrange equations at the classical level read
\begin{eqnarray}
  && \partial_t\Big[ m \bar{\gamma} \bar{v}_i(t) \Big] = 
	   q\bar{F}_{i\mu}\big(\bar{z}(t)\big) \bar{v}^\mu(t) , \label{ELcharge} \\
  && \partial_\nu \bar{F}^{\nu\mu}(x) +\bar{\alpha} \partial^\mu\partial_\nu \bar{A}^\nu(x) 
	   = - \mu^{}_0 q 
		     \bar{v}^\mu(t) \delta^3 ({\bf x}-\bar{\bf z}(t)). \label{ELEM}
\end{eqnarray}
In the Lorentz gauge $\partial_\nu \bar{A}^\nu(x)=0$, 
Eq. (\ref{ELEM}) reduces to
\begin{equation}
  \Box \bar{A}^\mu_{\bf x}(t) = -\mu_0 q 
	  \bar{v}^\mu(t) \delta^3({\bf x}-\bar{\bf z}(t)), 
		\label{ELEMFeyn}
\end{equation}
with $\Box \equiv \partial^\nu\partial_\nu = -\frac{1}{c^{2}}\partial_t^2 + \nabla^2$. One immediately obtains a solution
\begin{equation}
  \bar{A}^\mu_{\bf x}(t) =  \bar{A}^\mu_{[0]{\bf x}}(t) +\bar{A}^\mu_{[1]{\bf x}}(t) 
	\label{barAsol}
\end{equation}
where $\bar{A}^\mu_{[0]{\bf x}}(t)$ is the external field satisfying $\Box\bar{A}^\mu_{[0]{\bf x}}=0$, assumed to be regular around the worldline of the charged particle, and $\bar{A}^\mu_{[1]{\bf x}}(t)$ is the retarded field sourced from the charged particle, namely,
\begin{equation}
 \bar{A}^\mu_{[1]{\bf x}}(t) =\int_{-\infty}^{\infty} c\,
    dt'\int d^3 x' 
    G_{ret}(t,{\bf x}; t', {\bf x}') \mu^{}_0 q \bar{v}^\mu(t')\delta^3({\bf x}'-\bar{\bf z}(t'))
\label{barA1Gret}
\end{equation}
with the retarded Green's function defined by
\begin{equation}
  \left( \frac{1}{c^2}\partial_t^2 - \nabla^2\right)G_{ret}(t,{\bf x}; t', {\bf x}') = \delta(c t-c t')\delta^3({\bf x}-{\bf x}'),
\end{equation}
or formally,
\begin{equation}
  G_{ret}(t,{\bf x}; t', {\bf x}') = \frac{1}{4\pi}\delta(\sigma)\theta(t-t') \label{Gretdef}
\end{equation}
with the Synge's world function $\sigma(x,x')\equiv -\frac{1}{2} (x_\mu-x'_\mu)(x^\mu-x'^\mu)$\cite{BD82}. The above retarded Green's function diverges as $x'\to x$, implying that the self field established by the charged particle diverges around the particle as the particle size goes to zero. Following the same method as in \cite{JH02,LH06}, we regularize the retarded Green's function by a UV cutoff $\Lambda$ as
\begin{equation}
   G^\Lambda_{ret}(t,{\bf x}; t', {\bf x}')\equiv \frac{1}{4\pi}\sqrt{\frac{8}{\pi}}\Lambda^2 e^{-2\Lambda^4\sigma^2(x,x')}
	   \theta(t-t')\theta(\sigma), \label{GretL}
\end{equation}
which approaches $G_{ret}(t,{\bf x}; t', {\bf x}')$ as $\Lambda\to\infty$. 
Replacing the $G_{ret}$ in (\ref{barAsol}) by $G^\Lambda_{ret}$, we get
\begin{equation}
  \bar{A}^\mu_{\bf x}(t) = \bar{A}^\mu_{[0]{\bf x}}(t) + 
	  \mu^{}_0\frac{ qc}{4\pi} \left[ \frac{2^{\frac{7}{4}}\Gamma\left(\frac{5}{4}\right)\bar{\gamma}}{\sqrt{\pi} c} \Lambda 
		\bar{v}^\mu(t) - \frac{\bar{\gamma}^2}{c^2}	w_\nu^\mu \partial_t\bar{v}^\nu(t) \right]
		+ O(\Lambda^{-1}) \label{barALambda}
\end{equation} 
for large $\Lambda$, with
\begin{equation}
  w_\nu^\mu \equiv \delta_\nu^\mu + \frac{\bar{\gamma}^2}{c^2}\bar{v}_\nu \bar{v}^\mu =
	\delta_\nu^\mu + \frac{1}{c^2}\bar{u}_\nu \bar{u}^\mu, 
\end{equation}
implying $w_\nu^\mu \bar{v}_\mu = 0$. In (\ref{barALambda}), there seems to be a $\Lambda^1$ divergence corresponding to the EM potential established by a point charge ($\Phi \sim r^{-1}$) around itself ($r\to 0$). 
For electrons, however, the value of the whole coefficient proportional to $\Lambda^1$ in (\ref{barALambda}) can be very small compared with the electron mass [see discussions below (\ref{LADEqt})]. After inserting (\ref{barAsol}) with (\ref{barA1Gret}) and (\ref{GretL}) into (\ref{ELcharge}), a similar $\Lambda^1$ term arises in $q\bar{F}_{i\mu}$ with the coefficient 
\begin{equation}
  \Delta_m \equiv \frac{\mu^{}_0 q^2 2^{\frac{3}{4}}\Gamma\left(\frac{5}{4}\right)}{4\pi\sqrt{\pi}}\Lambda, \label{deltam}
\end{equation}
which can be absorbed by the mass of the charged particle. Then we obtain the Lorentz-Abraham-Dirac (LAD) equation \cite{Lo92a, Lo92b, Ab05, Di38},
\begin{equation}
  \bar{m} \partial_t\Big[  \bar{\gamma} \bar{v}_i(t) \Big] = q \bar{F}^{[0]}_{i\mu}
	  \big(\bar{z}(t)\big)\bar{v}^\mu(t)  + \bar{\Gamma}_i(t) 
		+ O( \Lambda^{-1}), \label{LADEqt}
\end{equation} 
where 
the bare mass $m$ has been considered as a constant of time,
\begin{equation}
  \bar{m} \equiv m + \Delta_m, \label{mren1}
\end{equation}
is the classical renormalized (or field-corrected) mass with contributions from the self field proportional to $\Lambda$, and $\bar{F}^{[0]}_{\mu \nu} \equiv\partial_\mu \bar{A}^{[0]}_\nu - \partial_\nu \bar{A}^{[0]}_\mu$ is the external EM field strengths. For single electrons, however, the value of 
$\Delta_m$ can be very small compared with the electron mass if we introduce a finite $\Lambda$ [see discussions at the end of this section].

In Eq.(\ref{LADEqt}), the radiation reaction force of order $\Lambda^0$ (the LAD force) reads
\begin{eqnarray}
\bar{\Gamma}_i &=&  \mu^{}_0 \frac{q^2 \bar{\gamma}^2}{4\pi c}
   \left[  \frac{2}{3} w_i^\mu \ddot{\bar{v}}_\mu
	 +2\frac{\bar{\gamma}^2}{c^2} \bar{v}^\rho \dot{\bar{v}}_\rho w_i^\mu \dot{\bar{v}}_\mu \right]\nonumber\\
&=& \frac{q^2 \mu^{}_0}{4\pi c}  \frac{2}{3} w_i^\mu \partial_t\big[ \bar{\gamma}	\partial_t \left(\bar{\gamma}\bar{v}_\mu\right)\big].
\end{eqnarray}
$\bar{\Gamma}_i$ contains the third derivative of particle position, $\partial_t\partial_t\bar{v}_i(t) = \dddot{\bar{z}}_i(t)$, which is contributed by the $\Lambda^6 (\Delta t)^5$ and $\Lambda^{10} (\Delta t)^9$ terms in the integral $\int_{-\infty}^t dt' = \int_0^\infty d\Delta t$ in (\ref{barA1Gret}) with $\Delta t = t-t'$. 

Up to the LAD force $\bar{\Gamma}_i$, Eq.(\ref{LADEqt}) can be directly derived from the LAD equation parametrized in proper time $\tau$ of the charged particle,
\begin{equation}
  \bar{m} \partial_\tau \bar{u}_\nu(\tau) = q\bar{F}^{[0]}_{\nu\mu}\left(\bar{z}\right)\bar{u}^\mu  +\bar{\sf \Gamma}_\nu	+O(\Lambda^{-1})
\label{LADEqTau}
\end{equation}
with the four-velocity $\bar{u}^\mu(\tau)\equiv \partial_\tau \bar{z}^\mu(\tau)$, by simply replacing $\partial_\tau$ by $\bar{\gamma}\partial_t$. Here,
\begin{eqnarray}
\bar{\sf \Gamma}_\nu &=& \frac{q^2\mu^{}_0}{4\pi c}  \frac{2}{3} 
	w_\nu^\mu \partial_\tau^2 \bar{u}_\mu  
= \frac{q^2 \mu^{}_0}{4\pi c}  \frac{2}{3} w_\nu^\mu \bar{\gamma} \partial_t\left[ \bar{\gamma}
	\partial_t \left(\bar{\gamma}\bar{v}_\mu\right)\right] .
\end{eqnarray}
Eq. (\ref{LADEqTau}) can also be obtained by the same regularization process for (\ref{LADEqt}). When comparing the above results parametrized in $t$ and $\tau$, we have used 
\begin{equation}
  \Delta \tau = \frac{\Delta t}{\bar{\gamma}}\left[ 1 + \frac{\Delta t}{2} \frac{\bar{\gamma}^2}{c^2} \dot{\bar{v}}_\rho \bar{v}^\rho
	-\frac{\Delta t^2}{3!} \left(\frac{\bar{\gamma}^4}{c^4} (\dot{\bar{v}}_\rho \bar{v}^\rho)^2 + 
	\frac{\bar{\gamma}^2}{c^2}(\dot{\bar{v}}_\rho \dot{\bar{v}}^\rho + \ddot{\bar{v}}_\rho \bar{v}^\rho) \right) + O(\Delta t^3)  \right] \label{dTauvsdt}
\end{equation}
with $\Delta\tau \equiv \tau-\tau'$ and $\Delta t \equiv t-t'$, and
\begin{eqnarray} 
&&-\int_{-\infty}^\tau d\tau' = \int_{-\infty}^0 d \Delta \tau = \int_{-\infty}^0 \frac{d\Delta t}{\bar{\gamma}(t')}= 
\int_{-\infty}^0 \frac{d\Delta t}{\bar{\gamma}(t)} \Big[ 1+ \nonumber\\ && \hspace{.5cm} \left. \Delta t \,\,
	\frac{\bar{\gamma}^2}{c^2} \dot{\bar{v}}_\rho \bar{v}^\rho - 
	\frac{\Delta t^2}{2} \left( \frac{\bar{\gamma}^4}{c^4} (\dot{\bar{v}}_\rho \bar{v}^\rho)^2 + 
	\frac{\bar{\gamma}^2}{c^2}(\dot{\bar{v}}_\rho \dot{\bar{v}}^\rho + \ddot{\bar{v}}_\rho \bar{v}^\rho) \right) + O(\Delta t^3) \right],
\end{eqnarray} 
where, without specifying the argument, $\bar{\gamma} = \bar{\gamma}(t)$ is understood.
We find that the ignored radiation reaction force of order $\Lambda^{-1}$ in (\ref{LADEqt}) has an extra term to the one in (\ref{LADEqTau}). 
Nevertheless, from the order of $\Lambda^{-1}$ on, the result is expected to depend on the shape or charge distribution of the particle if the particle has been regularized to an extended object, and also on the form of the regularized retarded Green's function of the field \cite{Sp99}. Such a dependence on regularization scheme is negligible in the regime of our interest, and should not be considered in our effective theory.

Recall that the action for relativistic particles $S_z$ in (\ref{SwithN}) is time-reparametrization invariant. We have {\it chosen} the parameter as the Minkowski time in (\ref{SwithN}) (the Minkowski-time gauge). 
For what we want here the proper-time gauge is not more convenient than the Minkowski-time gauge. Indeed, to quantize a theory for relativistic particles in terms of proper time, one needs to go through the BRST quantization procedure \cite{HT92}, which introduces ghosts and other complications.

By treating the charged particle quantum-mechanically in our effective theory, we have implicitly assumed that the probability of finding the charged particle in the Universe is conserved, i.e., there are no new particles created or any existing particles annihilated. We expect that our effective theory would break down beyond the charged particle production scale.  {Note that in our regularized retarded Green's function (\ref{GretL}), we have $e^{-2\Lambda^4\sigma^2(x,x')} \approx e^{-\Lambda^4 (c\Delta\tau)^4/2}[1+O(\Delta\tau^5)]$, where $\Lambda^{-1}$ sets the scale of proper length $c\Delta\tau$. Thus we choose the UV cutoff $\Lambda$ as the reciprocal of the {\it proper Compton wavelength} $\lambda^{}_C= h/\bar{m}c$ of the particle, which is reference frame independent. 
For an electron, 
$q = 1.6\times 10^{-19}$ C, $\bar{m}=9.1 \times 10^{-31}$ kg, 
$\lambda^{}_C\approx 2.4 \times 10^{-12}$ m, and so $\Delta_m \approx 9.1\times 10^{-34}$ kg $\sim 10^{-3}\bar{m}$ in (\ref{deltam}), which is a perturbative correction.}
Anyway, only the renormalized mass $\bar{m}$ is supposed to be physically measurable, while the bare mass $m$ and field correction $\Delta_m$ are not.

\subsection{Quantization and counterpart of LAD force at the quantum level}
\label{SecQLAD}

From (\ref{Stot}), the conjugate momenta of the deviations $\tilde{z}^i$ and $\tilde{A}_{\bf x}^\mu$ read
\begin{eqnarray}
  \tilde{p}_i &=& \frac{\delta S}{\delta\partial_t\tilde{z}^i} = m\bar{\gamma}\bar{M}_{ij}\dot{\tilde{z}}^j + 
	    q\left(\tilde{A}_i^{\bf z}+\tilde{z}^j\partial_j \bar{A}_i^{\bar{\bf z}}\right), \label{pi2zi} \\
	\tilde{\pi}^i_{\bf x} &=& \frac{\delta S}{\delta \partial_t \tilde{A}^{\bf x}_i} = 
	    \frac{1}{\mu^{}_0 c}\tilde{F}^{i0}_{\bf x}, \label{pii2Ai} \\ 
	\tilde{\pi}^0_{\bf x} &=& \frac{\delta S}{\delta \partial_t \tilde{A}^{\bf x}_0} =
	    \frac{\bar{\alpha}}{\mu^{}_0 c}\partial_\mu \tilde{A}^\mu_{\bf x}, \label{pi02A0}
\end{eqnarray}
and the Hamiltonian of the quadratic part of $S$ defined on the $t$-slice is given by
\begin{eqnarray}
  \tilde{H}_2 &=& \tilde{p}_i \dot{\tilde{z}}^i + c\sum_{\bf x} \left( \tilde{\pi}^i_{\bf x} \partial^{}_0 \tilde{A}_i^{\bf x}
	  + \tilde{\pi}^0_{\bf x} \partial^{}_0 \tilde{A}_0^{\bf x} \right) - L_2 \nonumber\\
	&=& \frac{\bar{M}^{ij}}{2m\bar{\gamma}} 
	\left[ \tilde{p}_i - q\left(\tilde{A}_i^{\bar{\bf z}}+\tilde{z}^k \partial_k \bar{A}_i^{\bar{\bf z}}\right) \right]
	\left[ \tilde{p}_j - q\left(\tilde{A}_j^{\bar{\bf z}}+\tilde{z}^l \partial_l \bar{A}_j^{\bar{\bf z}} \right)\right] \nonumber\\
	&& +\sum_{\bf x} \left\{ \frac{\mu^{}_0 c^2}{2} \tilde{\pi}_i^{\bf x}\tilde{\pi}^i_{\bf x}
	   -\frac{\mu^{}_0 c^2}{2\bar{\alpha}} \left(\tilde{\pi}^0_{\bf x}\right)^2 
	   + c\tilde{\pi}^i_{\bf x}\partial_i \tilde{A}_0^{\bf x}  + c\tilde{\pi}^0_{\bf x}\partial_i \tilde{A}^i_{\bf x} + 
			\frac{1}{4\mu^{}_0}\tilde{F}_{ij}^{\bf x}\tilde{F}^{ij}_{\bf x} \right\}  \nonumber\\
	&& -q\left( \bar{v}^\mu\tilde{z}^i \partial_i \tilde{A}_\mu^{\bar{\bf z}} +
	  \frac{\bar{v}^\mu}{2} \tilde{z}^i\tilde{z}^j \partial_i\partial_j \bar{A}_\mu^{\bar{\bf z}}\right), \label{Hamil2}
\end{eqnarray} 
where $\bar{M}^{ij}$ is the inverse matrix of $\bar{M}_{ij}$ in (\ref{Mijdef}), defined by $\bar{M}^{ij}\bar{M}_{jk}=\delta^i_k$. 

In our linearized theory described by the above quadratic Hamiltonian $\hat{H}_2$ [or equivalently, the quadratic action 
$S_2\equiv\int dt L_2$ in (\ref{L2})], the Hamilton equations (or the Euler-Lagrange equations) yield
\begin{eqnarray}
&& \partial_t\left( m\bar{\gamma} \bar{M}_{ij}\dot{\tilde{z}}^j \right) =
    q\left[\tilde{F}_{i\mu}^{\bar{\bf z}} \bar{v}^\mu + \left(\tilde{z}^j\partial_j \bar{F}_{i\mu}^{\bar{\bf z}}\right)\bar{v}^\mu 
	+\bar{F}_{ij}^{\bar{\bf z}}\dot{\tilde{z}}^j\right],\label{EOMParticle}\\
&& \partial_\mu \tilde{F}^{\mu \nu}_{\bf x} +\bar{\alpha} \partial^\nu \partial_\mu \tilde{A}^\mu_{\bf x} =
  - \mu^{}_0 q \tilde{V}^\nu \delta^3({\bf x}-\bar{\bf z}), \label{EOMField}
\end{eqnarray} 
where $\tilde{V}^\nu \equiv ( -c\tilde{z}^j \partial_j, \, \dot{\tilde{z}}^i -\bar{v}^i \tilde{z}^j \partial_j)$.

We quantize our linearized theory by promoting the perturbative variables (deviations from their classical values) $\tilde{z}^i$ and $\tilde{A}^i_{\bf x}$  to the operators $\hat{z}^i$ and $\hat{A}^i_{\bf x}$, and introducing the equal-time commutation relations
\begin{equation}
  [ \hat{z}^i, \hat{p}_j ] = i\hbar \delta^i_j,  \label{zpCR}
\end{equation}
and
\begin{equation}
	[ \hat{A}^\mu_{\bf x}, \hat{\pi}^\nu_{\bf y} ] = i\hbar \eta^{\mu\nu} \delta^3({\bf x}-{\bf y}). \label{ApiCR}
\end{equation}
The evolution of the system is governed by the quadratic Hamiltonian $\hat{H}_2$ with all the deviations $\tilde{\cal O}$ in (\ref{Hamil2}) promoted to the operators $\hat{\cal O}$. Since our effective theory is linear, the Heisenberg equations for the operators have the same form as Eqs. (\ref{EOMParticle}) and (\ref{EOMField}) while the deviations there are replaced by the operators.

We assume the charged particle and EM fields are not coupled until the moment $t=t_0$.
Then for $t\le t_0$, we have $q=0$, and Eqs. (\ref{pi2zi}), (\ref{EOMParticle}), and (\ref{EOMField}) give
\begin{equation}
 \hat{p}_i = m\bar{\gamma}\bar{M}_{ij}\dot{\hat{z}}^j, \hspace{1cm} 
   \dot{\hat{p}}^j = 0, \label{HEfreeParticle}
\end{equation}
\begin{equation}
\partial_\mu \hat{F}^{\mu \nu}_{\bf x} + \bar{\alpha} \partial^\nu \partial_\mu \hat{A}^\mu_{\bf x}=0,\label{HEfreeField}
\end{equation}
where $\hat{z}^j$ and $\hat{A}^\mu_{\bf x}$ are operators.
Thus, before the coupling is switched on, $\hat{p}_i$ are independent of time, and the time dependence of the deviation from the classical worldline of the particle,
\begin{equation}
  \hat{z}^i(t) = \hat{z}^i(\bar{t}_0)+\hat{p}_j\int_{\bar{t}_0}^t d\tilde{t}\,\frac{\bar{M}^{ij}(\tilde{t}\,)}{m\bar{\gamma}(\tilde{t}\,)}
	\label{ziFree}
\end{equation}
with a constant $\bar{t}_0\le t_0$, is purely from the coefficients of $\hat{p}_j$. 

As for EM fields, following the canonical quantization in the Lorentz-Feynman gauge ($\bar{\alpha}=1$) \cite{IZ80}, we write
\begin{equation}
	\hat{A}^\mu_{[0]{\bf x}}(t) = \sum_{\bf k}\sum_{\lambda=0}^3 \left[
	  \epsilon^\mu_{(\lambda){\bf k}}e^{-i \omega t + i {\bf k}\cdot {\bf x}} \, \hat{b}^{(\lambda)}_{\bf k} + 
		\epsilon^{\mu*}_{(\lambda){\bf k}}e^{i \omega t - i {\bf k}\cdot {\bf x}}\,\hat{b}^{(\lambda)\dagger}_{\bf k}\right] \label{AmuFree}
\end{equation}
for $t\le t_0$, with $\omega=|{\bf k}| c$, 
\begin{equation} 
  \sum_{\bf k} \equiv \int \frac{d^3 k}{(2\pi)^3}\sqrt{\frac{\hbar}{2\omega\varepsilon^{}_0}}  \label{Sumkdef}
\end{equation} 
with vacuum permittivity $\varepsilon^{}_0 = (\mu^{}_0 c^2)^{-1} \approx 8.85\times 10^{-12}\,{\rm C}^2\cdot {\rm N}^{-1}\cdot {\rm m}^{-2}$ 
and
\begin{equation}
	[ \hat{b}^{(\lambda)}_{\bf k}, \hat{b}^{(\lambda')\dagger}_{\bf k'} ] = (2\pi)^3 
	\eta^{(\lambda)(\lambda')}\delta^3({\bf k}-{\bf k'}). \label{bbcommutator}
\end{equation}
Here $\epsilon^\mu_{(0){\bf k}}=\delta^\mu_0$ is the unit temporal vector in the $t$ direction, $\epsilon^\mu_{(3){\bf k}} =(0, {\bf k}/|{\bf k}|)$ is the longitudinal unit vector, and $\epsilon^\mu_{(1){\bf k}}$ and $\epsilon^\mu_{(2){\bf k}}$ are the two orthonormal polarization vectors perpendicular to $\epsilon^\mu_{(3){\bf k}}$ and $\epsilon^\mu_{(0){\bf k}}$. 
Following the Gupta-Bleuler formalism, we will adopt the physical states $|\psi_{phys}\rangle$ of EM fields in the Lorentz gauge as those satisfying $\left(\hat{b}^{(0)}_{\bf k}-\hat{b}^{(3)}_{\bf k}\right)|\psi_{phys}\rangle = 0$ \cite{IZ80}. Note that the Minkowski vacuum state $|0^{}_M\rangle$ of EM fields is a physical state. 

For $t>t_0$, the particle-field interaction has been switched on. One can write
\begin{eqnarray}
 	\hat{z}^i(t) &=& \sum_{j=1}^3 \left[ {\cal Z}^i_{z^j_{}}(t)\hat{z}^j +{\cal Z}^i_{p^{}_j}(t) \hat{p}^{}_j \right]+ 
		\sum_{\bf k}\sum_{\lambda=0}^3 \left[{\cal Z}^{i}_{(\lambda){\bf k}}(t)\hat{b}^{(\lambda)}_{\bf k} + 
		{\cal Z}^{i *}_{(\lambda) {\bf k}}(t)\hat{b}^{(\lambda)\dagger}_{\bf k}\right], \label{ziInt} \\
  \hat{A}^\mu_{\bf x}(t) &=& 
	  \sum_{j=1}^3 \left[ {\cal A}^\mu_{z^j}(t,{\bf x})\hat{z}^j +{\cal A}^\mu_{p_j}(t,{\bf x}) \hat{p}^{}_j \right]+
	  \nonumber\\ && \hspace{4cm}
		\sum_{\bf k}\sum_{\lambda=0}^3 \left[{\cal A}^\mu_{(\lambda){\bf k}}(t,{\bf x})\hat{b}^{(\lambda)}_{\bf k} 
		+ {\cal A}^{\mu *}_{(\lambda){\bf k}}(t,{\bf x})\hat{b}^{(\lambda)\dagger}_{\bf k}\right],\label{AmuInt}
\end{eqnarray}
and the conjugate momenta
\begin{eqnarray}
  \hat{p}^{}_i(t) &=& m\bar{\gamma}\bar{M}_{ij}(t)\dot{\hat{z}}^j(t) + 
	                 q\left[\hat{A}_i^{\bar{\bf z}(t)}(t)+\hat{z}^j\partial_j \bar{A}_i^{\bar{\bf z}(t)}(t)\right], \label{hatpi2zi} \\
	\hat{\pi}^i_{\bf x}(t) &=& \frac{1}{\mu^{}_0 c}\hat{F}^{i0}_{\bf x}(t), \\ 
	\hat{\pi}^0_{\bf x}(t) &=& \frac{\bar{\alpha}}{\mu^{}_0 c}\partial_\mu \hat{A}^\mu_{\bf x}(t), 
\end{eqnarray}
from (\ref{pi2zi})-(\ref{pi02A0}).
Here and below, $\hat{z}^j \equiv \hat{z}^j(\bar{t}_0)$, $\hat{p}^{}_i\equiv \hat{p}^{}_i(\bar{t}_0)$ and $\hat{b}^{(\lambda)}_{\bf k}\equiv \hat{b}^{(\lambda)}_{\bf k}(\bar{t}_0)$ for some $\bar{t}_0\le t_0$ [see (\ref{ziFree})] are understood, and we set $\bar{t}_0 \to t_0-$ for simplicity.

Compare (\ref{ziInt}) and (\ref{AmuInt}) with the free operators (\ref{ziFree}) and (\ref{AmuFree}), one can read off
the free mode functions as
\begin{equation}
   {\cal Z}^{i}_{[0]z_{}^j}(t) = \delta^i_j, \hspace{1cm}
   {\cal Z}^{i}_{[0]p^{}_j}(t) = \int_{\bar{t}_0}^t d\tilde{t} \, \frac{\bar{M}^{ij}(\tilde{t}\,)}{m\bar{\gamma}(\tilde{t}\,)}\hspace{1cm}
	 {\cal Z}^{i}_{[0](\lambda){\bf k}}(t) = 0, \label{freeModeZ}
\end{equation}
and
\begin{equation}
	 {\cal A}^\mu_{[0] z_{}^j} = {\cal A}^\mu_{[0]p^{}_j} = 0, \hspace{1cm} {\cal A}^\mu_{[0](\lambda){\bf k}}(t,{\bf x})=
	  \epsilon^\mu_{(\lambda){\bf k}} e^{-i\omega t+i{\bf k}\cdot{\bf x}}. \label{freeModeA}
\end{equation}
Before the coupling is switched on ($t\le t_0$), one has ${\cal Z}^i_{\Omega}(t) = {\cal Z}^{i}_{[0]\Omega}(t)$ and 
${\cal A}^\mu_{\Omega}(t,{\bf x}) = {\cal A}^\mu_{[0]\Omega}(t,{\bf x})$ labeled by the collective index $\Omega \equiv ( z^j_{}, p^{}_j, (\lambda){\bf k})$.

Applying the commutation relations (\ref{zpCR}) and (\ref{ApiCR}) to the Heisenberg equations, one can see that the evolution equations for the mode functions are again in the same form as the Euler-Lagrange equations of the deviations Eqs. (\ref{EOMParticle}) and (\ref{EOMField})
after the coupling is switched on,
namely,
\begin{eqnarray}
&& \partial_t\left( m\bar{\gamma} \bar{M}_{ij}\dot{\cal Z}^j_{\Omega} \right) = q 
	\Big[{\cal F}_{\Omega i\mu}^{\bar{\bf z}}\bar{v}^\mu + 
	{\cal Z}^j_{\Omega}\,\big(\partial_j \bar{F}_{i\mu}^{\bar{\bf z}}\big)\bar{v}^\mu +
	\bar{F}_{ij}^{\bar{\bf z}} \dot{\cal Z}^j_{\Omega} 
	\Big] \label{EOMParticleMode} \\
&& \partial_\mu {\cal F}^{\mu \nu}_{\Omega} (t,{\bf x}) + \bar{\alpha} \partial^\nu \partial_\mu {\cal A}^\mu_{\Omega} (t,{\bf x})=
  -\mu^{}_0 q {\cal V}^\nu_{\Omega} \delta^3({\bf x}-\bar{\bf z}), \label{EOMFieldMode}
\end{eqnarray}
by virtue of the linearity of our quadratic Hamiltonian $\hat{H}_2$. 
Here ${\cal F}^\Omega_{\mu\nu}= \partial_\mu {\cal A}^\Omega_\nu - \partial_\nu {\cal A}^\Omega_\nu$,
\begin{eqnarray}
&& {\cal V}^0_{\Omega}(t) \equiv -c {\cal Z}^j_{\Omega} \partial_j, \hspace{.5cm}
	{\cal V}^i_{\Omega}(t) \equiv \partial_t {\cal Z}^i_{\Omega} - 
	    \bar{v}^i {\cal Z}^j_{\Omega}  \partial_j. \label{bigV}
\end{eqnarray}
Note that, for simplicity, we assume that the particle-field coupling in the classical equations (\ref{ELcharge}), (\ref{ELEM}), and (\ref{LADEqt}) has been switched on at past infinity without runaways. This would allow us to use the simple classical trajectory of a uniformly accelerated charge in our discussion of the Unruh effect in Section \ref{SecUACinUE}.

In the Lorentz-Feynman gauge $\bar{\alpha}=1$, (\ref{EOMFieldMode}) can be written as 
\begin{equation}
  \Box {\cal A}^\mu_{\Omega}(t,{\bf x}) = -\mu^{}_0 q {\cal V}^\mu_\Omega (t) \delta^3( {\bf x} - \bar{\bf z}(t)), \label{EOMcalA}
\end{equation}
where the four components $\mu=0,1,2,3$ decouple, similar to the classical equation (\ref{ELEMFeyn}). 
Note that here we cannot impose the condition $\partial_\mu {\cal A}^\mu_\Omega = 0$ analogous to the Lorentz gauge $\partial_\mu \bar{A}^\mu= 0$ at the classical level, otherwise from (\ref{AmuInt}) we will have $\partial_\mu \hat{A}^\mu = 0$, which is inconsistent with the commutation relation (\ref{ApiCR}) for $\nu=0$.

Eq. (\ref{EOMcalA}) has a general solution similar to (\ref{barAsol}), reading
\begin{equation}
  {\cal A}^\mu_\Omega (t, {\bf x}) = {\cal A}^\mu_{[0]\Omega}(t,{\bf x})+{\cal A}^\mu_{[1]\Omega}(t,{\bf x}), \label{AmuFull}
\end{equation}
where
\begin{equation}
  {\cal A}^\mu_{[1]\Omega}(t,{\bf x}) =\mu^{}_0 q\int_{t^{}_0}^\infty c\, 
  dt'\int d^3 x'\, G_{ret}(t,{\bf x};t', {\bf x}'){\cal V}^\mu_\Omega(t') 
	  \delta^3( {\bf x}' - \bar{\bf z}(t')). 	\label{AmodeInhom}
\end{equation}
Using the same regularization and renormalization as those in Section \ref{SecCED}, we obtain
\begin{eqnarray}
&& 	\hspace{-1cm}\bar{m}\partial_t\left\{\bar{\gamma}\bar{M}_{ij}\partial_t{\cal Z}^j_\Omega (t) \right\} = 
 \sum_{n=1}^3\frac{\partial\bar{\Gamma}_i}{\partial \left(\partial_t^n \bar{z}^j\right)}\partial_t^n {\cal Z}^j_\Omega + 
\nonumber\\ 
&&	q\left\{ \bar{v}^\mu {\cal F}_{i\mu}^{[0]\Omega}(t,\bar{\bf z}(t)) +\bar{v}^\mu{\cal Z}^j_\Omega \partial_j 
  \bar{F}_{i\mu}^{[0]}(t,\bar{\bf z}) +\bar{F}_{ij}^{[0]}(t,\bar{\bf z})
	    \partial_t  {\cal Z}^j_\Omega\right\} + O(\Lambda^{-1}),
\label{QLADEqt}
\end{eqnarray}
for $t - t^{}_0 \gg (c\Lambda)^{-1}$. Here, 
\begin{eqnarray}
&& \hspace{-1.5cm}\sum_{n=1}^3\frac{\partial\bar{\Gamma}_i}{\partial \left(\partial_t^n \bar{z}^j\right)}\partial_t^n 
	     {\cal Z}^j_\Omega = \mu^{}_0 \frac{q^2\bar{\gamma}^4}{4\pi c^3} \times \nonumber\\
&&\left\{ 
	\frac{2c^2}{3\bar{\gamma}^2}
	\bar{M}_{ij} \partial_t^3{\cal Z}^j_\Omega
	+2\left[ \bar{v}^k \dot{\bar{v}}_k \eta^{}_{ij} + \left(\dot{\bar{v}}_i + 2
	  \frac{\bar{\gamma}^2}{c^2}\bar{v}^k \dot{\bar{v}}_k\bar{v}_i\right)\bar{v}_j\right] 
		\partial_t^2{\cal Z}^j_\Omega \right. \nonumber\\
&& \hspace{.2cm} 
  + 2\left[\frac{1}{3}\bar{v}^k \ddot{\bar{v}}_k\eta^{}_{ij} + \frac{\bar{\gamma}^2}{c^2}\left(\bar{v}^k\dot{\bar{v}}_k\right)^2 \eta^{}_{ij}
		 + \frac{1}{3}\bar{v}_i \ddot{\bar{v}}_j + \frac{2}{3}\ddot{\bar{v}}_i \bar{v}_j + 
		\frac{4\bar{\gamma}^2}{3c^2}\bar{v}^k\ddot{\bar{v}}_k \bar{v}_i\bar{v}_j  \right. \nonumber\\
&& \hspace{.7cm}
  \left. \left.+ \dot{\bar{v}}_i \dot{\bar{v}}_j + \frac{\bar{\gamma}^2}{c^2}\bar{v}^k\dot{\bar{v}}_k\left( 4\dot{\bar{v}}_i \bar{v}_j 
	+ 2\bar{v}_i \dot{\bar{v}}_j \right) +6\left(\frac{\bar{\gamma}^2}{c^2}\bar{v}^k\dot{\bar{v}}_k \right)^2 \bar{v}_i \bar{v}_j \right]
	\partial_t{\cal Z}^j_\Omega \right\},
\label{QLADEqtLong}
\end{eqnarray}
and the renormalized mass $\bar{m}$ is the same as (\ref{mren1})
for the classical LAD equation (\ref{LADEqt}). 

Expanding the canonical momentum of the particle (\ref{hatpi2zi}) in the same way as (\ref{ziInt}), we have
\begin{equation}
 	\hat{p}^i(t) = \sum_{j=1}^3 \left[ {\cal P}^i_{z^j_{}}(t)\hat{z}^j +{\cal P}^i_{p^{}_j}(t) \hat{p}^{}_j \right]+ 
		\sum_{\bf k}\sum_{\lambda=0}^3 
		\left[{\cal P}^{i}_{(\lambda){\bf k}}(t)\hat{b}^{(\lambda)}_{\bf k} + 
		{\cal P}^{i *}_{(\lambda) {\bf k}}(t)\hat{b}^{(\lambda)\dagger}_{\bf k}\right]. \label{piInt}
\end{equation}
Before the coupling is switched on, one has 
\begin{equation}
   {\cal P}^{i}_{[0]z_{}^j}(t) = 0, \hspace{1cm}
   {\cal P}^{i}_{[0]p^{}_j}(t) = \eta^{ij}, 
	 \hspace{1cm} {\cal P}^{i}_{[0](\lambda){\bf k}}(t) = 0, 
	 \label{freeModeP}
\end{equation}
from (\ref{freeModeZ}) and (\ref{HEfreeParticle}). When $t>t^{}_0$, the mode functions ${\cal P}^i_\Omega$ become
\begin{eqnarray}
&& \hspace{-1cm} {\cal P}_i^\Omega(t) = m\bar{\gamma}\bar{M}_{ij}(t)\dot{\cal Z}^j_\Omega(t) + q \Big[ 
  {\cal A}_i^\Omega(t, \bar{\bf z}(t))+ {\cal Z}^j_\Omega(t)\partial_j \bar{A}_i^{\bar{\bf z}}(t)\Big] \nonumber\\ 
&=& \bar{\gamma}\bar{M}_{ij} \bar{m}' \partial_t {\cal Z}^j_\Omega 
  +q \left[ {\cal A}_i^{[0]\Omega}(t,\bar{\bf z}(t))+ 
   {\cal Z}^j_\Omega(t)\partial_j \bar{A}_i^{[0]\bar{\bf z}}(t)\right]   \nonumber\\		
&&  - \mu^{}_0 \frac{q^2 c}{4\pi} \left[ \partial_t\left( \frac{\bar{\gamma}^2}{c^2}\bar{M}_{ij}\partial_t{\cal Z}^j_\Omega\right) + 
  \frac{\bar{\gamma}^4}{c^4}\dot{\bar{v}}_i \bar{v}_j\partial_t{\cal Z}^j_\Omega\right]  + O(\Lambda^{-1})
\label{modeFnP}
\end{eqnarray} 
with the self fields of ${\cal A}^i_\Omega(t,{\bf x})$ and ${\bar A}^i_{\bf x}(t)$
around the position of the charged particle ${\bf x} \to \bar{\bf z}$ from (\ref{AmodeInhom}) and (\ref{barALambda}), respectively, inserted. 
Here
\begin{equation}
  \bar{m}' \equiv m + 2\Delta_m = \bar{m}+\Delta_m ,
\end{equation}
in which a physically non-measurable parameter $\Delta_m$ defined in (\ref{deltam}) is explicitly present. This may not be a problem. Since (\ref{modeFnP}) contains ${\cal A}_i^\Omega$, which is gauge dependent, the canonical momentum of the particle itself would not be physically measurable. 

\section{Wavepacket of charged particle at rest}
\label{SecRestParticle}

Consider a charged particle situated at rest in the Minkowski vacuum with zero background fields $\bar{F}^{\mu\nu}_{[0]}=0$. Let the particle's worldline be $\bar{z}^\mu (t) = \left(ct, 0,0,0\right)$. Then its four velocity is $\bar{v}^\mu=(c,0,0,0)$, which yields $\bar{M}_{ij}=\eta_{ij}$ from (\ref{Mijdef}), and so Eq. (\ref{QLADEqt}) reads
\begin{equation}
  \bar{m}\partial_t^2{\cal Z}_i^\Omega(t) =  q c {\cal F}_{i0}^{[0]\Omega}(t,{\bf 0}) + s\bar{m} 
	\partial_t^3{\cal Z}_i^\Omega(t) + O( \Lambda^{-1}) \label{QLADErest}
\end{equation}
after the particle-field coupling is switched on at $t=t_0$, and let $t_0\equiv 0$ for simplicity. Here the small parameter $s$ is defined as
\begin{equation}
  s \equiv \frac{q^2 \mu^{}_0}{6\pi c \bar{m}} \label{defofs}
\end{equation}
with unit of time. For electrons, the time scale $s \approx 6.3 \times 10^{-24}$  s corresponds to a length scale $sc = 2r_0/3$, where $r_0 \equiv q^2/(4\pi \varepsilon^{}_0 \bar{m}c^2) \approx 2.8 \times 10^{-15}$ m is the classical electron radius. 
When we apply our effective theory to electrons at rest,
the length resolution $\sim \Lambda^{-1}=\lambda_C \approx 2.4\times 10^{-12}$ m is much greater than $r_0$ and $sc$, namely, $s \ll 1/(c\Lambda) = t^{}_C \approx 8.1\times 10^{-21}$ s (the electron Compton time).
Note that 
the regularization-scheme-dependent $O(\Lambda^{-1})$ terms in (\ref{QLADEqt}) are those with $\bar{m} s/(c\Lambda)$ times the fourth proper-time derivatives of ${\cal Z}^j_\Omega$ or some product of lower derivatives of ${\cal Z}^j_\Omega$ and $\bar{z}^\mu$ having the same dimensions. 
In the approximations with the $O(\Lambda^{-1})$ terms 
in (\ref{QLADErest}) neglected, therefore, the $\bar{m} s^2$ terms should also be negligible \cite{Ro01}.
In other words, the $s/(c\Lambda)$-, $s^2$-, and higher-order corrections to the mode functions will be neglected in this paper.

\subsection{Mode functions}
\label{ModeFnRest}

The third-derivative term in Eq. (\ref{QLADErest}) will produce unphysical self-accelerating solutions in the absence of `external force' $q c {\cal F}_{i0}^{[0]\Omega}(t,{\bf 0})$. To avoid those runaway solutions, assume that in (\ref{QLADErest}), $s\bar{m}\partial_t^3{\cal Z}_i^\Omega$ is always a small correction to the other terms. Then inserting $\bar{m}\partial_t^2{\cal Z}_i^\Omega(t) =  q c {\cal F}_{i0}^{[0]\Omega}(t,{\bf 0}) + O(\bar{m}s)$ to the quantum LAD force $s\partial_t \left[\bar{m}\partial_t^2{\cal Z}_i^\Omega\right]$ in (\ref{QLADErest}), we obtain 
\begin{equation}
  \bar{m}\partial_t^2{\cal Z}_i^\Omega(t) =  q c \left(1 + s\partial_t \right) {\cal F}_{i0}^{[0]\Omega}(t,{\bf 0}) + 
	O( \bar{m}s/(c\Lambda), \bar{m}s^2) \label{QLLErest}
\end{equation}
which is the counterpart of the Landau-Lifshitz (LL) equation \cite{LL75}, now at the quantum level. The difference between (\ref{QLLErest}) and (\ref{QLADErest}) are beyond $O(\Lambda^{-1})$ and so negligible here \cite{Ro01}.
Solutions to (\ref{QLLErest}) for $t > t_0$ with $O( \Lambda^{-1})$ neglected can be formally written as
\begin{equation} 
  {\cal Z}_{
	j}^\Omega(t) = {\cal Z}_{j}^{[0]\Omega}(t) + \frac{1}{\bar{m}}
	\int^t_{t_0} d\tilde{t} \,  K(t, \tilde{t}\,) q c \left(1 + s\partial_{\tilde{t}} \right) {\cal F}_{i0}^{[0]\Omega}(\tilde{t},{\bf 0}),
 \label{ZfmSolrests0}
\end{equation} 
where $\bar{m}{\cal Z}_{j}^{[0]\Omega}(t) \sim C+ \tilde{C}\, t $ with  constants $C$ and $\tilde{C}$ from homogeneous solutions, and the evolution kernel
\begin{equation}
  K (t, \tilde{t}\,) = t-\tilde{t} \label{Krest}
\end{equation}
is the solution to $\partial_t^2 K(t, \tilde{t}\,) = 0$ satisfying the boundary conditions $K(\tilde{t},\tilde{t}\,)=0$ and $\partial_t K(t,\tilde{t}\,)= 1$.

From Eq.(\ref{freeModeA}), we have
\begin{eqnarray}
  {\cal F}_{(\lambda)j0}^{[0]{\bf k}}(t,{\bf x}) &=&
	\partial^{}_j {\cal A}_{(\lambda)0}^{[0]{\bf k}}(t,{\bf x})-\partial^{}_0 {\cal A}_{(\lambda)j}^{[0]{\bf k}}(t,{\bf x})
	\label{Fj0general}\\
	&=& {\cal E}_{(\lambda)j0}^{{\bf k}} e^{-i \omega t+i{\bf k}\cdot{\bf x}}, \label{Fj0}
\end{eqnarray}
where 
\begin{equation}
{\cal E}^{{\bf k}}_{(0)j0} = -i k_j, \hspace{.7cm}
{\cal E}^{{\bf k}}_{(1)j0} = i\frac{\omega}{c}\epsilon^{{\bf k}}_{(1)j}, \hspace{.7cm}
{\cal E}^{{\bf k}}_{(2)j0} = i\frac{\omega}{c}\epsilon^{{\bf k}}_{(2)j}, \hspace{.7cm}
{\cal E}^{{\bf k}}_{(3)j0} =	i k_j, \label{calEj}
\end{equation}
and ${\cal F}_{j0}^{[0]z_{}^l} = {\cal F}_{j0}^{[0]p^{}_l} =0$.
Assume that the transient during the switching-on of the particle-field interaction and the minimal time scale $(c\Lambda)^{-1}$ of the quantum LAD or LL equation does not change the form of the mode functions.
Then, after inserting the above ${\cal F}_{j0}^{[0]\Omega}$ into Eq. (\ref{QLLErest}) and matching the initial condition to the free mode functions (\ref{freeModeZ}) at $t=t_0=\bar{t}_0$, we find
\begin{equation}
  {\cal Z}^{j}_{z^{j'}} = \frac{m}{\bar{m}} \delta^j_{j'}, \hspace{1cm} 
	{\cal Z}^{j}_{p^{j'}} = \frac{\eta^{jj'}}{\bar{m}} ( t -t_0 ), 
	\label{Zjzjs2} 
\end{equation}
and 
\begin{equation}
  {\cal Z}^{j}_{(\lambda){\bf k}} = \frac{q c}{\bar{m}}{\cal E}^{{\bf k}\,\,j}_{(\lambda)\,\,0} \left( 1-i s \omega\right) f_\omega(t,t_0)	\label{Zjlks2}
\end{equation}
with
\begin{eqnarray}
	f_\omega(t,t_0) &\equiv& \int^t_{t_0} d\tilde{t} \,  K(t, \tilde{t}\,)  e^{-i \omega \tilde{t}} e^{-\omega\epsilon/2} \nonumber\\
	&=&  e^{-\omega\epsilon/2}\left[- \frac{1}{\omega^2}\left( e^{-i \omega t}-e^{-i\omega t_0}\right) 
	-\frac{i}{\omega}e^{-i\omega t_0}(t-t_0)\right],\label{ZjlkrestTheta}
\end{eqnarray}
which goes to $(t-t_0)^2/2$ as $\omega \to 0$. Here we have introduced a regulator $c\epsilon > 0$ to suppress the contribution from short-wavelength fluctuations of EM fields to the correlators, which will be calculated below.

\subsection{Particle correlators}

Suppose at $t=t_0 = 0$ the initial state of the combined system is $\rho^{\,{}_{\bf I}} = \rho_{{}_P}^{\,{}_{\bf I}} \otimes
\rho_{{}_F}^{\,{}_{\bf I}}$, which is a direct product of a Gaussian state $\rho_{{}_P}^{\,{}_{\bf I}}$ 
of the charged particle and the Minkowski vacuum state $\rho_{{}_F}^{\,{}_{\bf I}} = |0^{}_M\rangle \langle 0^{}_M |$ 
of EM fields. Then, by virtue of the linearity of our effective theory, the symmetrized two-point correlators of the renormalized charged particle's deviation from classical trajectory split into two parts, which are labeled as $P$-part and $F$-part\footnote{The `$P$-part' and `$F$-part' correlators are the counterparts of the `a-part' and `v-part' correlators, respectively, in \cite{LH06} and our series of papers on the Unruh-DeWitt detectors afterward.}: 
\begin{equation}
\langle\hat{z}^j(t),\hat{z}^{j'}(t)\rangle \equiv 
{\rm Tr}\, \left[  \rho_{}^{\,{}_{\bf I}} \, \{ \hat{z}^j(t),\hat{z}^{j'}(t)\} \right]
=\langle\hat{z}^j(t),\hat{z}^{j'}(t)\rangle^{}_P + \langle\hat{z}^j(t),\hat{z}^{j'}(t)\rangle^{}_F ,
\label{zizjrestAll}
\end{equation}
where $\{ A, B \} \equiv \frac{1}{2}(AB+BA)$,
under the expansion of $\hat{z}^j(t)$ given in (\ref{ziInt}).  The $P$-part 
\begin{eqnarray}
  \langle\hat{z}^j(t),\hat{z}^{j'}(t)\rangle^{}_P &\equiv& 
  {\rm Tr}\,\left[ \rho_{{}_P}^{\,{}_{\bf I}} \sum_{l, l'}
    \left\{ \left( {\cal Z}^j_{z_{}^l}(t)\hat{z}_{}^l + {\cal Z}^j_{p^{}_l}(t)\hat{p}^{}_l \right),
    \left( {\cal Z}^{j'}_{z_{}^{l'}}(t)\hat{z}_{}^{l'} + {\cal Z}^{j'}_{p^{}_{l'}}(t)\hat{p}^{}_{l'} \right)\right\} \right]
  \nonumber\\ &=& \sum_{l, l'} \left[  
    \langle \hat{z}_{}^l, \hat{z}_{}^{l'} \rangle^{}_{\rm I} \, {\cal Z}^j_{z_{}^l}(t){\cal Z}^{j'}_{z_{}^{l'}}(t)+
    \langle \hat{p}_{}^l, \hat{p}_{}^{l'} \rangle^{}_{\rm I} \, {\cal Z}^j_{p_{}^l}(t){\cal Z}^{j'}_{p_{}^{l'}}(t) 
  \right. + \nonumber\\ && \hspace{0.85cm} \left.
    \langle \hat{z}_{}^l, \hat{p}_{}^{l'} \rangle^{}_{\rm I}\, {\cal Z}^j_{z_{}^l}(t){\cal Z}^{j'}_{p_{}^{l'}}(t)+
    \langle \hat{p}_{}^l, \hat{z}_{}^{l'} \rangle^{}_{\rm I}\, {\cal Z}^j_{p_{}^l}(t){\cal Z}^{j'}_{z_{}^{l'}}(t)\right]
\label{zizjPrest}
\end{eqnarray}
with $\langle \hat{\cal O}_{}^l, \hat{\cal O}_{}^{l'} \rangle^{}_{\rm I} \equiv 
{\rm Tr}\, \left[\rho^{\,{}_{\bf I}}_{{}_P} \{ \hat{\cal O}_{}^l, \hat{\cal O}_{}^{l'} \}\right]$
depends on the initial state of the particle $\rho^{\,{}_{\bf I}}_{{}_P}$ 
only, and the $F$-part
\begin{equation}
  \langle\hat{z}^j(t),\hat{z}^{j'}(t)\rangle^{}_F \equiv \lim_{t'\to t}\sum_{\bf k, k'} 
\frac{1}{2}\left({\cal Z}^j_{(\lambda){\bf k}}(t) {\cal Z}^{j'*}_{(\lambda'){\bf k'}}(t') + 
{\cal Z}^{j'}_{(\lambda){\bf k}}(t') {\cal Z}^{j*}_{(\lambda'){\bf k'}}(t) \right)
\langle 0^{}_M| \hat{b}^{(\lambda)}_{\bf k} \hat{b}^{(\lambda')\dagger}_{\bf k'} |0^{}_M\rangle \label{zizjFrest} 
\end{equation}
depends on the initial state of the field $\rho_{{}_F}^{\,{}_{\bf I}}$ only. 
In (\ref{zizjFrest}), $\sum_{\bf k}$ has been defined in (\ref{Sumkdef}), and $\langle 0^{}_M| \hat{b}^{(\lambda)}_{\bf k} \hat{b}^{(\lambda')\dagger}_{\bf k'} |0^{}_M\rangle = (2\pi)^3 \eta^{(\lambda)(\lambda')}\delta^3({\bf k}-{\bf k}')$ from (\ref{bbcommutator}).

Substituting the mode functions ${\cal Z}^j_\Omega$ obtained in Section \ref{ModeFnRest} to the above expressions, we can calculate the two-point correlators up to $O(s)$. 
For example, the $F$-part of the particle-motion deviation correlator is formally
\begin{eqnarray}
&&\langle\hat{z}^j(t),\hat{z}^{j'}(t)\rangle^{}_F 
\equiv \lim_{t'\to t}\int \frac{d^3 k}{(2\pi)^3}\sqrt{\frac{\hbar}{2\omega\varepsilon^{}_0}} 
\int \frac{d^3 k'}{(2\pi)^3}\sqrt{\frac{\hbar}{2\omega'\varepsilon^{}_0}} \times \nonumber\\
&& \hspace{1.6cm} 
  \frac{1}{2}\left({\cal Z}^{j}_{(\lambda){\bf k}}(t) {\cal Z}^{j'*}_{(\lambda'){\bf k'}}(t') + 
  {\cal Z}^{j'}_{(\lambda){\bf k}}(t') {\cal Z}^{j*}_{(\lambda'){\bf k'}}(t) \right) 
  (2\pi)^3 \eta^{(\lambda)(\lambda')}\delta^3({\bf k}-{\bf k}') \nonumber\\
&&= \frac{\hbar}{(2\pi)^3\varepsilon^{}_0}{\rm Re}\, \lim_{t'\to t} \int_0^\infty \frac{\omega^2}{c^2}\frac{d\omega}{2\omega c} 
  \int_0^{2\pi} d\varphi \int_0^{\pi}d\theta\, \sin\theta \times \nonumber\\ && \hspace{1.6cm}
  \frac{q c}{\bar{m}} {\cal E}^{(\lambda)j}_{{\bf k}\,\,\,\,\,\,\,\,\,0} \left( 1- i s \omega \right) f_\omega(t,t_0)
	\frac{q c}{\bar{m}}{\cal E}^{{\bf k}\,\,j'*}_{(\lambda)\,\,0}\left( 1 + i s \omega \right)  f^*_\omega(t',t_0) \label{zizjF0Int}
\end{eqnarray}
from (\ref{Zjlks2}).
Since $\epsilon^j_{(0){\bf k}}=\delta_0^j = 0$, one can see that 
\begin{equation} 
\epsilon^{(\lambda)j}_{\bf k} \epsilon^{j'*}_{(\lambda){\bf k}} \equiv 
\sum_{\lambda,\lambda'=0}^3 \eta^{(\lambda)(\lambda')}
\epsilon^j_{(\lambda){\bf k}} \epsilon^{j'*}_{(\lambda'){\bf k}}=
\sum_{\lambda=1}^3 \epsilon^j_{(\lambda){\bf k}} \epsilon^{j'*}_{(\lambda){\bf k}} = \eta^{jj'}. \label{eesum}
\end{equation} 
From (\ref{calEj}), one further has
\begin{equation} 
  {\cal E}^{(\lambda)j}_{{\bf k}\,\,\,\,\,\,\,\,\,0}\, \epsilon^{ j'*}_{(\lambda){\bf k}} = \frac{i\omega}{c}\eta^{jj'}, \hspace{1cm} 
	{\cal E}^{(\lambda)j}_{{\bf k}\,\,\,\,\,\,\,\,\,0}\, {\cal E}^{{\bf k}\,\,j'*}_{(\lambda)\,\,0}=\frac{\omega^2}{c^2}\eta^{jj'}- k^j k^{j'},
	\label{Eesum}
\end{equation} 
and so
\begin{equation}
  \int_0^{2\pi} d\varphi \int_0^\pi d\theta \, \sin\theta \, 
	{\cal E}^{(\lambda)j}_{{\bf k}\,\,\,\,\,\,\,\,\,0}\, {\cal E}^{{\bf k}\,\,j'*}_{(\lambda)\,\,0} 
	= \frac{8\pi}{3}  \frac{\omega^2}{c^2} \eta^{jj'}, \label{EEsum}
\end{equation}
where ${\bf k} = ( \frac{\omega}{c} \sin\theta\cos\varphi, \frac{\omega}{c} \sin\theta\sin\varphi, \frac{\omega}{c} \cos\theta) = 
\frac{\omega}{c} \vec{\epsilon}_{(3){\bf k}}$. 
Thus, 
\begin{equation}
\langle\hat{z}^j(t),\hat{z}^{j'}(t)\rangle^{}_F
= \frac{\hbar q^2 \eta^{jj'}}{6\pi^2 c^3 \varepsilon^{}_0 \bar{m}^2}
   \lim_{t'\to t}\lim_{t'_0\to t_0} {\rm Re}\int_0^\infty d\omega\,\omega^3
   \left(1+ s^2 \omega^2\right) f_\omega(t,t_0) f^*_\omega(t',t'_0), \label{zjzjF0rest2}
\end{equation}
where we have distinguished the lower limits of the time integrations, namely, $t_0$ in $f_\omega$ and $t'_0$ in $f^*_\omega$, to control the divergence.

\subsubsection{Coincidence limit}
\label{SecCoinLim}

In the Unruh-DeWitt detector theory, the coincidence limits $t'\to t$ and $t'_0\to t_0$ produce logarithmic divergences in the two-point correlators \cite{LH06, LH07, LH10}. Similar divergences arise here. For example,
inserting (\ref{ZjlkrestTheta}) to (\ref{zjzjF0rest2}), $\langle\hat{z}^j(t),\hat{z}^{j'}(t)\rangle^{}_F$ will have a term proportional to
\begin{eqnarray}
	&& \lim_{t'\to t}\lim_{t'_0\to t_0}
	\int_0^\infty \frac{d\omega}{\omega}  \left[ e^{-\omega[\epsilon + i(t-t')]} + e^{-\omega[\epsilon -i(t'_0-t_0)]}
	- e^{-\omega[\epsilon + i(t-t'_0)]}-e^{-\omega[\epsilon - i(t'-t_0)]} \right] \nonumber\\
	&=& \lim_{t'\to t}\lim_{t'_0\to t_0}
	\left[ 
	 {\rm Ei}\Big(\omega[\epsilon + i(t-t')]\Big) +{\rm Ei}\Big(\omega[\epsilon -i(t'_0-t_0)]\Big)  
	      \right. \nonumber\\
	&& \hspace{2cm} \left. \left.-{\rm Ei}\Big(\omega[\epsilon +i(t-t'_0)]\Big)-
	   {\rm Ei}\Big(\omega[\epsilon -i(t'-t_0)]\Big)\right]\right|^{\infty}_{\omega=0+} \nonumber\\
	&=& -\lim_{\epsilon_0, \epsilon_1\to 0} 
	   \Big[ \ln \left|\omega_0(\epsilon + i\epsilon_1)\right| + \ln \big|\omega_0(\epsilon -i\epsilon_0)\big| \nonumber\\
	&& \hspace{2cm}   - \ln \left|\omega_0\big[\epsilon +i(t-t_0-\epsilon_0)\big]\right| - 
	\ln \big|\omega_0\left[\epsilon -i(t-t_0-\epsilon_1)\right]\big| \Big] \label{logdiv}
\end{eqnarray} 
contributed by the $\omega^{-2}$ terms in (\ref{ZjlkrestTheta}). Here $\omega_0$ is a constant with the unit of frequency, 
$\epsilon_1 \equiv t-t'$ and $\epsilon_0\equiv t'_0-t_0$ are the minimal time scales that the observer can resolve for the observation time and the initial time, respectively, in the history of the charged particle ($t-t_0 \ge \epsilon_0+\epsilon_1$ here and below.)
${\rm Ei}(x) \equiv \int_{-x}^\infty dy\, (e^{-y}/y)$ is the exponential integral function with the asymptotic behaviors ${\rm Ei}(x)= \gamma^{}_e + \ln |x| + x + O(x^2)$ as $|x|\to 0$ with the Euler's constant $\gamma^{}_e \approx 0.577216$, and $e^{-x}{\rm Ei}(x) = x^{-1} + x^{-2} +O(x^{-3})$ as $|x|\gg 1$. 

In Appendix A of Ref. \cite{LH10}, we learned that one should take $\epsilon\ll\epsilon_1, \epsilon_0$ to get the results with desired properties. There, the regulator $\epsilon$ should go to zero before further calculations with $\epsilon_1$ are done, and so we interpret $\epsilon$ as a `mathematical' cutoff in \cite{LH10}. And here,  following the argument by Bethe \cite{Be47, We48} in his calculation for the Lamb shift, we set $\epsilon$ as the electron Compton time $t_C$ 
{such that $c\epsilon = \lambda^{}_C$ for electrons at rest.} 
Suppose $\epsilon \ll \epsilon_1, \epsilon_0$ really holds, then the time-independent part in (\ref{logdiv}) becomes approximately $-\ln \omega_0\epsilon_1 - \ln \omega_0\epsilon_0$, which looks like logarithmic divergences as $\omega_0 \epsilon_0, \omega_0\epsilon_1 \to 0$. Nevertheless, the finite values of $\epsilon_0$ and $\epsilon_1$ would depend on experimental settings described below.

\subsubsection{The regulators}
\label{SecRegValue}

In Ref. \cite{TE89}, Tonomura {\it et al.} demonstrated in their experiment that a train of coherent single electrons can gradually form an interference pattern after passing through a biprism. While our wavepacket for a charged particle at rest in vacuum would not form any interference fringes since our wavepacket is neither for a moving particle (we have the longitudinal wavevector $k^{}_1=0$) nor for a superposition of two quantum states moving in non-parallel directions (we have transverse wavevector $k^{}_2 =k^{}_3=0$), the coherent lengths in transverse and longitudinal directions in Ref. \cite{TE89} can still help us to choose reasonable values for the regulators $\epsilon_0$ and $\epsilon_1$.

Due to the thermal and quantum fluctuations of the cathode, electrons just emitted by a field-emission electron gun (FEG) operating at room temperature typically have an energy spread $\Delta E \approx 0.3$ eV \cite{JEOL22}. This implies an uncertainty $\Delta t \sim h/\Delta E \approx 1.4 \times 10^{-14}$ s of time tagging to the history (or the characteristic longitudinal coherence time \cite{SSS95}) of an emitted electron. The same energy spread also corresponds to the longitudinal coherent length $\Delta z^{}_1 \sim v \Delta t \approx 1.7 \times 10^{-6}$ m \cite{SSS95}, which is close to the value of the longitudinal coherence length 1 $\mu{\rm m}$ (corresponding to $\Delta E \approx 0.5$ eV) given in Refs. \cite{TE89, To98} as the width of the electron wavepacket in the longitudinal direction.

The time resolution of single electron detectors can reach the order of 0.1 ns (e.g. TimePix4 \cite{TP4}), yet this is still much greater than the time scale of $\Delta t$. Considering $\Delta t$ as the ultimate time resolution in this experiment, and the single electron here is at rest (or in non-relativistic motion in electron microscopes),
let us set $\epsilon_0 \approx \epsilon_1= \Delta t = 1.4 \times 10^{-14}$ s 
\footnote{For accelerated single electrons in relativistic motion, the condition $\epsilon_0 \approx \epsilon_1$ may not hold; see Section \ref{SecUnruhEffect}.}, 
which is indeed much greater than $\epsilon = t^{}_C\approx 8.1 \times 10^{-21}$ s.
This fits our assumption learned from \cite{LH10}.

If the particle-field coupling is switched on at the initial moment $t_0$, the $\omega^{-1}$ term in (\ref{ZjlkrestTheta}) will produce a $t^2$-dependent term in the leading-order $F$-part correlator (\ref{zjzjF0rest2}) as 
\begin{eqnarray}
&&\frac{\hbar q^2 \eta^{jj'}}{6\pi^2 c^3 \varepsilon^{}_0 \bar{m}^2}
{\rm Re}\,\lim_{t'\to t}\lim_{t'_0\to t_0} (t-t_0)(t'-t'_0) \int_0^\infty d\omega\,\omega  e^{-\omega(\epsilon - i\epsilon_0)} \nonumber\\
&&\approx \frac{\hbar s}{\pi \bar{m}} \eta^{jj'} {\rm Re}\frac{(t-t_0)^2}{(\epsilon-i\epsilon_0)^2} 
	\label{zjzjt2noSwFn} 
\end{eqnarray} 
for charged particles from (\ref{zjzjF0rest2}) and (\ref{defofs}) (note that $\mu^{}_0 \varepsilon^{}_0 = c^{-2}$). 
When $\epsilon_0 \gg \epsilon$, the above term will be negative. One may worry whether it would dominate over the $t^2$-term in the $P$-part of the particle-motion deviation correlator,
\begin{equation}
 \langle\hat{z}^j(t),\hat{z}^{j'}(t)\rangle^{}_P \sim  
  \frac{\langle\hat{p}_{}^j, \hat{p}_{}^{j'}\rangle^{}_{\rm I}}{\bar{m}^2} \, (t-t_0)^2 \label{zjzjPt2}
\end{equation}
[inserting (\ref{Zjzjs2}) into (\ref{zizjPrest}).] 
If so, one would have a growing negative value of $\langle\hat{z}^j(t),\hat{z}^{j'}(t)\rangle$ at late times,
which is not a non-adiabatic transient only.
Fortunately, with the values of regulators we used earlier in this section, 
the negative (\ref{zjzjt2noSwFn}) is just a small correction to the positive $(t-t_0)^2$ term in (\ref{zjzjPt2}). 

{Indeed, since we have $\epsilon_0 \gg \epsilon$, the value of the coefficient of $(t-t_0)^2$ term in (\ref{zjzjt2noSwFn}) would be $C_F \equiv {\rm Re}\, \hbar s/[\pi\bar{m} (\epsilon-i\epsilon_0 )^2] \approx -\hbar s/(\pi \bar{m}\epsilon_0^2)\approx -1.18 \,{\rm m}^2/{\rm s}^2$. 
On the other hand, according to \cite{JEOL22}, an electron wavepacket emitted by a FEG and moving in the $x^1$ direction can be considered to have an initial width $\big[\langle \overline{\hat{z}_{\sf T}^2}
\rangle^{}_I\big]^{1/2} \approx 5$ nm in the transverse directions, ${\sf T}=2,3$, where $\langle \overline{\hat{z}_{\sf T}^2}
\rangle^{}_I$ will be defined in (\ref{z2Iren}). Then the coefficient of $(t-t_0)^2$ term in (\ref{zjzjPt2}) would have the value $C_P\equiv\langle\hat{p}_{\sf T}^2 \rangle^{}_{\rm I}/\bar{m}^2 = \left(\frac{\bar{m}}{\bar{m}'} \right)^2 \left[\hbar^2/(4\bar{m}^2 \langle \overline{\hat{z}_{\sf T}^2} \rangle^{}_I) + \frac{3\epsilon_0^2}{2\epsilon_1^2} |C_F| \right]
\approx 1.3 \times 10^{8} \, {\rm m}^2/{\rm s}^2$ from (\ref{p2Iren}) and (\ref{QSrenorm}) with $\epsilon_1\approx\epsilon_0$ and $\bar{m}'\approx \bar{m}$, 
and so $C_P+C_F$ is positive.} 

In the longitudinal direction, the initial width of the wavepacket $\big[\langle \overline{\hat{z}_1^2}\rangle^{}_I\big]^{1/2} \approx 1.7\,\mu{\rm m}$ as mentioned gives $C_P \approx 1.2 \times 10^3 \, {\rm m}^2/{\rm s}^2$, which is still much greater than $|C_F|$ for $\epsilon_0 \approx 1.4 \times 10^{-14}$ s. So $C_P+C_F$ keeps positive, too.

\subsubsection{Regularized correlators of particle-motion deviations}

With the regulators, (\ref{zjzjF0rest2}) and (\ref{ZjlkrestTheta}) yield
\begin{eqnarray}
  && \langle\hat{z}^j(t),\hat{z}^{j'}(t')\rangle^{}_F = 
	 \frac{\hbar s  \eta^{jj'}}{\pi\bar{m}}  {\rm Re}
	 \left\{ \ln \left[ \frac{(\eta-\epsilon_0-i\epsilon)(\eta-\epsilon_1+i\epsilon)}
		                  {(\epsilon_0 + i\epsilon)(\epsilon_1 -i\epsilon)}\right] + \right. \nonumber\\
	&& \hspace{1cm}\left. \frac{ \epsilon_0+\epsilon_1}{\epsilon_0+i\epsilon}			
		 -\frac{\eta}{\eta-\epsilon_1+i\epsilon} -\frac{\eta-\epsilon_0-\epsilon_1}{\eta-\epsilon_0-i\epsilon} 
	   - \frac{\eta(\eta-\epsilon_0-\epsilon_1)}{(\epsilon_0+i\epsilon)^2} 	+O(s^2)\right\}
		\label{zjzjF0rest3}
\end{eqnarray}
with $\eta \equiv t-t_0$ [cf. (\ref{logdiv}) and (\ref{zjzjt2noSwFn})]. Here we require that $t$ is no less than $t_0 +\epsilon_0+\epsilon_1$,
due to the uncertainty of time-tagging to the history of the charged particle.

The $P$-part correlators (\ref{zizjPrest}) can be obtained straightforwardly. Suppose
our initial state has $\langle\hat{z}_{}^j, \hat{p}_{}^{j'}\rangle^{}_{\rm I}=0$,
then (\ref{Zjzjs2}) and (\ref{zizjPrest}) yield
\begin{eqnarray}
\langle\hat{z}^j(t),\hat{z}^{j'}(t')\rangle^{}_P  &\approx&  \frac{m^2}{\bar{m}^2}  
\langle\hat{z}_{}^j, \hat{z}_{}^{j'}\rangle^{}_{\rm I} 
+\frac{\langle\hat{p}_{}^j, \hat{p}_{}^{j'}\rangle^{}_{\rm I}}{\bar{m}^2} \eta^2 \label{zjzjP01rest}
\end{eqnarray}
after the $\eta'\to\eta$ limit is taken.
Summing up (\ref{zjzjF0rest3}) and (\ref{zjzjP01rest}), we get the symmetrized two-point correlator of particle-motion deviation for $\eta \gg \epsilon_1 \approx \epsilon_0 \gg \epsilon\gg s$,
\begin{eqnarray}
  \langle\hat{z}^j(t),\hat{z}^{j'}(t')\rangle &\approx&
	\frac{m^2}{\bar{m}^2} \langle\hat{z}_{}^j, \hat{z}_{}^{j'}\rangle^{}_{\rm I} +
	\frac{\langle\hat{p}_{}^j, \hat{p}_{}^{j'}\rangle^{}_{\rm I} }
	     {\bar{m}^2} \eta^2 + \nonumber\\
&& \frac{\hbar s }{\pi \bar{m}} \eta^{jj'} \left[\ln \frac{\eta^2}{\epsilon_0\epsilon_1} -\frac{\eta^2}{\epsilon_0^2}
   + \left(\frac{\epsilon_1}{\epsilon_0}+1\right)\frac{\eta}{\epsilon_0}+ \frac{\epsilon_1}{\epsilon_0} -1 \right]+ 
	O(s^3).\label{zjzjPFrestAsmp}
\end{eqnarray}

\subsubsection{Correlators with canonical momentum of the particle} 

The symmetrized two-point correlators of the renormalized charged particle's canonical momentum 
from (\ref{hatpi2zi}) is 
\begin{equation}
\langle\hat{p}^j(t),\hat{p}^{j'}(t)\rangle \equiv 
{\rm Tr}\, \left[  \rho_{}^{\,{}_{\bf I}} \, \{ \hat{z}^j(t),\hat{z}^{j'}(t)\} \right]
= \langle\hat{p}^j(t),\hat{p}^{j'}(t)\rangle^{}_P + \langle\hat{p}^j(t),\hat{p}^{j'}(t)\rangle^{}_F , \label{pipjrestAll}
\end{equation}
where
\begin{eqnarray}
  \langle\hat{p}^j(t),\hat{p}^{j'}(t)\rangle^{}_P &\equiv& 
  {\rm Tr}\,\left[ \rho_{{}_P}^{\,{}_{\bf I}} \sum_{l, l'}
    \left\{ \left( {\cal P}^j_{z_{}^l}(t)\hat{z}_{}^l + {\cal P}^j_{p^{}_l}(t)\hat{p}^{}_l \right),
    \left( {\cal P}^{j'}_{z_{}^{l'}}(t)\hat{z}_{}^{l'} + {\cal P}^{j'}_{p^{}_{l'}}(t)\hat{p}^{}_{l'} \right)\right\} \right]
  \nonumber\\ &=& \sum_{l, l'} \left[  
    \langle \hat{z}_{}^l, \hat{z}_{}^{l'} \rangle^{}_{\rm I} \, {\cal P}^j_{z_{}^l}(t){\cal P}^{j'}_{z_{}^{l'}}(t)+
    \langle \hat{p}_{}^l, \hat{p}_{}^{l'} \rangle^{}_{\rm I} \, {\cal P}^j_{p_{}^l}(t){\cal P}^{j'}_{p_{}^{l'}}(t) 
  \right. + \nonumber\\ && \hspace{0.85cm} \left.
    \langle \hat{z}_{}^l, \hat{p}_{}^{l'} \rangle^{}_{\rm I}\, {\cal P}^j_{z_{}^l}(t){\cal P}^{j'}_{p_{}^{l'}}(t)+
    \langle \hat{p}_{}^l, \hat{z}_{}^{l'} \rangle^{}_{\rm I}\, {\cal P}^j_{p_{}^l}(t){\cal P}^{j'}_{z_{}^{l'}}(t)\right],
\label{pipjPrest}
\end{eqnarray}
\begin{eqnarray}
  \langle\hat{p}^j(t),\hat{p}^{j'}(t)\rangle^{}_F \equiv \lim_{t'\to t}\sum_{\bf k, k'} 
  \frac{1}{2}\left[{\cal P}^j_{(\lambda){\bf k}}(t) {\cal P}^{j'*}_{(\lambda'){\bf k}'}(t') + 
  {\cal P}^{j'}_{(\lambda){\bf k}}(t') {\cal P}^{j*}_{(\lambda'){\bf k}'}(t) \right]
  \langle 0^{}_M| \hat{b}^{(\lambda)}_{\bf k} \hat{b}^{(\lambda')\dagger}_{\bf k'} |0^{}_M\rangle .\nonumber\\ \label{ppksum}
\end{eqnarray}
From (\ref{modeFnP}), the particle at rest has the momentum mode functions 
\begin{equation}
  {\cal P}^j_\Omega(t) = \bar{m}' \partial_t {\cal Z}^j_\Omega(t) - \frac{3}{2}s \bar{m}\partial_t^2 {\cal Z}^j_\Omega(t)
	 + q {\cal A}^{[0]j}_\Omega(t, {\bf 0}) \label{modePrest}
\end{equation} 
with ${\cal A}^{[0]j}_\Omega(t, {\bf 0})$ given by (\ref{freeModeA}). From (\ref{Zjzjs2}) and (\ref{Zjlks2}), ${\cal P}^j_\Omega$ read
\begin{equation}
 {\cal P}^j_{z^{j'}}= 0, \hspace{1cm} {\cal P}^j_{p_{j'}}= \frac{\bar{m}'}{\bar{m}}
    \eta^{jj'}, \label{PjzjrestAsmp}
\end{equation} 
and 
\begin{equation} 
  {\cal P}^j_{(\lambda){\bf k}}= 
	q c {\cal E}^{{\bf k}\,\,j}_{(\lambda)\,\,0}(1-i s \omega)\left[ \frac{\bar{m}'}{\bar{m}} \dot{f_\omega}
	 -\frac{3}{2} s \ddot{f_\omega}\right] + q\epsilon^j_{(\lambda){\bf k}}e^{-i\omega t}e^{-\omega\epsilon/2}. \label{Pjlks2}
\end{equation}
Substituting the above mode functions ${\cal P}^j_\Omega$ to (\ref{pipjPrest}) and (\ref{ppksum}), with (\ref{eesum})-(\ref{EEsum}), the closed form of $\langle\hat{p}^j(t),\hat{p}^{j'}(t)\rangle$
up to $O(s)$ can be obtained by the same techniques for $\langle\hat{z}^j(t), \hat{z}^{j'}(t)\rangle$. 
The $F$-part of the particle-momentum correlator reads
\begin{eqnarray}
&&  \langle\hat{p}^j(t),\hat{p}^{j'}(t')\rangle^{}_F = \frac{\hbar s \bar{m}}{\pi}\eta^{jj'} {\rm Re}
	\int_0^\infty d\omega\, \omega 
	\left\{ \omega^2 \left[\frac{\bar{m}'}{\bar{m}}\right]^2 \dot{f}_\omega(t) \dot{f}^*_\omega(t')\right. \nonumber\\
&&\hspace{3cm}+\left. \frac{3\bar{m}'}{2\bar{m}} \omega e^{-\omega\epsilon/2}
	\left[ ie^{i\omega t'} \dot{f}_\omega(t)- ie^{-i\omega t} \dot{f}^*_\omega(t')\right]
	+ \frac{3}{2}e^{-i\omega(t-t')-\omega\epsilon} + O(s) \right\} \nonumber\\
&&= \frac{\hbar s \bar{m}}{\pi}\eta^{jj'}{\rm Re} \left\{ \left[\frac{\bar{m}'}{\bar{m}}\right]^2 
		\left[-\frac{1}{(\epsilon^{}_1-i\epsilon)^2}-\frac{1}{(\epsilon^{}_0 +i\epsilon)^2} 
		+\frac{1}{(\eta-\epsilon^{}_0-i\epsilon)^2} +\frac{1}{(\eta -\epsilon_1+i\epsilon)^2} 
		\right]\right.  \nonumber\\  
&&\hspace{2.7cm} + \frac{3\bar{m}'}{2\bar{m}}
	    \left[\frac{1}{(\epsilon^{}_1-i\epsilon)^2} - \frac{1}{(\eta-\epsilon^{}_1+i\epsilon)^2} 
			+\frac{1}{(-\epsilon^{}_1+i\epsilon)^2} - \frac{1}{(\eta-\epsilon^{}_0-i\epsilon)^2} \right]\nonumber\\
&&\hspace{2.7cm}\left. 
  -\frac{3}{2}\frac{1}{(\epsilon_1-i\epsilon)^2} + O(s^2) \right\}, \label{pjpjF0rest}
\end{eqnarray} 
where the $O(s^1)$ corrections in the curly brackets cancel out. 
At late times ($\eta\equiv t-t_0 \gg \epsilon_0 \approx \epsilon_1 \gg \epsilon$), from (\ref{PjzjrestAsmp}), (\ref{pipjPrest}), and (\ref{pjpjF0rest}), one has the corrected particle-momentum correlator
\begin{eqnarray}
\langle\hat{p}^j(t),\hat{p}^{j'}(t')\rangle &\to& \left[\frac{\bar{m}'}{\bar{m}}\right]^2 
  \langle \hat{p}_{}^j, \hat{p}_{}^{j'} \rangle^{}_{\rm I} +\nonumber\\
&&\frac{\hbar s \bar{m}}{\pi}\eta^{jj'} \left[ 
-\left[\frac{\bar{m}'}{\bar{m}}\right]^2\left(\frac{1}{\epsilon_1^2}+\frac{1}{\epsilon_0^2}\right)
+ \frac{\bar{m}'}{\bar{m}} \frac{3}{\epsilon_1^2} -\frac{3}{2\epsilon_1^2} \right],  
\label{pjpjPFrestAsmp}
\end{eqnarray} 
and at the `initial' moment $\eta\equiv t-t_0\approx \epsilon_0+\epsilon_1\gg \epsilon$ 
for the observer, one has
\begin{equation}
\langle\hat{p}^j(t'_0+\epsilon_1),\hat{p}^{j'}(t'_0)\rangle \approx
\left[\frac{\bar{m}'}{\bar{m}}\right]^2 \langle \hat{p}_{}^j, \hat{p}_{}^{j'} \rangle^{}_{\rm I} 
+ \frac{3\hbar s \bar{m}}{2\pi}\eta^{jj'}
\left[\frac{\bar{m}'}{\bar{m}}\left(\frac{1}{\epsilon_1^2}-\frac{1}{\epsilon_0^2}\right)-\frac{1}{\epsilon_1^2} \right].
\label{pjpjPFrestInit}
\end{equation}

It is also straightforward to obtain the correlators
\begin{equation}
\langle\hat{p}^j(t),\hat{z}^{j'}(t)\rangle 
= \langle\hat{p}^j(t),\hat{z}^{j'}(t)\rangle^{}_P + \langle\hat{p}^j(t),\hat{z}^{j'}(t)\rangle^{}_F ,\label{pizjrestAll}
\end{equation}
of the particle momentum and deviation with
\begin{eqnarray}
\langle\hat{p}^j(t),\hat{z}^{j'}(t')\rangle^{}_P &=&  \sum_{l, l'} \left[  
\langle\hat{z}_{}^l, \hat{z}_{}^{l'}\rangle^{}_{\rm I} \, {\cal P}^j_{z_{}^l}(t){\cal Z}^{j'}_{z_{}^{l'}}(t')+
\langle\hat{p}_{}^l, \hat{p}_{}^{l'}\rangle^{}_{\rm I} \, {\cal P}^j_{p_{}^l}(t){\cal Z}^{j'}_{p_{}^{l'}}(t') +
\right.\nonumber\\ &&  \left.
\langle\hat{z}_{}^l, \hat{p}_{}^{l'}\rangle^{}_{\rm I} \, {\cal P}^j_{z_{}^l}(t){\cal Z}^{j'}_{p_{}^{l'}}(t')+
\langle\hat{p}_{}^l, \hat{z}_{}^{l'}\rangle^{}_{\rm I} \, {\cal P}^j_{p_{}^l}(t){\cal Z}^{j'}_{z_{}^{l'}}(t')\right], 
\label{pizjPrest}
\end{eqnarray}
\begin{equation}
\langle\hat{p}^j(t),\hat{z}^{j'}(t')\rangle^{}_F 
= \frac{\hbar}{16\pi^3\varepsilon^{}_0 c^3} {\rm Re}  
\int_0^\infty \omega d\omega e^{-\omega \epsilon} \int_0^{2\pi} d\varphi \int_0^\pi d\theta \,\sin\theta\, 
  {\cal P}^{(\lambda)j}_{\bf k}(t) {\cal Z}^{j'*}_{(\lambda){\bf k}}(t'). 
\end{equation}
After some algebra, we find 
\begin{eqnarray}
  \langle\hat{p}^j(t),\hat{z}^{j'}(t')\rangle^{}_F &=& \frac{\hbar s}{\pi}\eta^{jj'} {\rm Re}\left\{
     \frac{\bar{m}'}{\bar{m}}\left[ \frac{-1}{\epsilon^{}_1-i\epsilon}+\frac{1}{\epsilon^{}_0+i\epsilon} 
		+\frac{1}{\eta -\epsilon^{}_0-i\epsilon}-\frac{1}{\eta -\epsilon^{}_1+i\epsilon} \right.\right. \nonumber\\
&&	\hspace{2.7cm}\left. + (\eta -\epsilon^{}_0-\epsilon^{}_1)\left(\frac{-1}{(\epsilon^{}_0-i\epsilon)^2}+
    \frac{1}{(\eta-\epsilon^{}_0+i\epsilon)^2}\right)\right]\nonumber\\
&& \hspace{.3cm} \left. +\frac{3}{2}\left[ \frac{1}{\epsilon^{}_1-i\epsilon} - \frac{1}{\eta -\epsilon^{}_0-i\epsilon} -
    \frac{\eta-\epsilon^{}_0-\epsilon^{}_1}{(\eta -\epsilon^{}_0-i\epsilon)^2}\right] + O(s^2) \right\},
\end{eqnarray}
which vanishes at the `initial' moment $\eta=\epsilon^{}_0+\epsilon^{}_1$.
The $O(s)$ corrections in the above braced brackets cancel out again.
Suppose initially $\langle \hat{p}_{}^j, \hat{z}_{}^{j'} \rangle^{}_{\rm I}
= 0$, then at late times ($\eta \gg \epsilon_0$, $\epsilon_1$),
\begin{eqnarray}
  \langle\hat{p}^j(t),\hat{z}^{j'}(t')\rangle &\to& \left[ \frac{\langle \hat{p}_{}^j, \hat{p}_{}^{j'}\rangle^{}_{\rm I}}{\bar{m}}
	+ \frac{\hbar s}{\pi} \eta^{jj'} \left( -\frac{1}{\epsilon_0^2} + O(s^2)\right) \right]\frac{\bar{m}'}{\bar{m}}
		\eta . 
	\label{pjzjPFrestAsmp}
\end{eqnarray}
The contribution from the $F$-part of the correlator is negative and proportional to $1/\epsilon_0^2$ in the above result.

\subsection{Quantum coherence of particle state}
\label{SecQCoherenceRest}

The reduced state of the charged particle can be obtained by tracing out the field degrees of freedom in the quantum state of the combined system. In the presence of   particle-field coupling, the reduced state of the charged particle is a mixed state. For Gaussian states of a single charged particle considered in this paper, the purity, von Neumann entropy, and the effective temperature of it can be computed using the uncertainty function \cite{LH07} 
\begin{equation}
  {\cal  U}(\tau) \equiv \sqrt{\left|\det {\bf C}\right|} \label{UncFnDef}
\end{equation}
where ${\bf C}$ is the $6\times 6$ covariance matrix 
\begin{equation}
{\bf C} = \left( \begin{array}{cc}
\langle\hat{p}_i(\tau),\hat{p}_j(\tau)\rangle & \langle\hat{z}^i(\tau),\hat{p}_j(\tau)\rangle \\
\langle\hat{p}_i(\tau),\hat{z}^j(\tau)\rangle & \langle\hat{z}^i(\tau),\hat{z}^j(\tau)\rangle 
\end{array} \right).
\end{equation}
The purity of the reduced state of the particle then reads
\begin{equation}
  {\bf P} = \frac{(\hbar/2)^3}{\cal U}. \label{Purity}
\end{equation}

Since one can always diagonalize the particle sector of the quadratic Hamiltonian (\ref{Hamil2}) and the Gaussian states of the particle $|z^{}_I\rangle$ simultaneously, without loss of generality, we consider the cases where the three degrees of freedom of the particle motion are decoupled. Assume $\langle \hat{\cal O}_i, \hat{\cal O}_j\rangle_{\rm I}\equiv {\rm Tr} 
\left[ \rho_{}^{\rm I}  \{ \hat{\cal O}_i(\bar{t}^{}_0),\hat{\cal O}'_j(\bar{t}^{}_0)\} \right]\propto \eta^{}_{ij}$ for ${\cal O}, {\cal O}' = z, p$. Then, after the particle-field coupling has been switched on, one still has $\langle \hat{\cal O}_i(t),\hat{\cal O}'_j(t)\rangle = \langle \hat{\cal O}_i(t),\hat{\cal O}'_j(t)\rangle^{}_P + \langle \hat{\cal O}_i(t),\hat{\cal O}'_j(t)\rangle^{}_F \propto \eta_{ij}$ from (\ref{zjzjPFrestAsmp}), (\ref{pjpjPFrestAsmp}), and (\ref{pjzjPFrestAsmp}) for the particle at rest. Thus,
\begin{equation}
  {\cal U} = \prod_{i=1}^3 \sqrt{u_i}, \label{Uui}
\end{equation}
where
\begin{equation}
  u_i(t) \equiv \langle \hat{z}_i(t),\hat{z}_i(t)\rangle \langle\hat{p}_i(t),\hat{p}_i(t)\rangle -
		        \langle\hat{z}_i(t),\hat{p}_i(t)\rangle \langle\hat{p}_i(t),\hat{z}_i(t)\rangle  \label{uidef}
\end{equation}
for each $i=1,2,3$ (no sum).
To evaluate the quantum coherence of a particle, one may perform interference experiments separately in different directions or independent degrees of freedom. To compare with  experimental results, it would be convenient to define the purity for a single, independent degree of freedom as
\begin{equation}
  {\bf P}_i \equiv \frac{\hbar/2}{\sqrt{u_i}}, \label{Puri}
\end{equation}
whence the purity of the system is ${\bf P} = \prod_{i=1}^3  {\bf P}_i$.
For a general covariant matrix ${\bf C}$, the values of $\sqrt{u_i}$ and ${\bf P}_i$ could be taken after the covariant matrix is transformed into three $2\times 2$ blocks of independent degrees of freedom, though their connection with experimental results may become indirect.

Right before the particle-field coupling is switched on, one has 
\begin{equation}
u_i(t^{}_0)	=	\langle\hat{z}_i^2\rangle^{}_{\rm I}\, \langle\hat{p}_i^2\rangle^{}_{\rm I} -
	    \langle \hat{z}_i, \hat{p}_i\rangle_{\rm I}\langle \hat{p}_i, \hat{z}_i\rangle_{\rm I}.
\end{equation}
for $\bar{t}^{}_0\to t^{}_0-$ from (\ref{zizjrestAll}), (\ref{pipjrestAll}), and (\ref{pizjrestAll}) with (\ref{freeModeZ}), (\ref{freeModeP}). 
If we assume that the particle at that time was in a pure state, the condition
\begin{equation}
  \langle\hat{z}_i^2\rangle^{}_{\rm I}\, \langle\hat{p}_i^2\rangle^{}_{\rm I} - \big(\langle \hat{z}_i, \hat{p}_i\rangle_{\rm I} \big)^2 = \left(\frac{\hbar}{2}\right)^2, \label{URinit}
\end{equation}
for each $i$ would be introduced to give ${\bf P}=1$ initially. 
Around the `initial' moment for the observer $\eta \approx \epsilon_0+\epsilon_1$ after the coupling is switched on, however, one has $\langle\hat{z}^j(t),\hat{z}^{j'}(t)\rangle \approx \frac{m^2}{\bar{m}^2}\langle\hat{z}_{}^j, \hat{z}_{}^{j'}\rangle^{}_{\rm I}$ from (\ref{zjzjF0rest3}) 
and (\ref{zjzjP01rest}). Together with (\ref{pjpjPFrestInit}), we would have
\begin{equation} 
  u_i \approx \left[ \left(\frac{m}{\bar{m}}\right)^2\langle\hat{z}_i^2\rangle^{}_{\rm I}\right] 
	\left[ \left(\frac{\bar{m}'}{\bar{m}}\right)^2
	\langle\hat{p}_i^2\rangle^{}_{\rm I} - \frac{3\hbar s \bar{m}}{2\pi\epsilon_1^2} \right],\label{uiInit}
\end{equation}
if $\epsilon_0 \approx \epsilon_1$, 
and $\langle\hat{p}^j(t),\hat{z}^{j'}(t)\rangle$ is vanishing at that moment. Suppose $\langle \hat{p}_i, \hat{z}_i \rangle^{}_{\rm I} = 0$, which implies $\langle \hat{z}_i^2\rangle^{}_{\rm I} \, \langle \hat{p}_i^2\rangle^{}_{\rm I} = \hbar^2/4$. Then (\ref{uiInit}) 
becomes 
\begin{equation}
  u_i \approx \frac{m^2 \bar{m}'{}^2}{\bar{m}^4} \frac{\hbar^2}{4} -\frac{3\hbar s \bar{m}}{2\pi}
	\frac{m^2}{\bar{m}^2}\frac{\langle\hat{z}_i^2\rangle^{}_{\rm I}}{\epsilon_1^2} ,
	\label{uiInit2}
\end{equation}
where $\frac{m^2 \bar{m}'{}^2}{\bar{m}^4} = \left[ 1 - (\Delta_m/\bar{m})^2\right]^2 < 1$ in the first term,
and the second term in (\ref{uiInit2}) is negative. Thus, the $u_i$ in (\ref{uiInit2}) must be less than $\hbar^2/4$, indicating that the reduced state of the particle with these parameter values would be unphysical. 

To avoid this, we noticed that the bare correlators $\langle \hat{z}_i^2\rangle^{}_{\rm I}$ and $\langle \hat{p}_i^2\rangle^{}_{\rm I}$ are not physically measurable. 
Also, since FEGs produce single electrons via quantum tunneling, which is a purely quantum effect, it is reasonable to assume that the electron just emitted by a FEG is in a nearly pure state. Even if we put in an artificial switching function for the particle-field interaction in our description, the non-adiabatic effect during the period of switching-on, if any, should not hurt the purity significantly. For this reason, we   assign $u_i \equiv \hbar^2/4$ in (\ref{uiInit}) at the `initial' moment $\eta \equiv t-t_0 = \epsilon_0+\epsilon_1 \approx 2\epsilon_0$ right after the first appearance of the single electron. 
In other words, {\it we renormalize the quantum state such that}
\begin{equation}
\langle\overline{\hat{z}_i^2}\rangle^{}_{\rm I}
\langle\overline{\hat{p}_i^2}\rangle^{}_{\rm I} \equiv \frac{\hbar^2}{4} \label{QSrenorm}
\end{equation}
with 
\begin{eqnarray}
&& \langle\overline{\hat{z}_i^2} \rangle^{}_{\rm I}\equiv \langle z_i(t'_0+\epsilon_1),z^i(t'_0)\rangle \approx \frac{m^2}{\bar{m}^2} \langle \hat{z}_i^2\rangle^{}_{\rm I}, \label{z2Iren}\\
&& \langle\overline{\hat{p}_i^2} \rangle^{}_{\rm I}\equiv \langle p_i(t'_0+\epsilon_1),p^i(t'_0)\rangle \approx \frac{\bar{m}'{}^2}{\bar{m}^2}\langle \hat{p}_i^2\rangle^{}_{\rm I} -
	\frac{3\hbar s \bar{m}}{2\pi\epsilon_1^2}, \label{p2Iren}
\end{eqnarray}
and $\langle \overline{\hat{z}_i^2} \rangle^{}_{\rm I}$ is supposed to be physically measurable in experiments. 
We will not ask whether the bare quantum state of an electron satisfies $\langle z_i^2 \rangle^{}_{\rm I}\, \langle p_i^2\rangle^{}_{\rm I} = \hbar^2/4$, or equivalently, whether the electrons before the particle-field coupling is switched-on was in a pure state or in a mixed state. 

At late times ($\eta \gg \epsilon_0\approx \epsilon_1$), (\ref{zjzjPFrestAsmp}), (\ref{pjpjPFrestAsmp}), and (\ref{pjzjPFrestAsmp}) {yield 
\begin{eqnarray}
&& u_i 
	\approx \frac{\hbar s \bar{m}'}{\pi\epsilon_1^2}\left(\frac{3}{2}-\frac{\bar{m}'}{\bar{m}}
	\right) \left[ \langle\overline{\hat{z}_i^2}\rangle^{}_{\rm I} + 
	  \frac{\hbar s}{\pi\bar{m}}\left( \ln\frac{\eta^2}{\epsilon_0\epsilon_1} +\frac{\eta}{\epsilon_0^2}(\epsilon_0+\epsilon_1) + \frac{\epsilon_1}{\epsilon_0}-1\right)
		\right] + \nonumber\\
&&\left[ \langle\overline{\hat{p}_i^2}\rangle^{}_{\rm I} + \frac{\hbar s\bar{m}}{\pi} 
	    \left( \frac{3}{2\epsilon_1^2} - \frac{\bar{m}'{}^2}{\bar{m}^2\epsilon_0^2} \right)\right]
		\left[ \langle\overline{\hat{z}_i^2}\rangle^{}_{\rm I} + 
			\frac{\hbar s}{\pi\bar{m}}\left( \ln\frac{\eta^2}{\epsilon_0\epsilon_1} +\frac{\eta}{\epsilon_0^2}(\epsilon_0+\epsilon_1) + \frac{\epsilon_1}{\epsilon_0}-1\right) +\right.\nonumber\\ 
&& \hspace{5.8cm} \left. \frac{\hbar s}{\pi \bar{m}'}
	  \left( \frac{3}{2}-\frac{\bar{m}'}{\bar{m}}\right)
		\frac{\eta^2}{\epsilon_1^2}	\right]. \label{uiAsmp}
\end{eqnarray}
For electrons at rest, $\Delta_m \sim 10^{-3} \bar{m}$ and so $\bar{m}'/\bar{m} \approx 1 + O(10^{-3})$.} One can see that the above $u_i$ will never be less than $\langle\overline{\hat{z}_i^2}\rangle^{}_{\rm I} \langle\overline{\hat{p}_i^2}\rangle^{}_{\rm I}$ for $\eta\gg \epsilon_0$.

Since $\eta \gg \epsilon_0 \gg \epsilon \gg s$ at late times, the $\eta^2/\epsilon_0^2$ term in (\ref{uiAsmp}) dominates over the $\big(\ln\frac{\eta^2}{\epsilon_0\epsilon_1} + \cdots 
\big)$ term. 
Suppose the initial width of the wavepacket $\big(\langle\overline{\hat{z}_i^2}\rangle^{}_{\rm I}\big)^{\frac{1}{2}} \ll \sqrt{\hbar\pi \epsilon_1^2/(2s\bar{m})} \approx 75 \,\mu$m 
such that $\langle\overline{\hat{p}_i^2}\rangle^{}_{\rm I} = \hbar^2/\big(4\langle\overline{\hat{z}_i^2}\rangle^{}_{\rm I}\big) \gg \hbar s\bar{m}/(2\pi \epsilon_1^2)$
and so the first line of the right-hand-side of (\ref{uiAsmp}) is small compared to the other terms for $\bar{m}'\approx \bar{m}$ and $\epsilon_1\approx \epsilon_0$.
Then (\ref{uiAsmp}) can be reduced to 
\begin{eqnarray}
  u_i &\approx& \left[ \langle\overline{\hat{p}_i^2}\rangle^{}_{\rm I} +\frac{\hbar s \bar{m}}{\pi}\left(\frac{3}{2\epsilon_1^2}-\frac{1}{\epsilon_0^2}\right) \right] \left[ \langle\overline{\hat{z}_i^2}\rangle^{}_{\rm I} + \frac{\hbar s}{2\pi  \epsilon_1^2 \bar{m}}\eta^2\right]. \label{uiAsmp2}
\end{eqnarray}
since $\bar{m}'\approx \bar{m}$.
The purity in the $x^i$ direction, ${\bf P}_i$ in (\ref{Puri}), drops from $1$ significantly when the $\eta^2$ term 
in the above $u_i$ becomes comparable with the first term $\langle\overline{\hat{z}_i^2}\rangle^{}_{\rm I}$, namely, when $\eta^2 \hbar s/(2\pi  \epsilon_1^2 \bar{m}) \sim \langle\overline{\hat{z}_i^2}\rangle^{}_{\rm I}$.
Accordingly, we estimate the decoherence time in this direction as 
\begin{equation}
   T^{\rm dc}_i \equiv \sqrt{\frac{6\pi\bar{m} \epsilon_1^2}{\hbar s}}\,\sqrt{\langle\overline{\hat{z}_i^2}\rangle^{}_{\rm I}}, \label{Tdecoh}
\end{equation}
which is proportional to the initial width of the electron wavepacket in the $x^i$ direction, $\big(\langle\overline{\hat{z}_i^2}\rangle^{}_{\rm I}\big)^{1/2}$.  
At $\eta\equiv t-t_0 = T^{\rm dc}_i$, we have ${\bf P}_i \approx 1/2$. 

Now we apply our formula to the electron interference experiment in Ref. \cite{TE89}.
According to \cite{TE89}, the distance from the source to the screen in their electron microscope is about 1.5 m, and in most of the journey the single electrons are approximately in inertial motion. The coherent electrons tunneling out of the FEG are accelerated to a speed $v \approx 50$ keV $= 1.2 \times 10^8$ m/s $\approx 0.4\,c$, corresponding to a de Broglie wavelength $5.4 \times 10^{-12}$ m and a flying time $t^{}_F \approx 1.2 \times 10^{-8}$ s from the source to the image plane. At the speed $v$, an electron's Lorentz factor $\bar{\gamma}(t^{}_F)=1/\sqrt{1-v^2/c^2}$ is about $1.1$ only, and so the electron's time dilation is not significant for laboratory observers. Since the interference pattern does form on the image plane, the electron wavepacket in \cite{TE89} must have sufficiently high quantum coherence. This means that its purity must still be of order 1 after flying for $t^{}_F$, and the short acceleration stage at early times as well as the passages through the electron lenses should not have too much impact on the electrons' coherence. 

The diameter of the virtual source of a FEG applied in Ref. \cite{TE89} is about 5 nm \cite{To98, JEOL22}, so we set the initial width of the wavepacket in the transverse directions $\big(\langle\overline{\hat{z}_{\sf T}^2}\rangle^{}_{\rm I}\big)^{1/2} = 5 \times 10^{-9}$ m (${\sf T}=2$ or $3$), which is well above $\lambda^{}_C/\bar{\gamma}(t'_0) \approx 2.2 \times 10^{-12}$ m and justifies our Gaussian-wavepacket approximation \cite{HHL23}. At the same time we set $\langle \hat{p}_i, \hat{z}_i\rangle^{}_{\rm I}=0$ for simplicity. 
These lead to a transverse spreading rate of the wavepacket $v_{}^{\sf T} = \bar{m}^{-1} \big(\langle\overline{\hat{p}_{\sf T}^2}\rangle^{}_{\rm I}\big)^{1/2} \approx 1.2 \times 10^4$ ${\rm m}/{\rm s}$. 
From (\ref{zizjPrest}) with (\ref{Zjzjs2}) inserted, we find $\langle \hat{z}_{}^{\sf T}(t^{}_F), \hat{z}_{}^{\sf T}(t^{}_F) \rangle \approx (141\,\,\mu{\rm m})^2$ when the wavepacket is arriving at the image plane. This is consistent with the transverse coherence length of electrons, $140$ $\mu{\rm m}$ at the screen, given in \cite{TE89}. 

Inserting these parameter values as well as those mentioned in Section \ref{SecRegValue} to the formula (\ref{uiAsmp}) and then (\ref{Puri}),
we find that the purity in the transverse directions for an electron wavepacket around the image plane is {about \footnote{In the preliminary result in Ref. \cite{Lin23}, the $F$-part correlators are under-estimated by a factor of $4\pi$, so that the values of ${\bf P}^{}_{i}$ in \cite{Lin23} are higher than what we obtain  here.}
\begin{equation}
   {\bf P}^{}_{\sf T} \equiv \frac{\hbar}{2\sqrt{u^{}_{\sf T}(t^{}_F)}} \approx 0.47 . 
\end{equation}
It turns out that the flying time $t^{}_F$ here is quite close to the decoherence time 
$T^{\rm dc}_{\sf T}\approx 1.13\times 10^{-8}$ s estimated by (\ref{Tdecoh}).
The above result suggests that {\it quantum decoherence by vacuum fluctuations of EM fields may be a major source of the blurring in the interference pattern reported in Ref.} \cite{TE89}. For a slightly larger initial width, we get a slightly better purity 
which would still be compatible with the contrast of the interference pattern measured in Ref. \cite{TE89}.}

In the longitudinal direction, almost all the parameters are the same except that the initial width of the wavepacket is now 
$\big(\langle\overline{\hat{z}_{1}^2}\rangle^{}_{\rm I}\big)^{1/2} 
= 1.7 \times 10^{-6}$ m, which is much greater than $\big(\langle\overline{\hat{z}_{\sf T}^2}\rangle^{}_{\rm I}\big)^{1/2} $ and $\lambda^{}_C/\bar{\gamma}(t^{}_0)$ but much less than $75\mu$m so that our Gaussian approximation and the approximated formula (\ref{uiAsmp2}) are still good. We get
\begin{equation}
   {\bf P}^{}_1 \equiv \frac{\hbar}{2\sqrt{u^{}_{1}}} \approx 0.9995, 
\end{equation}
and then the purity of the single electron wavepacket centered around the image plane of the electron microscope in \cite{TE89} is ${\bf P} = {\bf P}_1 {\bf P}_2 {\bf P}_3 \approx 0.9995 \times (0.47)^2 \approx 0.22$. 

If we have a larger time-tagging uncertainty (or a lower time resolution) for the history of electrons in the experiment, corresponding to larger values for the regulators $\epsilon_1$ and $\epsilon_0$ 
(or a smaller value of energy uncertainty $\Delta E$ for the electrons tunneling out of the FEG), 
then we will get a longer decoherence time $T^{\rm dc}_i$ from (\ref{Tdecoh}), and the value of purity ${\bf P}^{}_i$ will be closer to 1 for the same flying proper time $\tau^{}_F$ of single electrons.

\section{Accelerated wavepacket in a uniform electric field}
\label{SecUACinUE}

Consider a uniform electric field with field tensor $\bar{F}_{[0]}^{01} = -\bar{F}_{[0]}^{10} = {\cal E}/c$, which is constant in spacetime. A charged particle moving with uniform acceleration in this uniform electric field follows a classical worldline 
\begin{equation}
   \bar{z}^\mu (\tau) = \left(\frac{c^2}{a}\sinh \frac{a\tau}{c}, \frac{c^2}{a}\cosh \frac{a\tau}{c}, 0,0 \right) \label{zintauUAC}
\end{equation}
with  proper acceleration $a = q{\cal E}/\bar{m}$. This is a solution to the LAD equation (\ref{LADEqTau}) parametrized by the proper time $\tau$ of the particle. Define the scaled acceleration $\alpha\equiv a/c$ (in the unit of frequency), the corresponding 4-velocity of the particle reads
\begin{equation}
  \bar{u}^\mu (\tau) = \partial_\tau\bar{z}^\mu(\tau) = \left(c \cosh \alpha \tau, c \sinh \alpha \tau,0,0\right) = 
	(\bar{\gamma} c, \bar{\gamma}\bar{\bf v}) = \bar{\gamma}\bar{v}^\mu
\end{equation}
such that $\bar{u}_\mu \bar{u}^\mu = -c^2$. In terms of the Minkowski time $t=\frac{1}{c}z^0(\tau)= \alpha^{-1}\sinh \alpha\tau$, one has
\begin{eqnarray}
  \bar{z}^\mu (t) &=& \left(ct, c\sqrt{ \alpha^{-2}+ t^2},0,0\right),\label{zintUAC}\\
  \bar{\gamma}(t) &=& \cosh \alpha\tau(t) = \sqrt{1+\sinh^2 \alpha\tau(t)} = \sqrt{1 + \left(\alpha t\right)^2}, 
	  \label{gammaUAC} \\
  \bar{v}^\mu(t) &=& \left( c, c \tanh \alpha\tau(t), 0,0\right) = \left(c, c\frac{\alpha t}{\bar{\gamma}(t)}, 0,0\right),\label{vintUAC}
\end{eqnarray}
which yields $\bar{M}_{11}=\cosh^2 \alpha\tau(t)=\bar{\gamma}^2(t)$ and $\bar{M}_{22}=\bar{M}_{33}=1$ with other elements vanishing, 
as defined in (\ref{Mijdef}) (so the nonvanishing elements of the inverse matrix $M^{ij}$ are $\bar{M}^{11}=1/\cosh^2 \alpha\tau(t)$ and $\bar{M}^{22}=\bar{M}^{33}=1$.) Then Eq. (\ref{QLADEqt}) reads
\begin{eqnarray}
	\bar{m} \partial_t \left[\bar{\gamma}^3 \partial_t {\cal Z}^\Omega_1\right] &=&
	   q c\, {\cal F}_{10}^{[0]\Omega}\big(\bar{z}(t)\big) + 
	   s\bar{m}\bar{\gamma}\partial_t^2 \left[ \bar{\gamma}^3 \partial_t  {\cal Z}^\Omega_1 \right] + O(\Lambda^{-1}), \label{UACmodeL}\\
  \bar{m} \partial_t \left[\bar{\gamma} \partial_t {\cal Z}^\Omega_{\sf T}\right] &=&
	   q c\,{\cal F}^{[0]\Omega}_{{\sf T}0}\big(\bar{z}\big)+
		 q c \frac{\alpha t}{\bar{\gamma}}{\cal F}^{[0]\Omega}_{{\sf T}1}\big(\bar{z}\big) \nonumber\\ & &
	+ s\bar{m}\left[ \bar{\gamma}^2 \partial_t^3 {\cal Z}^\Omega_{\sf T}
	+ 3\bar{\gamma}(\partial_t\bar{\gamma}) \partial_t^2{\cal Z}^\Omega_{\sf T}\right] +O(\Lambda^{-1}) \label{UACmodeT}
\end{eqnarray}
with $s$ defined in (\ref{defofs}) and ${\sf T}=2,3$ in the directions transverse to the acceleration and velocity of the particle. Below we use `longitudinal' and `transverse' correlators for the two-point correlators of $z_{}^j$ and $p_{}^j$ with respect to the direction of acceleration of the particle, which should not be confused with the longitudinal and transverse degrees of freedom of EM fields $A_{}^\mu$.

\subsection{Mode functions}
\label{modefnUAC}

\subsubsection{Particle-motion deviation in the direction of acceleration}

Assume $s$ is sufficiently small. Applying the same iteration in obtaining (\ref{QLLErest}) to (\ref{UACmodeL}), we have 
\begin{equation}
  \bar{m} \partial_\tau\Big[ \bar{\gamma}^2(\tau)\partial_\tau {\cal Z}_{1}^\Omega(\tau)\Big] =
	  qc\bar{\gamma}(\tau)\left(1+s\partial^{}_{\tau} \right) {\cal F}_{10}^{[0]\Omega}\big(\bar{z}(\tau)\big) +O(\Lambda^{-1}),  
		\label{EqZUACs0}
\end{equation}
where $\tau= \alpha^{-1}\sinh^{-1} \alpha t$ from (\ref{zintauUAC}) and (\ref{zintUAC}), $\partial_\tau = \bar{\gamma}\partial_t$, 
and $\bar{\gamma}(\tau)=\cosh \alpha \tau $ from (\ref{gammaUAC}).
Note that $\tau$ here, though looks like the proper time for the particle, is simply a convenient {\it parameter} for solving the above differential equations. Our theory is formulated in Minkowski coordinates with the time slices parametrized by Minkowski time $t$.   

The formal solution to Eq. (\ref{EqZUACs0}) with $O(\Lambda^{-1})$ neglected for $\tau\equiv \tau(t) > \tau^{}_0 \equiv \tau(t^{}_0)$ reads 
\begin{equation}
  \bar{m} {\cal Z}_{1}^\Omega (\tau) = C_{1}^\Omega + \tilde{C}_{1}^\Omega \tanh \alpha \tau + 
	 	q c \int^{\tau}_{\tau_0} 
	d\tilde{\tau} K_\parallel \left(\tau, \tilde{\tau} \right)
	\bar{\gamma}(\tilde{\tau})\left(1+s\partial^{}_{\tilde{\tau}} \right)   {\cal F}_{10}^{[0]\Omega}\big(\bar{z}(\tilde{\tau})\big),
\end{equation}
where $C_{1}^\Omega$ and $\tilde{C}_{1}^\Omega$ are constants to be determined by the initial condition at the moment when the interaction is switched on, and the kernel 
\begin{equation}
  K_\parallel(\tau, \tau') \equiv \alpha^{-1}\left( \tanh \alpha\tau - \tanh \alpha\tau' \right) \label{Kparallel}
\end{equation}
is the solution to 
\begin{equation}
  \partial_\tau \Big[ \bar{\gamma}^2(\tau)\, \partial_\tau K_\parallel(\tau, \tau') \Big] = 0
\end{equation}
satisfying the condition $K_\parallel(\tau',\tau')= 0$ and $\bar{\gamma}^2(\tau) \partial_\tau K_\parallel(\tau, \tau') = 1$. When $\alpha\tau, \alpha\tau'\ll 1$, $K_\parallel(\tau, \tau') \approx \tau-\tau' = K(\tau, \tau')$ in (\ref{Krest}) for the particle at rest.

Matching the initial conditions (\ref{freeModeZ}) and (\ref{freeModeA}) for $\tau = \tau^{}_0$, we find
\begin{eqnarray}
  && {\cal Z}^1_{z_{}^j}(\tau) = \frac{m}{\bar{m}}\delta^1_j, 
	\hspace{1cm} {\cal Z}^1_{p^{}_j}(\tau) = \frac{\eta^{1j}}{\bar{m}} K_\parallel (\tau, \tau^{}_0),
		 \label{Z1zjUAC} \\  
	&& {\cal Z}^1_{(\lambda){\bf k}}(\tau) = \frac{qc}{\bar{m}}{\cal E}^{{\bf k}\,\,1}_{(\lambda)\,\,0}f^\parallel_{\bf k} (\tau)		\label{Z1lkUAC}
\end{eqnarray}
with ${\cal E}^{{\bf k}\,\,1}_{(\lambda)\,\,0} = \eta^{1\mu}{\cal E}^{{\bf k}}_{(\lambda)\mu 0}$ given in (\ref{calEj}), and
\begin{eqnarray}
f^\parallel_{\bf k}(\tau) &=& \int^\tau_{\tau^{}_0} 
	d\tilde{\tau} K_\parallel(\tau,\tilde{\tau})\bar{\gamma}(\tilde{\tau})\big[ 1+ s\, i k_\nu \dot{\bar{z}}^\nu(\tilde{\tau})\big]
		e^{ik_\mu \bar{z}^\mu(\tilde{\tau})} e^{-\omega\epsilon/2}. \label{fparakDef} 
\end{eqnarray}
Here $k_\mu \equiv (-\omega/c, {\bf k})$ with $\omega \equiv c |{\bf k}|$, and so $i k_\mu \bar{z}^\mu(\tilde{\tau}) = 
-i\frac{\omega}{c}\bar{z}^0(\tilde{\tau})+i k^{}_1 \bar{z}^1(\tilde{\tau}) = -i\frac{\omega}{\alpha}\sinh\alpha\tilde{\tau} 
+ i k^{}_1 \frac{c}{\alpha}\cosh\alpha\tilde{\tau}$ and $i k_\nu \dot{\bar{z}}^\nu(\tilde{\tau}) = -i\omega\cosh\alpha\tilde{\tau}
+i k^{}_1 c \sinh\alpha\tilde{\tau}$ from (\ref{zintauUAC}).

Substituting (\ref{zintauUAC})-(\ref{vintUAC}) to (\ref{modeFnP}), the canonical momentum of the uniformly accelerated particle in the direction of acceleration reads
\begin{eqnarray}
{\cal P}^\Omega_1(\tau) &=& q {\cal A}^{[0]\Omega}_1\big(\bar{z}(\tau)\big)+
  \bar{m}' \cosh^2 \alpha\tau\, \partial^{}_\tau {\cal Z}^1_\Omega(\tau)  - \nonumber\\
&& \frac{3}{2}s \bar{m}\left[ \cosh^2\alpha\tau\, \partial_\tau^2 {\cal Z}^1_\Omega  + 
	\alpha\left(\frac{3}{2}\sinh2\alpha\tau +\tanh\alpha\tau\right)\,\partial^{}_\tau {\cal Z}^1_\Omega \right]
\label{P1UAC}
\end{eqnarray}
with $O(\Lambda^{-1})$ neglected.
Inserting the above results for ${\cal Z}^1_{\Omega}$, we get
\begin{eqnarray}
&&{\cal P}^{z_{}^j}_{1}(\tau) =0, \hspace{2cm}
  {\cal P}^{p^{}_j}_{1}(\tau) = \eta^{1j}\Phi_\parallel(\tau), \label{P1zjUAC}\\
&&{\cal P}^{(\lambda){\bf k}}_{1}(\tau) = q\epsilon^1_{(\lambda){\bf k}}e^{ik_\mu \bar{z}^\mu(\tau)}e^{-\omega\epsilon/2} + 
  qc{\cal E}^{{\bf k}\,\,1}_{(\lambda)\,\,0} \times \nonumber\\
&&	\hspace{.5cm} \left\{ \Phi_\parallel(\tau) \cosh^2\alpha\tau \, \dot{f}^\parallel_{\bf k} (\tau) -\frac{3}{2}s\cosh\alpha\tau 
  \big[ 1+ s\, i k_\nu \dot{\bar{z}}^\nu(\tau)\big] e^{ik_\mu \bar{z}^\mu(\tau)}e^{-\omega\epsilon/2}\right\}, 	\label{P1lkUAC}
\end{eqnarray}
where
\begin{equation}
	\Phi_{\parallel} (\tau)\equiv \frac{\bar{m}'}{\bar{m}} -\frac{3}{2} s\alpha 
	\tanh\alpha\tau \big( 1 + {\rm sech}^2 \alpha\tau\big). 
\end{equation}

\subsubsection{Particle-motion deviation transverse to acceleration}
\label{SecZTransverse}

Eq. (\ref{UACmodeT}) can be written as
\begin{equation}
  \bar{m} \partial_\tau^2 {\cal Z}^\Omega_{\sf T}(\tau)= qc{\cal F}^\Omega_{\sf T}(\tau) + 
	 s \bar{m} \partial_\tau \Big[ \left(\partial_\tau^2 -\alpha^2 \right){\cal Z}^\Omega_{\sf T}(\tau)\Big] + O(\Lambda^{-1}),  	
	\label{UACmodeTtau}
\end{equation}
where 
\begin{equation}
  {\cal F}^\Omega_{\sf T}(\tau) \equiv {\cal F}^{[0]\Omega}_{{\sf T}0}\big(\bar{z}(\tau)\big) \, \cosh \alpha\tau
	+ {\cal F}^{[0]\Omega}_{{\sf T}1}\big(\bar{z}(\tau)\big)\, \sinh\alpha\tau.
\end{equation}
While the third derivative of ${\cal Z}^\Omega_{\sf T}$ in the above equation will be treated as a perturbation, the first derivative of ${\cal Z}^\Omega_{\sf T}$ may be not. 
In fact, $s \alpha^2 =s a^2/c^2 > 1$ Hz for $a > 4.3 \times 10^{20}$ m$/{\rm s}^2$, which is achievable in laboratories for electrons accelerated by intense laser fields (e.g. \cite{ELINP, SULF}). 
So here we treat the first derivative term in (\ref{UACmodeTtau}) as of  leading order, and insert $\bar{m} \partial_\tau^2 {\cal Z}^\Omega_{\sf T}= qc{\cal F}^\Omega_{\sf T}(\tau) -s\alpha^2 \bar{m}\partial_\tau{\cal Z}^\Omega_{\sf T}+ O(sc)$ to the third derivative term $s \partial^{}_\tau \left( \bar{m}\partial_\tau^2 {\cal Z}^\Omega_{\sf T}\right)$. Then we obtain
\begin{equation}
\bar{m}\Big[ \left( 1+s^2\alpha^2 \right) \partial^2_\tau + s \alpha^2 \partial^{}_\tau \Big]{\cal Z}^\Omega_{\sf T}  = qc(1+s\partial^{}_\tau){\cal F}^\Omega_{\sf T} + O(\Lambda^{-1}). \label{UACmodeTtaus0}
\end{equation} 
The formal solutions to the above equation with $O(\Lambda^{-1})$ neglected for $\tau > \tau^{}_0$ read
\begin{eqnarray}
  \bar{m}{\cal Z}^\Omega_{\sf T}(\tau) &=& C^\Omega_{\sf T} + \tilde{C}^\Omega_{\sf T} 
	  e^{-\frac{s \alpha^2}{\varsigma} (\tau-\tau^{}_0)} + q c \int^\tau_{\tau^{}_0}
	  d\tilde{\tau} K_\perp(\tau, \tilde{\tau})\left( 1 + s \partial^{}_{\tilde{\tau}} \right) {\cal F}^\Omega_{\sf T}(\tilde{\tau})
\end{eqnarray}
with constants $C^\Omega_{{\sf T}}$ and $\tilde{C}^\Omega_{{\sf T}}$, $\varsigma \equiv 1+ (s \alpha)^2$, and the kernel 
\begin{equation}
  K_\perp(\tau, \tilde{\tau}) \equiv \frac{1}{s\alpha^2}\left( 1- e^{-\frac{s\alpha^2}{\varsigma} (\tau-\tilde{\tau})}\right),  \label{Kperpdef}
\end{equation}
which is the solution to 
\begin{equation}
  \varsigma \partial_\tau^2 K_\perp(\tau,\tilde{\tau}) + s\alpha^2\partial_\tau K_\perp(\tau,\tilde{\tau}) = 0
\end{equation}
satisfying $K_\perp(\tau,\tau)=0$ and $\left( \varsigma\partial^{}_\tau + s\alpha^2\right) K_\perp(\tau,\tilde{\tau})=1$. 

Matching the initial conditions (\ref{freeModeZ}) and (\ref{freeModeA}) for $\tau = \tau^{}_0$, we find 
\begin{eqnarray}
  && {\cal Z}^{\sf T}_{z_{}^j}(\tau) = \frac{m}{\bar{m}}\delta^{\sf T}_j, \hspace{1cm} 
	{\cal Z}^{\sf T}_{p^{}_j}(\tau) = \frac{\eta^{{\sf T}j}}{\bar{m}}\varsigma K_\perp(\tau, \tau^{}_0), \label{ZTzjUAC} \\
	&& {\cal Z}^{\sf T}_{(\lambda){\bf k}}(\tau)=  
	\frac{qc}{\bar{m}}\sum_{B=0,1}{\cal E}^{{\bf k}\,\,{\sf T}}_{(\lambda)\,\,B} f^{\perp}_{{\bf k}B}(\tau) \label{ZTlkUAC}
\end{eqnarray}
for $\tau>\tau^{}_0$, where
\begin{equation}
  f^{\perp}_{{\bf k}B}(\tau) \equiv \int^\tau_{\tau^{}_0} d\tilde{\tau} K_\perp(\tau,\tilde{\tau})\big( 1+ s\partial^{}_{\tilde{\tau}}\big)
	\frac{1}{2}\Big[ e^{\alpha\tilde{\tau}}+(-1)_{}^B e^{-\alpha\tilde{\tau}}\Big]e^{ik_\mu \bar{z}^\mu(\tilde{\tau})}e^{-\omega\epsilon/2},
	\label{fperpkDef}
\end{equation}
${\cal E}^{(0){\bf k}}_{{\sf T}1} = {\cal E}^{(3){\bf k}}_{{\sf T}1} \equiv 0$, and
\begin{equation}
{\cal E}^{(1){\bf k}}_{{\sf T}1}\equiv i\left( k^{}_{\sf T}\epsilon^{(1){\bf k}}_{1}-k^{}_{1}\epsilon^{(1){\bf k}}_{\sf T}\right),
\hspace{.7cm}
{\cal E}^{(2){\bf k}}_{{\sf T}1}\equiv i\left( k^{}_{\sf T}\epsilon^{(2){\bf k}}_{1}-k^{}_{1}\epsilon^{(2){\bf k}}_{\sf T}\right),
\label{calEj1}
\end{equation}
such that 
\begin{equation}
  {\cal F}_{{\sf T}1}^{[0](\lambda){\bf k}}(t,{\bf x}) =
	\partial^{}_{\sf T} {\cal A}_{1}^{[0](\lambda){\bf k}}(t,{\bf x})-\partial^{}_1 {\cal A}_{{\sf T}}^{[0](\lambda){\bf k}}(t,{\bf x})
	= {\cal E}_{{\sf T}1}^{(\lambda){\bf k}} e^{-i \omega t + i{\bf k}\cdot{\bf x}}.
\end{equation}
One may say ${\cal E}_{{\sf T}1}^{(\lambda){\bf k}}$ corresponds to magnetic-field-type fluctuations, while ${\cal E}_{{\sf T}0}^{(\lambda){\bf k}}$ corresponds to electric-field-type fluctuations from (\ref{Fj0}).

From (\ref{zintauUAC})-(\ref{vintUAC}) and (\ref{modeFnP}), the canonical momentum of the uniformly accelerated particle perpendicular to acceleration reads
\begin{equation} 
{\cal P}^\Omega_{\sf T}(\tau) = q{\cal A}^{[0]\Omega}_{\sf T}(\bar{z}(\tau))+\bar{m}' \partial^{}_\tau {\cal Z}^{\sf T}_\Omega(\tau)  
	- \frac{3}{2}s \bar{m}\Big( \partial_\tau^2 +\alpha\tanh\alpha\tau\, \partial^{}_\tau \Big){\cal Z}^{\sf T}_\Omega .
\end{equation}
Inserting the above result for ${\cal Z}^{\sf T}_\Omega (\tau)$ for $\tau > \tau_0$, we have
\begin{eqnarray}
&&{\cal P}^{z_{}^j}_{\sf T} = 0, \hspace{2cm} 
{\cal P}^{p^{}_j}_{\sf T} = \eta^{{\sf T}j} e^{-\frac{s \alpha^2}{\varsigma}(\tau-\tau_0)} \Phi_\perp(\tau),\\
&& {\cal P}^{(\lambda){\bf k}}_{\sf T} = q \epsilon_{(\lambda){\bf k}}^{\sf T} e^{ik_\mu\bar{z}^\mu(\tau)} e^{-\omega\epsilon/2} + 
  qc \sum_{B=0,1} {\cal E}^{{\bf k}\,\,{\sf T}}_{(\lambda)\,\,B}\times \nonumber\\
&& \hspace{.5cm} \left\{ \Phi_\perp(\tau) \dot{f}^{\perp}_{{\bf k}B}(\tau) -\frac{3s}{2\varsigma}\big( 1+ s\partial^{}_{\tau}\big)
	\frac{1}{2}\Big[ e^{\alpha\tau}+(-1)_{}^B e^{-\alpha\tau}\Big]e^{ik_\mu \bar{z}^\mu(\tau)} e^{-\omega\epsilon/2} \right\}
\end{eqnarray}
with
\begin{equation}
  \Phi_\perp (\tau) \equiv \frac{\bar{m}'}{\bar{m}}-\frac{3}{2}s \alpha\tanh\alpha\tau + \frac{3}{2}\frac{s^2\alpha^2}{\varsigma}. \label{Phiperpdef}
\end{equation}

Since $sc = 2r_0/3$ with the classical electron radius $r_0 \approx 2.8 \times 10^{-15}$ m much shorter than the electron Compton wavelength $\lambda^{}_C\approx 2.4 \times 10^{-12}$ m, even if the electric field is very close to the Schwinger limit ${\cal E}=\frac{\bar{m}^2 c^3}{q\hbar}$, with which $\alpha = \frac{a}{c} =  \frac{q{\cal E}}{\bar{m}c} = \frac{\bar{m}c^2}{\hbar} = \frac{1}{t^{}_C}$ (electron Compton frequency), one will still have $s \alpha = s a/c \approx 3.87\times 10^{-4} \ll 1$.  
Actually, the most powerful lasers in the world to date can achieve an intensity of $10^{23}$ ${\rm W}/{\rm cm}^2$ 
\cite{ELINP, SULF}, 
the electric fields of which produce roughly a scaled acceleration $\alpha \sim 10^{-3} t_C^{-1}$ at most.
This means that $s\alpha$ is always a small, dimensionless parameter in our effective theory for uniformly accelerated charges, and so in the following we will neglect the $O\left((s\alpha)^2\right)$ terms and treat $\varsigma \approx 1$.

\subsection{Unruh effect}
\label{SecUnruhEffect}

The symmetric two-point correlation function of the uniformly accelerated particle's position deviation can be computed by inserting the mode functions obtained in Section \ref{modefnUAC} to (\ref{zizjrestAll}), (\ref{pipjrestAll}), and (\ref{pizjrestAll}). For example, inserting (\ref{Z1lkUAC}) into (\ref{zizjFrest}), we obtain 
\begin{eqnarray}
&&\langle \hat{z}^1(t), \hat{z}^1(t') \rangle^{}_F = \frac{\hbar}{(2\pi)^3\varepsilon^{}_0}\times \nonumber\\ 
&&{\rm Re} \int_0^\infty \frac{\omega^2}{c^2}\frac{d\omega}{2\omega c} 
\int_0^{2\pi} d\varphi\int_0^{\pi}d\theta\, \sin\theta \,\, \frac{qc}{\bar{m}}{\cal E}^{(\lambda)j}_{{\bf k}\,\,\,\,\,\,\,\,0}\,\,
\frac{qc}{\bar{m}} {\cal E}^{{\bf k}\,\,j'*}_{(\lambda)\,\,0}\,\, f^\parallel_{\bf k}(\tau)f^{\parallel*}_{\bf k}(\tau'),
\label{z1z1FUAC1}
\end{eqnarray}
where $\tau = \alpha^{-1}\sinh^{-1}\alpha t$, $\tau'\equiv \tau(t')$.

To find the result to the leading order of $s$, we need to calculate the integral
\begin{equation} 
{\cal I}_{BB'}^{\,j\,j'}(\tilde{\tau},\tilde{\tau}') \equiv \int \omega d\omega d\varphi \, 
	d\cos\theta \, {\cal E}^{(\lambda)j}_{{\bf k}\,\,\,\,\,\,\,\,B} \, {\cal E}^{{\bf k}\,\,j'*}_{(\lambda)\,\,B'} \,
	e^{-i\frac{\omega}{c}\left[\bar{z}^0(\tilde{\tau})-\bar{z}^0(\tilde{\tau}')\right]+
	i k_1\left[\bar{z}^1(\tilde{\tau})-\bar{z}^1(\tilde{\tau}')\right]-\omega\epsilon} \label{calIjjBB}
\end{equation}	
with $B,B'=0,1$. Following the same method in Ref. \cite{LH06}, let
\begin{equation}
  e^{-i\frac{\omega}{c}\bar{z}^0(\tilde{\tau})+i k_1\bar{z}^1(\tilde{\tau})} \equiv 
	\int_{-\infty}^\infty d\kappa e^{-i\kappa\tilde{\tau}}\varphi_{\bf k}(\kappa). \label{varphikDef}
\end{equation}
Then we get 
\begin{eqnarray}
 {\cal I}_{00}^{11}(\tilde{\tau},\tilde{\tau}') &=& 
	2\pi\int_0^{\infty}\omega d\omega e^{-\omega\epsilon} \int_{-1}^1 d\cos\theta \times\nonumber\\ && 
  \frac{\omega^2}{c^2}\left(1-\cos^2\theta\right) \int_{-\infty}^\infty d\kappa \int_{-\infty}^\infty d\kappa' 
  e^{-i \kappa \tilde{\tau}+i \kappa'\tilde{\tau}'}\varphi^{}_{\bf k}(\kappa)\varphi^*_{\bf k}(\kappa'). \label{I1UAC0}
\end{eqnarray}
where we have substituted ${\cal E}^{(\lambda)1}_{{\bf k}\,\,\,\,\,\,\,\,\,0} \, {\cal E}^{{\bf k}\,\,1*}_{(\lambda)\,\,0}= \left(\frac{\omega}{c}\right)^2 \eta^{11}-(k^1)^2 =\left(\frac{\omega}{c}\right)^2 (1-\cos^2\theta)$ from (\ref{calEj}) and (\ref{eesum}).
Inserting (\ref{zintauUAC}), writing
\begin{equation}
  \varphi_{\bf k}(\kappa) = \int_{-\infty}^\infty \frac{d\check{\tau}}{2\pi} e^{i\kappa \check{\tau}} 
	e^{-i\frac{\omega}{c}\bar{z}^0(\check{\tau})+i k_1\bar{z}^1(\check{\tau})}  \label{FTeikzUAC}
\end{equation}
which is simply the inverse Fourier transform of (\ref{varphikDef}), and integrating $\cos\theta$ and $\omega$ in turn, we obtain
\begin{eqnarray}
&&{\cal I}_{00}^{11}(\tilde{\tau},\tilde{\tau}')=\frac{4c^2}{\pi}\int \frac{d\kappa d\kappa' e^{-i\kappa\tilde{\tau}+i\kappa'\tilde{\tau}'}   
   d\check{\tau} d\check{\tau}' e^{i\kappa\check{\tau}-i\kappa'\check{\tau}'}}
	{\left[\left(\bar{z}^0(\check{\tau})-\bar{z}^0(\check{\tau}')-ic\epsilon \right)^2-
	 \left(\bar{z}^1(\check{\tau})-\bar{z}^1(\check{\tau}')\right)^2\right]^2} \nonumber\\
&&=\frac{4c^2}{\pi}\int \frac{d\bar{\kappa}\, d D \,\,
  e^{-i\bar{\kappa}(\tilde{\tau}-\tilde{\tau}')- \frac{i}{2}D(\tilde{\tau}+\tilde{\tau}')}\, 
   d\check{T}\,d\check{\Delta}\,\, e^{i D\check{T}+i\bar{\kappa}\check{\Delta}}}
	{\left[\left(\frac{2c}{\alpha}\sinh\alpha\check{T} \sinh\frac{\alpha\check{\Delta}}{2}\right)^2 -\left(
	  \frac{2c}{\alpha}\cosh\alpha \check{T} \sinh\frac{\alpha\check{\Delta}}{2}-ic\epsilon \right)^2\right]^2}
\end{eqnarray}
with $\check{T} \equiv (\check{\tau}+\check{\tau}')/2$, $\check{\Delta}\equiv \check{\tau}-\check{\tau}'$,
$\bar{\kappa}\equiv (\kappa+\kappa')/2$, and $D \equiv \kappa-\kappa'$. The $D$-Integration gives a Dirac delta function
$2\pi \delta(\check{T}-\tilde{T})$ with $\tilde{T}\equiv (\tilde{\tau}+\tilde{\tau}')/2$ (so $\tau^{}_0 < \tilde{T} < \tau$). Then the $\check{T}$ integration yields
\begin{equation}
{\cal I}_{00}^{11}(\tilde{\tau},\tilde{\tau}') = 8c^2\int \frac{d\bar{\kappa} e^{-i\bar{\kappa}(\tilde{\tau}-\tilde{\tau}')} 
   d\check{\Delta} e^{i\bar{\kappa}\check{\Delta}}}
	{\left[\frac{4c^2}{\alpha^2}\sinh^2\frac{\alpha\check{\Delta}}{2} - i c\epsilon  \left(
	  \frac{4c}{\alpha}\cosh{\alpha \tilde{T}}\sinh\frac{\alpha\check{\Delta}}{2}-ic\epsilon \right)\right]^2}. \label{I1UACe}
\end{equation}
The poles of the above integrand are located at the solutions to $\sinh\frac{\alpha\check{\Delta}}{2} = i\epsilon \frac{\alpha}{2} e^{\pm\alpha\tilde{T}}$ in the complex $\check{\Delta}$-plane. Similar to the transition probability from time-dependent perturbation theory in Appendix A of Ref. \cite{LH10}, if $|\alpha \tilde{T}|$ is sufficiently large, the poles obtained from $\sinh\frac{\alpha\check{\Delta}}{2} = i\frac{\alpha}{2} \epsilon e^{+\alpha\tilde{T}}$ and those from $\sinh\frac{\alpha\check{\Delta}}{2} = i\frac{\alpha}{2}\epsilon e^{-\alpha\tilde{T}}$ will split significantly, and one will obtain a pole structure very different from \cite{BD82, LH10}
\begin{equation}
{\cal I}_{00}^{11}(\tilde{\tau},\tilde{\tau}') \approx 8c^2\int d\bar{\kappa} e^{-i\bar{\kappa}(\tilde{\tau}-\tilde{\tau}')} 
   \int d\check{\Delta} \frac{e^{i\bar{\kappa}\check{\Delta}}}
	{\left[\frac{4c^2}{\alpha^2}\sinh^2\frac{\alpha}{2}(\check{\Delta}-i\epsilon')\right]^2}, \label{I0011BD}
\end{equation}
which derives the Unruh effect, as will be shown later in (\ref{I1final}).

The form of (\ref{I0011BD}) suggests that one could simply replace the exponent of the integrand of a mode function, say, $f^\parallel$ in (\ref{fparakDef}), from $ik_\mu \bar{z}^\mu(\tilde{\tau})-\omega \epsilon/2$ to $ik_\mu \bar{z}^\mu \left(\tilde{\tau}- i\epsilon/2\right)$. 
However, one cannot obtain (\ref{I0011BD}) exactly by doing this.  
To bridge (\ref{I1UACe}) and (\ref{I0011BD}), alternatively, one may introduce an {\it ad hoc} assumption that $\epsilon \sim \epsilon'/\cosh[\alpha(\tilde{\tau}+\tilde{\tau}')/2]$ in (\ref{I1UACe}) with a small constant $\epsilon'$ to suppress the splitting of the poles we mentioned above. This requires $\epsilon$ in (\ref{I1UACe}) to be a function of $\tilde{\tau}$ and $\tilde{\tau}'$, and so the factor $e^{-\omega\epsilon(\tilde{\tau})/2}$ in mode function (\ref{fparakDef}) or (\ref{fperpkDef}) must be a part of the integrand. Thus, the factor $e^{-\omega\epsilon}$ in (\ref{calIjjBB}) contributed by $f^\parallel_{\bf k}(\tau)f^{\parallel *}_{\bf k}(\tau')$ in (\ref{z1z1FUAC1}) should be replaced by $\exp\big\{ -\frac{\omega}{2} \epsilon'\big[ F(\tilde{\tau}) + F^*(\tilde{\tau}') \big] \big\}$ with some function $F$. Unfortunately, we failed to find any function $F$ giving either $\frac{1}{2}\big[F(\tilde{\tau}) + F^*(\tilde{\tau}')\big] \sim 1/\cosh[\alpha(\tilde{\tau}+\tilde{\tau}')/2]$, or even $\frac{1}{2}\big[F(\tilde{\tau}) + F^*(\tilde{\tau}')\big] \times \cosh[\alpha(\tilde{\tau}+\tilde{\tau}')/2] \sim O(1)$ in general [e.g., for $F(\tilde{\tau})=1/\bar{\gamma}(\tilde{\tau}) = 1/\cosh\alpha\tilde{\tau}$ from (\ref{gammaUAC}), when $0 < \tilde{\tau} \ll \tilde{\tau}'$ and $\alpha\tilde{\tau}' \gg 1$, one has $\frac{1}{2}\big[1/\bar{\gamma}(\tilde{\tau})+ 1/\bar{\gamma}(\tilde{\tau}') \big] \cosh\frac{\alpha}{2}(\tilde{\tau}+\tilde{\tau}') \approx \frac{1}{2} (1/\cosh\alpha\tilde{\tau})\cosh\frac{\alpha}{2}\tilde{\tau}' \gg 1$.]

It has been demonstrated in Ref. \cite{HHL23} that, to keep the single-particle interpretation applicable and the charge density distribution nearly Gaussian for a long time, the minimal width of a Klein-Gordon wavepacket at the moment $\tau(t)=\tau^{}_0$ is about $\lambda^{}_C/\bar{\gamma}(\tau^{}_0)$. Inspired by this observation, we propose to replace the $e^{-\omega \epsilon/2}$ factor in the mode function (\ref{fparakDef}) 
by $e^{-\omega\epsilon'/[2\bar{\gamma}(\tau)]}$, namely, we set
\begin{equation}
  \epsilon(t) = \frac{\epsilon'}{\bar{\gamma}[\tau(t)]} \label{epsloft}
\end{equation}
with constant $\epsilon'$ chosen as the electron Compton time $t^{}_C$ in Minkowski coordinates such that 
$ c\epsilon(t)= \lambda^{}_C/{\bar{\gamma}(t)}$ in (\ref{I1UACe}) and 
the factor $e^{-\omega\epsilon}$ in (\ref{calIjjBB}) becomes $\exp \big\{-\frac{\omega}{2}\epsilon'\big[\bar{\gamma}^{-1}(\tau)+\bar{\gamma}^{-1}(\tau')\big] \big\} \approx e^{-\omega\epsilon'/\bar{\gamma}(\tau)}$. Since $\tilde{\tau} \in [\tau^{}_0, \tau]$ (and $\tilde{\tau}' \in [\tau'_0, \tau']$) in the mode function (\ref{fparakDef}), from (\ref{gammaUAC}), we have $\frac{1}{2}\big[ 1/\bar{\gamma}(\tau) + 1/\bar{\gamma}(\tau')] \,\times \cosh \frac{\alpha}{2}(\tilde{\tau}+\tilde{\tau}') \le 1 + O(\epsilon^{}_1)$ for $\tau > |\tau_0|$, and so
\begin{equation}
{\cal I}_{00}^{11}(\tilde{\tau},\tilde{\tau}') = 8c^2\int \frac{d\bar{\kappa} e^{-i\bar{\kappa}(\tilde{\tau}-\tilde{\tau}')} 
   d\check{\Delta} e^{i\bar{\kappa}\check{\Delta}}}
	{\left[\frac{4c^2}{\alpha^2}\sinh^2\frac{\alpha\check{\Delta}}{2} - i c\epsilon' \left(
	  \frac{4c}{\alpha}\vartheta \sinh\frac{\alpha\check{\Delta}}{2}-\frac{i \epsilon' c}{\bar{\gamma}^2(\tau)}
		\right)\right]^2}, \label{I1UACegamma}
\end{equation}
with $\vartheta = \cosh(\alpha\tilde{T})/\cosh(\alpha\tau)$ no greater than $O(1)$ because $\tau^{}_0 < \tilde{T} < \tau$. Then, for $\tau >|\tau_0|$, (\ref{I1UACegamma}) can be approximated by (\ref{I0011BD}). 
Summing over the contributions from the poles at $\check{\Delta}= i(\epsilon'+2\pi n/\alpha)$ with integer $n$ in (\ref{I0011BD}), we obtain
\begin{eqnarray}
{\cal I}_{00}^{11}(\tilde{\tau},\tilde{\tau}') 
&\approx&  \frac{8\pi}{3c^2} \int d\kappa e^{-i\kappa(\tilde{\tau}-\tilde{\tau}')-\kappa \epsilon'}
	\frac{\kappa(\kappa^2+\alpha^2)}{1-e^{-2\pi\kappa/\alpha}}, \label{I1final}
\end{eqnarray}
where $\bar{\kappa}$ has been renamed to $\kappa$, and $\epsilon'=t^{}_C$ is assumed to be much smaller than $2\pi/\alpha$, namely, $\alpha \ll 2\pi/t^{}_C$.
In the above integrand, one can find a Planck factor corresponding to a bosonic bath at the Unruh temperature $T^{}_U \equiv \hbar a/(2\pi c k^{}_B)$. Thus we say the behavior of the correlator $\langle \hat{z}^1(t), \hat{z}^1(t') \rangle$ of a charged particle uniformly accelerated in Minkowski vacuum is analogous to the one moving in a thermal bath of photons at temperature $T_U$ proportional to its proper acceleration $a$, which is the Unruh effect.

Associated with (\ref{epsloft}), we let 
\begin{eqnarray}
  \epsilon_0(t^{}_0) &=& \tau(t^{}_0+\epsilon'_0)-\tau(t^{}_0), \hspace{.5cm}
	\epsilon_1(t) = \tau(t)-\tau(t-\epsilon'_1) \label{e0e1UAC}
\end{eqnarray}
with constants $\epsilon'_0$ and $\epsilon'_1$ determined by specific experimental settings in the laboratory frame. These are consistent with the regulators  in Section \ref{SecRestParticle} for the single electrons at rest. 
For $\alpha\epsilon'_0$, $\alpha\epsilon'_1 \ll 1$, we have $\epsilon_0 \approx \epsilon'_0/\bar{\gamma}[\tau(t^{}_0)]$ and $\epsilon_1 \approx \epsilon'_1/\bar{\gamma}[\tau(t)]$ from (\ref{dTauvsdt}). 
Also, $\epsilon_1(t)\gg \epsilon(t)$ for all $t$ if $\epsilon'_1 \gg \epsilon'$ $(= t^{}_C)$.

A few remarks are in order.
First, by setting $\epsilon \to \epsilon'/\bar{\gamma}(\tau)$ in mode function (\ref{fparakDef}) rather than $\epsilon'/\bar{\gamma}(\tilde{\tau})$, we have assumed that the frequency cutoff in a particle correlator at some moment $\tau$ depends on the particle motion at that moment only, but not on the history of its motion at $\tilde{\tau} \in [\tau^{}_0, \tau)$ in the past. 
So the factor $e^{-\omega\epsilon}$ can be moved out of the $\tilde{\tau}$ integrals in (\ref{fparakDef}) and (\ref{fperpkDef}).

Second, $e^{-\omega\epsilon} = e^{-\omega\epsilon'/\bar{\gamma}(\tau)}$ corresponds to the UV cutoff $\bar{\gamma}(\tau)/t^{}_C$ in frequency, or $\lambda^{}_C/\bar{\gamma}(\tau)$ in wavelength, which implies that the vacuum fluctuations of wavelengths much shorter than the electron Compton wavelength in the laboratory frame can have significant contribution to the particle correlators when $\bar{\gamma}(\tau) \gg 1$. Does this alert a breakdown of our effective theory? No: A charged particle in uniform motion in Minkowski space should not trigger pair-productions or be recoiled significantly by field fluctuations, no matter how close the particle's speed is to the speed of light relative to a laboratory observer. Although the UV cutoff for a moving electron observed in the laboratory frame is blue-shifted from the one for an electron at rest, this does not mean a breakdown of single-particle interpretation for the electron. 

Third, as we mentioned, the choice (\ref{epsloft}) is inspired by the observation in Ref. \cite{HHL23}, which is about the quasi-(1+1) dimensional wavepackets in the context of relativistic quantum mechanics only.
Accordingly, in the longitudinal direction (parallel to the acceleration), we set the regulator $c\epsilon^{}_\parallel(t) = c\epsilon'/\bar{\gamma}(t)=\lambda^{}_C/\bar{\gamma}(t)$ reflecting length contraction of the wavepackets observed in the laboratory frame. How about the regulators in the transverse directions? A reasonable choice is $c\epsilon^{}_\perp(t)=c[\tau(t+\epsilon')-\tau(t)]$ due to time dilation. It turns out that for $\alpha \epsilon' \ll 1$ ($\alpha t^{}_C = 1$ is the Schwinger limit), one has $c\epsilon^{}_\perp(t) \approx c\epsilon'/\bar{\gamma}(t) = c\epsilon^{}_\parallel(t)$. Thus it is sufficient to take the same $c\epsilon(t) = \lambda^{}_C/\bar{\gamma}(t)$ in all directions for single electrons.  

Fourth, not every uniformly accelerated charge with the choice of regulator (\ref{epsloft}) can be interpreted as experiencing the Unruh effect in (\ref{I1final}), since $\vartheta$ in (\ref{I1UACegamma}) can be larger than $O(1)$ when $|\tilde{T}| > |\tau|$ for $\tau < |\tau^{}_0|$ . 
If a photoelectron is ejected with $\bar{\gamma}(\tau_0)\gg 1$ initially
and then decelerated by a negative voltage such that $\bar{\gamma}(\tau)\ll\bar{\gamma}(\tau^{}_0)$ for a period, then in this period its (\ref{I1UACegamma}) cannot be approximated by (\ref{I0011BD}) and so the electron may behave differently from those with the Unruh effect in (\ref{I1final}) \cite{LH10}.  
In this paper, nevertheless, we only consider the electrons of initial speed in the laboratory frame much less than the speed of light. A full justification of our choice of the UV cutoff (\ref{epsloft}) and the Unruh effect is left to further experiments. 

Finally, the Unruh effect is often taken to be an equivalence of a uniformly accelerated detector/atom in the Minkowski vacuum of a bosonic field and a stationary detector/atom immersed in a bosonic thermal bath. We shall show that there is a fallacy. When one calculates the $F$-part of the correlators of the canonical momentum of the particle, e.g., $\langle \hat{p}^1(t), \hat{p}^1(t') \rangle$, one will find that in the product of ${\cal P}_1{\cal P}_1$ the cross terms of ${\cal A}^1$ and $\dot{\cal Z}^1$ contain integrals in the form
\begin{equation} 
  {\cal I}_2(\tilde{\tau},\tilde{\tau}') \equiv 2\pi c\int \omega d\omega \int_{-1}^{1} d\cos\theta\, 
	\epsilon^{(\lambda)1}_{{\bf k}} \, {\cal E}^{{\bf k}\,\,1*}_{(\lambda)\,\,0} \,
	e^{-i\frac{\omega}{c}\left[\bar{z}^0(\tilde{\tau})-\bar{z}^0(\tilde{\tau}')\right]+
	i k_1\left[\bar{z}^1(\tilde{\tau})-\bar{z}^1(\tilde{\tau}')\right]-\omega\epsilon}
\end{equation}
[see (\ref{P1UAC}), (\ref{P1lkUAC}), and (\ref{ppksum}).] Introducing (\ref{Eesum}), (\ref{FTeikzUAC}), 
and the same regulators in obtaining (\ref{I1final}), for $\tau > |\tau^{}_0|$, one has
\begin{eqnarray}
	{\cal I}_2(\tilde{\tau},\tilde{\tau}') &=& -2\pi i \int \omega^2 d\omega e^{-\omega\epsilon'/\bar{\gamma}(\tau)} 
	\int_{-1}^{1} d\cos\theta\, \int d\kappa\, d\kappa' 
  e^{-i \kappa \tilde{\tau}+i \kappa'\tilde{\tau}'}\varphi^{}_{\bf k}(\kappa)\varphi^*_{\bf k}(\kappa')\nonumber\\
	&=& \frac{\alpha^3}{4\pi} \int d\kappa \, d\kappa' e^{-i\kappa\tilde{\tau}+i\kappa'\tilde{\tau}'}
	\int d\check{T}\, d\check{\Delta} e^{i(\kappa-\kappa')\check{T}+\frac{i}{2}(\kappa+\kappa')\check{\Delta}} \,
	  \frac{\cosh\alpha \check{T}}{\sinh^3 \frac{\alpha}{2}\check{\Delta} + O(\epsilon')}\nonumber\\
	&\approx& -i\pi \cosh \alpha\tilde{T} \int d\kappa \,
	e^{-i \kappa(\tilde{\tau}-\tilde{\tau}')-\kappa\epsilon'} \frac{4 \kappa^2+\alpha^2}{1+e^{-2\pi \kappa/\alpha}}, \label{I2final}
\end{eqnarray}
where $\bar{\kappa}=(\kappa+\kappa')/2$ in the last line has been renamed to $\kappa$.
The above integrand turns out to have a Planck factor corresponding to a {\it fermionic} bath at the same Unruh temperature $T^{}_U \equiv \hbar a/(2\pi c k^{}_B)$.

\subsection{Particle correlators}
\label{calcUACcorr}

From (\ref{calEj}) and (\ref{calEj1}), one can see that ${\cal E}^{(\lambda)1}_{{\bf k}\,\,\,\,\,\,\,\,0} \, {\cal E}^{{\bf k}{\sf T}*}_{(\lambda)\,0} = -k^1 k^{\sf T}$, ${\cal E}^{(\lambda)1}_{{\bf k}\,\,\,\,\,\,\,\,0} \, {\cal E}^{{\bf k}{\sf T}*}_{(\lambda)\,1}=\frac{\omega}{c} k^{\sf T}$, and $\epsilon^{(\lambda)1}_{\bf k} \, {\cal E}^{{\bf k}{\sf T}*}_{(\lambda)\,1}=-i k^{\sf T}$ are all proportional to $\sin\varphi$ or $\cos\varphi$, which will be averaged out after the $\varphi$-integration in the mode sum ($\int d^3 k$) for a correlator is done. Moreover, $\epsilon^{(\lambda)1}_{\bf k} \epsilon^{{\sf T}*}_{(\lambda){\bf k}} =0$ from (\ref{eesum}), and
${\cal E}^{(\lambda)1}_{{\bf k}\,\,\,\,\,\,\,\,0} \, \epsilon^{ {\sf T}'*}_{(\lambda){\bf k}}=
 {\cal E}^{(\lambda){\sf T}}_{{\bf k}\,\,\,\,\,\,\,\,0} \, \epsilon^{1*}_{(\lambda){\bf k}} =0$ from (\ref{Eesum}).
Thus the cross correlators between the longitudinal and transverse motional degrees of freedom of a single electron, such as $\langle \hat{z}^1(t), \hat{z}^{\sf T}(t')\rangle$ and $\langle \hat{p}^1(t), \hat{p}^{\sf T}(t')\rangle$, are all vanishing.

\subsubsection{Correlators of longitudinal deviations}
\label{LongCorrUAC}

Inserting (\ref{Z1zjUAC}) into (\ref{zizjPrest}), we have
\begin{equation}
\langle \hat{z}_{}^1(t), \hat{z}_{}^1(t') \rangle^{}_P = \langle\overline{\hat{z}_1^2}\rangle^{}_{\rm I}
+ \frac{\langle p_{1}^{2}\rangle^{}_{\rm I}}{\bar{m}^2} K_\parallel\big[\tau(t), \tau(t^{}_0)\big] K_\parallel\big[\tau(t'),\tau(t'_0)\big]		\label{z1z1PUAC}
\end{equation}
where $\langle\overline{\hat{z}_1^2}\rangle^{}_{\rm I}\equiv \langle \hat{z}^{}_1(t'_0+\epsilon'_1), \hat{z}^{}_1(t'_0)\rangle = \frac{m^2}{\bar{m}^2} \langle z_1^2\rangle^{}_{\rm I}$, and we will require
\begin{equation}
  \langle p_{1}^{2}\rangle^{}_{\rm I} =\left[\frac{\bar{m}\,}{\bar{m}'} \right]^2 \left[
	\frac{\hbar^2}{4 \langle\overline{\hat{z}_1^2}\rangle^{}_{\rm I}} - 
	\langle \hat{p}^{}_{1}(t'_0+\epsilon'_1), \hat{p}^{}_{1}(t'_0)\rangle^{}_F\right] \label{p1p1IUAC}
\end{equation}
with $\langle \hat{p}^{}_{1}(t'_0+\epsilon'_1), \hat{p}^{}_{1}(t'_0)\rangle^{}_F\approx -3\hbar s\bar{m}/(2\pi \epsilon'_1{}^2)$
for an electron initially at rest [$\epsilon^{}_1(t^{}_0) = \epsilon'_1 \approx \epsilon'_0$,
cf. (\ref{pjpjPFrestInit}) and (\ref{QSrenorm}).]
From (\ref{Kparallel}), one can see that 
$\langle \hat{z}^1(t), \hat{z}^1(t') \rangle^{}_P$ with $\tau_0=0$ saturates to a finite constant 
$\langle\overline{\hat{z}_1^2}\rangle^{}_{\rm I} + \frac{\langle p_{1}^{2}\rangle^{}_{\rm I}}{\bar{m}^2\alpha^2}$ 
when $\alpha\tau(t)=\sinh^{-1}\alpha t \gg 1$. This is consistent with the observation in relativistic quantum mechanics of a charged particle moving in uniform electric fields \cite{HHL23}.

For the $F$-part, inserting the $s^0$-terms of $f^\parallel_{\bf k}(\tau)$ in (\ref{fparakDef}) with $K_\parallel$ in (\ref{Kparallel}) as well as ${\cal I}_{00}^{11}$ in (\ref{I1final}) into (\ref{z1z1FUAC1}), then the leading order of $\langle \hat{z}^1(t), \hat{z}^1(t') \rangle^{}_F$ reads
\begin{eqnarray}
\langle \hat{z}^1(t), \hat{z}^1(t') \rangle^{\{0\}}_F 
&\approx& \frac{3\hbar s}{8\pi^2\bar{m}} {\rm Re}\, \frac{8\pi}{3} 
  \int_{-\infty}^\infty d\kappa\frac{\kappa(\kappa^2+\alpha^2)}{1-e^{-2\pi\kappa/\alpha}} \int_{\tau_0}^\tau d\tilde{\tau} 
	\int_{\tau'_0}^{\tau'} d\tilde{\tau}' \times \nonumber\\ && 
\left[\zeta^{}_+(\tau) e^{\alpha\tilde{\tau}}+\zeta_-(\tau)e^{-\alpha\tilde{\tau}}\right] e^{-i\kappa\tilde{\tau}}
\left[\zeta^{}_+(\tau') e^{\alpha\tilde{\tau}'}+\zeta_-(\tau')e^{-\alpha\tilde{\tau}'}\right] e^{i\kappa\tilde{\tau}'}\nonumber\\
&=& \frac{\hbar s}{\pi \bar{m}} {\rm Re}\,
  \int_{-\infty}^\infty d\kappa\frac{\kappa(\kappa^2+\alpha^2)}{1-e^{-2\pi\kappa/\alpha}} \times\nonumber\\
&& \left[\zeta^{}_+(\tau) \frac{e^{(-i\kappa+\alpha)\tau}-e^{(-i\kappa+\alpha)\tau^{}_0}}{-i\kappa+\alpha} 
	+\zeta_-(\tau)\frac{e^{(-i\kappa-\alpha)\tau}-e^{(-i\kappa-\alpha)\tau^{}_0}}{-i\kappa -\alpha} \right] \times\nonumber\\
&& \left[\zeta^{}_+(\tau') \frac{e^{(i\kappa+\alpha)\tau'}-e^{(i\kappa+\alpha)\tau'_0}}{i\kappa+\alpha} 
	+\zeta_-(\tau')\frac{e^{(i\kappa-\alpha)\tau'}-e^{(i\kappa-\alpha)\tau'_0}}{i\kappa -\alpha} \right] 	,
 \label{z1z1FUAC2}
\end{eqnarray}
where $\zeta_\pm(\tau) \equiv \alpha^{-1}(\tanh\alpha\tau\mp 1)$. At late times, $\zeta_+(\tau) \approx -2\alpha^{-1} e^{-2\alpha\tau}$ goes to zero and $\zeta_-(\tau)\approx 2\alpha^{-1}$ goes to a finite constant. Let $\tau-\tau'=\epsilon_1 > 0$ and $\tau'_0-\tau_0=\epsilon_0 > 0$, then performing the above integral on the complex $\kappa$-plane and summing up the contribution from the poles (possibly at $\kappa = \pm i \alpha n$, $n\in {\bf Z}$), we obtain
\begin{eqnarray}
&&\langle \hat{z}^1(t), \hat{z}^1(t') \rangle^{\{0\}}_F \approx \frac{\hbar s \alpha^2}{4\pi\bar{m}}\times \nonumber\\ 
&&\Big\{ g(\epsilon_1) \left[ -\zeta^{}_+\zeta'_+ e^{\alpha(\tau+\tau')} - \zeta^{}_-\zeta'_- e^{-\alpha(\tau+\tau')}+
   \zeta^{}_+\zeta'_- h_4(\epsilon_1)-\zeta^{}_-\zeta'_+h_2(\epsilon_1)	\right]\nonumber\\ 
&& +g(\epsilon_0)\left[ -\zeta^{}_+\zeta'_+ e^{\alpha(\tau^{}_0+\tau'_0)}-\zeta^{}_-\zeta'_- e^{-\alpha(\tau^{}_0+\tau'_0)}+
	\zeta^{}_-\zeta'_+h_4(\epsilon_0)	-\zeta^{}_+\zeta'_-h_2(\epsilon_0)\right]\nonumber\\
&& +g(\tau-\tau'_0) \left[ \zeta^{}_+\zeta'_+ e^{\alpha(\tau+\tau'_0)} +\zeta^{}_-\zeta'_- e^{-\alpha(\tau+\tau'_0)}-
   \zeta^{}_+\zeta'_- h_4(\tau-\tau'_0)+\zeta^{}_-\zeta'_+h_2(\tau-\tau'_0)	\right]\nonumber\\
&& +g(\tau'-\tau^{}_0) \left[ \zeta^{}_+\zeta'_+ e^{\alpha(\tau'+\tau^{}_0)} +\zeta^{}_-\zeta'_- e^{-\alpha(\tau'+\tau^{}_0)}-
   \zeta^{}_-\zeta'_+ h_4(\tau'-\tau^{}_0)+\zeta^{}_+\zeta'_-h_2(\tau'-\tau^{}_0)	\right]\nonumber\\
&& - \left. 2\left( \zeta^{}_+\zeta'_- + \zeta^{}_-\zeta'_+ \right)\left[ \alpha\left(\tau'-\tau'_0\right)-
    \ln \frac{\left(1-e^{-\alpha\epsilon_0}\right)\left(1-e^{-\alpha\epsilon_1}\right)}
		  {\left(1-e^{-\alpha(\tau-\tau'_0)}\right)\left(1-e^{-\alpha(\tau'-\tau^{}_0)}\right)} \right] \right\} \label{z1z1FUAC3}
\end{eqnarray}
with $\zeta_\pm \equiv \zeta_\pm(\tau)$, $\zeta'_\pm \equiv \zeta_\pm(\tau')$, $h_n(x)\equiv 3e^{-\alpha x}-n$, and
\begin{equation}
   g(x) \equiv \frac{e^{-\alpha x}}{\left(1-e^{-\alpha x}\right)^2}.
\end{equation}
When $\alpha\tau\gg 1$,
\begin{eqnarray}
   \langle \hat{z}^1(t), \hat{z}^1(t') \rangle^{\{0\}}_F \to -\frac{\hbar s}{\pi\bar{m}}g(\epsilon_0)
	 e^{-\alpha(\tau^{}_0+\tau'_0)}, \label{z1z1FUAClate}
\end{eqnarray}
which is a negative and finite constant, 
contributed by the $-g(\epsilon_0) \zeta^{}_-\zeta'_- e^{-\alpha(\tau^{}_0+\tau'_0)}$ term in the third line of (\ref{z1z1FUAC3}).

The next-to-leading order in $s$ of $\langle \hat{z}^1(t), \hat{z}^1(t') \rangle^{}_F$ turns out to be $O(s \epsilon')$ with $\epsilon'= \bar{\gamma}/(c\Lambda) = t^{}_C$ compared with the leading order. Thus, according to the discussion below Eq.(\ref{defofs}), we have $\langle \hat{z}^1(t), \hat{z}^1(t') \rangle^{}_F \approx \langle \hat{z}^1(t), \hat{z}^1(t') \rangle^{\{0\}}_F$ with higher-order corrections negligible. 
As $\alpha\to 0$, $\langle \hat{z}^1(t), \hat{z}^1(t') \rangle^{}_F$ for $t^{}_0=0$ goes back to the result (\ref{zjzjF0rest3}) for a charged particle at rest. 

With (\ref{P1zjUAC}) and (\ref{P1lkUAC}), calculations for $\langle \hat{p}^1(t), \hat{z}^1(t') \rangle$ and $\langle \hat{p}^1(t), \hat{p}^1(t') \rangle$ are straightforward using similar techniques. 
From (\ref{P1zjUAC}), one has ${\cal P}_1^{z_{}^j} =0$, and ${\cal P}_1^{p^{}_j}$ goes to a positive finite constant at late times since $\Phi^{}_\parallel \to \frac{\bar{m}'}{\bar{m}}-\frac{3}{2}s\alpha$ as $\alpha \tau\to \infty$. Thus, the $P$-parts 
\begin{eqnarray}
 \langle \hat{p}^1(t), \hat{z}^1(t') \rangle^{}_P &=& \frac{\langle p_{1}^{2}\rangle^{}_{\rm I}}{\bar{m}} 
     \Phi_\parallel\big[\tau(t)\big] K_\parallel\big[\tau(t'),\tau(t'_0)\big] \to
		\frac{\langle p_{1}^{2}\rangle^{}_{\rm I}}{\bar{m}\alpha} \left(\frac{\bar{m}'}{\bar{m}}-\frac{3}{2}s\alpha\right), 
		\label{p1z1PUAC} \\
 \langle \hat{p}^1(t), \hat{p}^1(t') \rangle^{}_P &=& \langle p_{1}^{2}\rangle^{}_{\rm I} \,
    \Phi_\parallel\big[\tau(t)\big]\Phi_\parallel\big[\tau(t')\big] \to 
		\langle p_{1}^{2}\rangle^{}_{\rm I} \left(\frac{\bar{m}'}{\bar{m}}-\frac{3}{2}s\alpha\right)^2, \label{p1p1PUAC}
\end{eqnarray} 
saturate to finite constants at late times (here $t^{}_0=0$), similar to the behavior of $\langle \hat{z}^1(t), \hat{z}^1(t') \rangle^{}_P$.

\begin{figure}
\includegraphics[width=4.9cm]{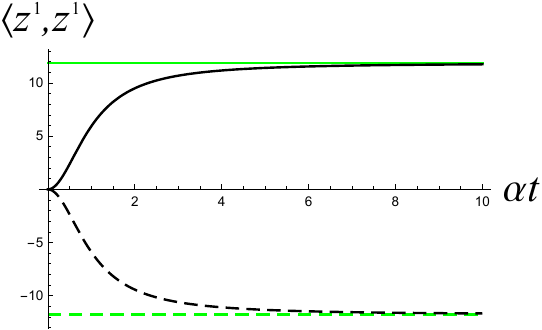}
\includegraphics[width=4.9cm]{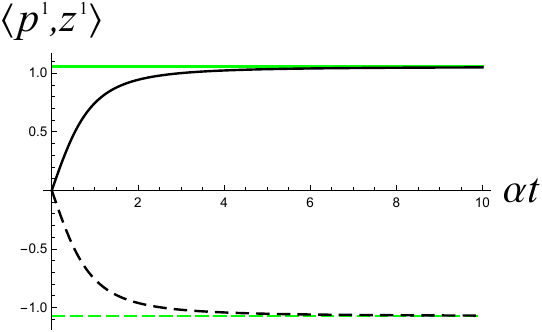}
\includegraphics[width=4.9cm]{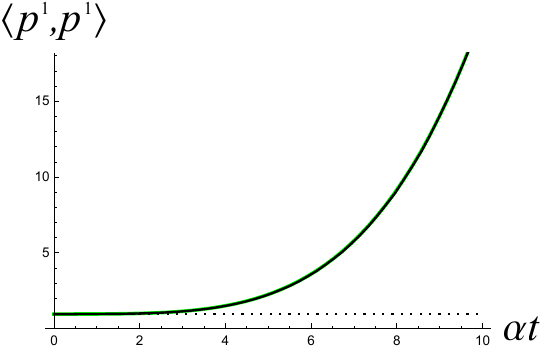}
\caption{(Left) Time evolution of $\langle \hat{z}^1(t), \hat{z}^1(t) \rangle$ (black solid curve), which is indistinguishable from $\langle \hat{z}^1(t), \hat{z}^1(t) \rangle^{}_P$ [Eq. (\ref{z1z1PUAC})] here. The black dashed curve represents $10^3 \times \langle \hat{z}^1(t), \hat{z}^1(t)\rangle^{}_F$ [Eq. (\ref{z1z1FUAC3})]. The green lines represent their late-time values. (Middle) Time evolution of $10^{28}\times\langle \hat{p}^1(t), \hat{z}^1(t) \rangle$ (black) and $10^{31}\times\langle \hat{p}^1(t), \hat{z}^1(t) \rangle^{}_F$ (black dashed). (Right) Time evolution of $10^{57}\times\langle \hat{p}^1(t), \hat{p}^1(t) \rangle$ (black) and $10^{57}\times\langle \hat{p}^1(t), \hat{p}^1(t) \rangle^{}_P$ (black dotted). The green curve represents the scaled ($10^{57}\times$) sum of (\ref{p1p1FUAClate}) and the late-time value of $\langle \hat{p}^1(t), \hat{p}^1(t) \rangle^{}_P$. Here the physical parameters are in the SI units, $t^{}_0=0$, $\alpha=10\,\,{\rm s}^{-1}$, 
$\epsilon_0 \approx \epsilon'_0=1.4\times 10^{-14}$ s, $\epsilon_1 \approx \epsilon'_1/\bar{\gamma}(t)$ with $\epsilon'_1=\epsilon'_0$, the initial values 
$\langle\overline{\hat{z}_1^2}\rangle^{}_{\rm I} = (1.7\,\,\mu{\rm m})^2$, 
$\langle \hat{p}^1(t'_0+\epsilon'_1), \hat{p}^1(t'_0)\rangle 
= \hbar^2/\big( 4 \langle\overline{\hat{z}_1^2}\rangle^{}_{\rm I} \big)$, and 
$\langle \hat{p}^1, \hat{z}^1\rangle^{}_{\rm I}=0$.} 
\label{FIGcorrUACL}
\end{figure}

Some numerical results of time evolution of the particle correlators are shown in Figure \ref{FIGcorrUACL}, based on their closed-form expressions. One can see that all the $P$-parts of the correlators as well as $\langle \hat{z}^1(t), \hat{z}^1(t') \rangle^{}_F$ indeed saturate at late times.
{However, the behaviors of $\langle \hat{p}^1(t), \hat{z}^1(t') \rangle^{}_F$ and  $\langle \hat{p}^1(t), \hat{p}^1(t') \rangle^{}_F$ are different from them.}

In the middle plot of Figure \ref{FIGcorrUACL}, $\langle \hat{p}^1(t), \hat{z}^1(t') \rangle^{}_F$ seems to saturate to a negative constant
$-\frac{\hbar s \alpha \bar{m}'}{\pi \bar{m}} e^{-\alpha\epsilon_0}/(1-e^{-\alpha\epsilon_0})^2$ when we choose $\tau'_0=0$, $\epsilon'_1=\epsilon'_0$, and $\alpha\epsilon_0 \ll 1$. But actually, there is a negative growing term $-\alpha [\tau(t)-\tau^{}_0 ]$ in $\langle \hat{p}^1(t), \hat{z}^1(t') \rangle^{}_F$ at late times, though it is not significant in the example with $\alpha\epsilon_0 \ll 1$ here.

For the $F$-part of the momentum correlator, 
we find that the next-to-leading order correction to $\langle \hat{p}^1(t), \hat{p}^1(t') \rangle^{}_F$ is $O(s\alpha)$ compared with the leading-order result $\langle \hat{p}^1(t), \hat{p}^1(t') \rangle^{(0)}_F$, and
the momentum correlator for $\alpha\tau\gg 1$ and $\tau\gg\epsilon_1$ is approximately
\begin{eqnarray}
  &&\langle \hat{p}^1(t), \hat{p}^1(t') \rangle^{}_F \approx \frac{3\hbar \bar{m}}{8\pi} s\alpha^2 e^{2\alpha\tau(t)}
  \times \nonumber\\
  && \left\{ \Phi_+ \, \Big[ 3-2\alpha\epsilon_1 +2 g(\epsilon_1)-4\ln(1-e^{-\alpha\epsilon_1})\Big] 
	-\frac{2}{3}\Phi_+^2 g(\epsilon_1) + O(s\alpha) \right\} \label{p1p1FUAClate}
\end{eqnarray}
with $\epsilon_1=\tau-\tau'$ given in (\ref{e0e1UAC}) and $\Phi_+ \equiv \lim_{\alpha\tau\gg1}[\Phi_\parallel(\tau)-s\alpha \frac{\bar{m}'}{\bar{m}} ] = \frac{\bar{m}'}{\bar{m}}(1-s\alpha)-\frac{3}{2}s\alpha >0$. 
Here, the $\Phi_+$ term is produced by the cross term of ${\cal A}^1$ and $\dot{\cal Z}^1$ in (\ref{P1UAC}). Its coefficient is positive since $-2\alpha \epsilon_1$ becomes very small for $\alpha\tau\gg 1$ even if $\alpha$ is large enough to make $\alpha\epsilon'_1 \gg 1$ 
\footnote{This justifies the choice of $\epsilon_1(t)$ in (\ref{e0e1UAC}). If $\alpha\epsilon_1$ is a constant of the Minkowski time $t$, once $\alpha=a/c$ gets so large that $\alpha\epsilon_1 \gg 3/2$ ($a \gg 3.2 \times 10^{22}$ m$/{\rm s}^2$ for $\epsilon_1 = 1.4 \times 10^{-14}$ s, which is achievable for electrons accelerated by a terawatt laser), $\langle \hat{p}^1(t), \hat{p}^1(t') \rangle^{(0)}_F$ will behave like $-(\alpha t)^2$ at late times and eventually make $\langle \hat{p}^1(t), \hat{p}^1(t') \rangle$ negative. This will violate the uncertainty relation and ruin the Gaussian state.}. 
The $\Phi_+$ term dominates over the $\Phi_+^2$ term, which is negative and derived from the $\dot{\cal Z}^1\dot{\cal Z}^1$ term. Thus, at late times ($\alpha t\gg 1$), $\langle \hat{p}^1(t), \hat{p}^1(t') \rangle^{}_F$ grows like $\alpha^2 g\big(\epsilon^{}_1(t)\big)e^{2\alpha\tau(t)}\approx \big(2\alpha t/\epsilon^{}_1(t)\big)^2$, as shown in Figure \ref{FIGcorrUACL} (right). In Section \ref{SecCoherenceUAC}, we will see that such a growing $\langle \hat{p}^1(t), \hat{p}^1(t') \rangle$ will continuously decrease the quantum coherence in the direction of acceleration for a laboratory observer at late times, as $\langle \hat{z}^1(t), \hat{z}^1(t') \rangle$ saturates to a constant and $\langle \hat{p}^1(t), \hat{z}^1(t') \rangle$ goes down relatively slowly. 

The wavepacket in the direction of acceleration is spreading in phase space after it's creation. At late times, the spreading rate in the $z_{}^1$ direction of phase space decreases, while the wavepacket keeps spreading as $\bar{\gamma}^2(t) t^2$ in the $p_{}^1$ direction. This is in contrast to the behavior of the wavepacket of a charged particle at rest, which spreads as $t^2$ in the $z_{}^1$ direction of phase space [see Eq.(\ref{zjzjPFrestAsmp})], while the width in the $p_{}^1$ direction goes to a constant at late times [Eq.(\ref{pjpjPFrestAsmp})].

\subsubsection{Correlators of transverse deviations}

Assume $\langle \hat{z}^{\sf T}, \hat{z}^{{\sf T}'}\rangle^{}_{\rm I} = \langle \hat{p}^{\sf T}, \hat{p}^{{\sf T}'}\rangle^{}_{\rm I} = 0$ for ${\sf T} \not= {\sf T}'$, and $\langle \hat{p}^{\sf T}, \hat{z}^{{\sf T}'}\rangle^{}_{\rm I} = 0$ for all ${\sf T}$ and ${\sf T}'$.
Inserting (\ref{ZTzjUAC}) 
to (\ref{zizjPrest}), one can see that the $P$-part of the transverse deviation correlator $\langle \hat{z}^{\sf T}(t), \hat{z}^{\sf T'}(t') \rangle^{}_P$ eventually saturates 
as $K_\perp(\tau, \tau_0)$ in (\ref{Kperpdef}) goes to a constant at late times. While this behavior looks similar to the one for $\langle \hat{z}^{1}(t), \hat{z}^{1}(t') \rangle^{}_P$ in the direction parallel to acceleration, the time scale of $\langle \hat{z}^{\sf T}(t), \hat{z}^{\sf T'}(t') \rangle^{}_P$ is $\varsigma/(s\alpha^2) \approx (s\alpha)^{-1} \alpha^{-1}$, which is much longer than its longitudinal counterpart's timescale $\alpha^{-1}$ since $s\alpha \ll 1$ in our effective theory as we mentioned below (\ref{Phiperpdef}). 

In calculating the $F$-parts of the transverse deviation correlator, the situation is more complicated 
than the longitudinal ones. 
There are two different types of fluctuations associated with the two different factors ${\cal E}^{(\lambda){\sf T}}_{{\bf k}\,\,\,\,\,\,\,\,\,0}$ and  ${\cal E}^{(\lambda){\sf T}}_{{\bf k}\,\,\,\,\,\,\,\,\,1}$ in (\ref{ZTlkUAC}). 
From (\ref{calEj1}) and (\ref{calEj}), it is straightforward to see
\begin{eqnarray}
  && {\cal E}^{(\lambda){\sf T}}_{{\bf k}\,\,\,\,\,\,\,\,\,1} \, {\cal E}^{{\bf k}\,\,{\sf T}'*}_{(\lambda)\,\,1} = 
		k_1^2\eta^{\sf{TT}'}- k^{\sf T} k^{{\sf T}'}, \hspace{1cm}
	{\cal E}^{(\lambda){\sf T}}_{{\bf k}\,\,\,\,\,\,\,\,\,0} \, {\cal E}^{{\bf k}\,\,{\sf T}'*}_{(\lambda)\,\,1} = 
		-\frac{\omega}{c} k^{}_1 \eta^{\sf{TT}'}, \nonumber\\ 
	&&{\cal E}^{(\lambda){\sf T}}_{{\bf k}\,\,\,\,\,\,\,\,\,1} \, \epsilon^{{\sf T}'*}_{(\lambda){\bf k}} = -i k^{}_1 \eta^{\sf{TT}'},
\label{E1E1sum}
\end{eqnarray}
in addition to (\ref{Eesum}). After some algebra similar to Section \ref{SecUnruhEffect}, the integrals (\ref{calIjjBB}) with $j, j' = {\sf T}, {\sf T}'$ are found to be
\begin{eqnarray}
{\cal I}_{00}^{{\sf TT}'}(\tilde{\tau},\tilde{\tau}') &=& \frac{8\pi}{3c^2} \eta^{{\sf TT}'}\cosh\alpha(\tilde{\tau}+\tilde{\tau}') 
  \int d\kappa\,e^{-i\kappa(\tilde{\tau}-\tilde{\tau}')}\frac{\kappa (\kappa^2+\alpha^2)}{1-e^{-2\pi \kappa/\alpha}},\\
{\cal I}_{01}^{{\sf TT}'}(\tilde{\tau},\tilde{\tau}') &=& -\frac{8\pi}{3c^2} \eta^{{\sf TT}'}\sinh\alpha(\tilde{\tau}+\tilde{\tau}') 
  \int d\kappa\,e^{-i\kappa(\tilde{\tau}-\tilde{\tau}')}\frac{\kappa (\kappa^2+\alpha^2)}{1-e^{-2\pi \kappa/\alpha}},
\end{eqnarray}
${\cal I}_{11}^{{\sf TT}'}(\tilde{\tau},\tilde{\tau}')= {\cal I}_{00}^{{\sf TT}'}(\tilde{\tau},\tilde{\tau}')$ in value, and
${\cal I}_{10}^{{\sf TT}'}(\tilde{\tau},\tilde{\tau}')= {\cal I}_{01}^{{\sf T'T}*}(\tilde{\tau}',\tilde{\tau})$.
Inserting (\ref{fperpkDef}) into (\ref{zizjFrest}), then
summing over $B=0,1$ with the above formulas, the $F$-part of the transverse deviation correlator becomes 
\begin{eqnarray}
&&\langle \hat{z}^{\sf T}(t), \hat{z}^{\sf T'}(t') \rangle^{}_F 
= \frac{\hbar s \eta_{}^{\sf TT'}}{\pi \bar{m}s^2\alpha^4} 
  {\rm Re}\,\int d\kappa\frac{\kappa(\kappa^2+\alpha^2)}{1-e^{-2\pi\kappa/\alpha}} 
  \int_{\tau^{}_0}^{\tau} d\tilde{\tau} \int_{\tau'_0}^{\tau'} d\tilde{\tau}' \times \nonumber\\
&&\hspace{.6cm} \left\{ \Big[ e^{-i\kappa\tilde{\tau}} - e^{-s\alpha^2\tau} 
	e^{(-i\kappa+ s\alpha^2)\tilde{\tau}}\Big]
	\Big[e^{-i\kappa\tilde{\tau}'} -  e^{-s\alpha^2\tau'} 
	e^{(-i\kappa+s\alpha^2)\tilde{\tau}'}\Big]+O(s^2)\right\}.
\end{eqnarray}
One can see the Planck factor corresponding to a bosonic bath at the Unruh temperature $T^{}_U \equiv \hbar a/(2\pi c k^{}_B)$ in the $\kappa$ integral.
The above $\tilde{\tau}$ and $\tilde{\tau}'$ integrals can be done straightforwardly to obtain an expression similar to (\ref{z1z1FUAC2}). Then one can sum over the contributions by the poles at $\kappa = 0$, $\kappa=\pm i n \alpha$  ($n=2,3,4,\cdots$), and $\kappa=\pm i s\alpha^2$. They are
combinations of elementary functions, the polylogarithm functions, and the hypergeometric functions. As $\alpha\to 0$, $\langle \hat{z}^{\sf T}(t), \hat{z}^{\sf T'}(t') \rangle^{}_F$ with $t^{}_0=0$ goes back to the result (\ref{zjzjF0rest3}) for the charged particle at rest. 

\begin{figure}
\includegraphics[width=4.9cm]{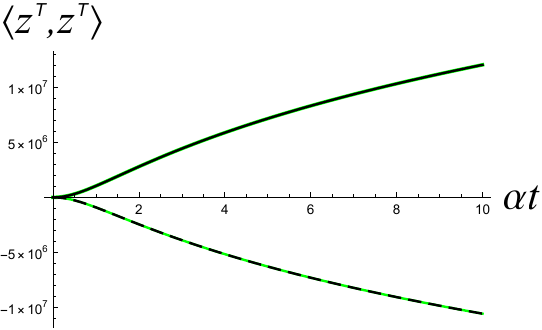}
\includegraphics[width=4.9cm]{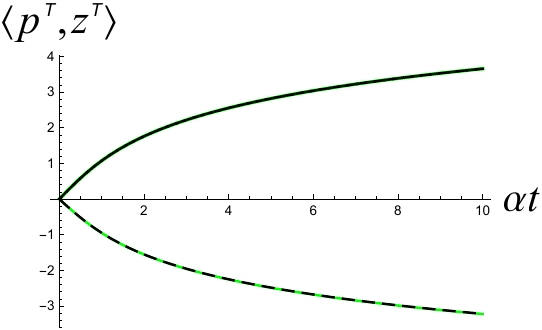}
\includegraphics[width=4.9cm]{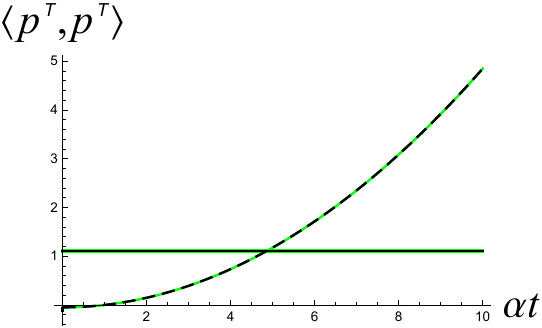}
\caption{(Left) Time evolution of electron's $\langle \hat{z}^{\sf T}(t), \hat{z}^{\sf T}(t) \rangle$ (black solid curve), ${\sf T}=2,3$, which is indistinguishable from $\langle \hat{z}^{\sf T}(t), \hat{z}^{\sf T}(t) \rangle^{}_P$ in this plot. The black dashed curve represents $10^8 \times \langle \hat{z}^{\sf T}(t), \hat{z}^{\sf T}(t)\rangle^{}_F$. (Middle) Time evolution of $10^{23}\times\langle \hat{p}^{\sf T}(t), \hat{z}^{\sf T}(t) \rangle$ (black) and $10^{31}\times\langle \hat{p}^{\sf T}(t), \hat{z}^{\sf T}(t) \rangle^{}_F$ (black dashed). (Right) Time evolution of $10^{52}\times\langle \hat{p}^{\sf T}(t), \hat{p}^{\sf T}(t) \rangle$ (black) and $10^{59}\times\langle \hat{p}^{\sf T}(t), \hat{p}^{\sf T}(t) \rangle^{}_F$ (black dashed). 
The green curves represent the approximated behaviors (\ref{zTzTapproxET})-(\ref{pTzTFapproxET}). 
Here the parameter values are the same as those in Figure \ref{FIGcorrUACL}. 
The initial values for $t_0=0$ are $\langle\overline{\hat{z}_{\sf T}^2}\rangle^{}_{\rm I} = (5\,\,{\rm nm})^2$ , $\langle \hat{p}^{\sf T}(\epsilon'_0+\epsilon'_1), \hat{p}^{\sf T}(\epsilon'_0)\rangle = \hbar^2/\big( 4 \langle\overline{\hat{z}_{\sf T}^2}\rangle^{}_{\rm I} \big)$, and $\langle \hat{p}^{}_{\sf T}, \hat{z}^{}_{\sf T}\rangle^{}_{\rm I}=0$.} 
\label{FIGcorrUACT}
\end{figure}

At very late times ($s\alpha^2 (\tau'-\tau'_0)\gg 1$), the $\tau'-\tau'_0$ term contributed by the pole at $\kappa=0$
dominates, as the other terms in $\langle \hat{z}^{\sf T}(t), \hat{z}^{\sf T'}(t') \rangle^{}_F$ settle to finite constant values. 
Then we have $\langle \hat{z}^{\sf T}(t), \hat{z}^{\sf T'}(t') \rangle^{}_F \approx \frac{\hbar s}{\pi \bar{m}(s\alpha)^2} \eta_{}^{{\sf TT}'} \alpha[\tau(t')-\tau(t'_0)]\,\, +$ constant.

In Figure \ref{FIGcorrUACT} (left), however, the transverse deviation correlators for a single electron with $\alpha=10\, {\rm s}^{-1}$ and $\alpha (t-t^{}_0) \le 10$ are still not at their very late times [$s\alpha^2 (\tau-\tau^{}_0) < 2 \times 10^{-22}]$. In that plot,
\begin{equation}
  \langle \hat{z}_{}^{\sf T}(t), \hat{z}_{}^{\sf T}(t') \rangle \approx 
  \langle \hat{z}_{}^{\sf T}(t), \hat{z}_{}^{\sf T}(t') \rangle^{}_P \approx 
	\langle\overline{\hat{z}_{\sf T}^2}\rangle^{}_{\rm I} +	
	\langle p_{\sf T}^2\rangle^{}_{\rm I} 
	  \left[ \frac{\tau(t)-\tau(t^{}_0)}{\bar{m}}\right]^2  \label{zTzTapproxET}
\end{equation}
with $\langle\overline{\hat{z}_{\sf T}^2}\rangle^{}_{\rm I}\equiv \langle\hat{z}^{}_{\sf T}(t'_0+\epsilon'_1),\hat{z}^{}_{\sf T}(t'_0)\rangle$ and
\begin{equation}
  \langle p_{\sf T}^2 \rangle^{}_{\rm I} =\left[\frac{\bar{m}\,}{\bar{m}'}\right]^2 
	\left[\frac{\hbar^2}{4 \langle\overline{\hat{z}_{\sf T}^2}\rangle^{}_{\rm I}} - 
	\langle \hat{p}^{}_{\sf T}(t'_0+\epsilon'_1), \hat{p}^{}_{\sf T}(t'_0)\rangle^{}_F\right],
\end{equation}
with $\langle \hat{p}^{}_{\sf T}(t'_0+\epsilon'_1), \hat{p}^{}_{\sf T}(t'_0)\rangle^{}_F \approx -3\hbar s \bar{m}/(2\pi \epsilon'_1{}^2)$, similar to its longitudinal counterpart (\ref{p1p1IUAC}).
The early-time $F$-part of the correlator in Figure \ref{FIGcorrUACT} (where $\alpha\epsilon_0\ll 1$) is about
\begin{equation}
  \langle \hat{z}_{}^{\sf T}(t), \hat{z}_{}^{\sf T'}(t') \rangle^{}_F 
\approx -\frac{\hbar s\eta_{}^{\sf TT'}}{\bar{m}\pi}
	 \frac{\big[\tau(t)-\tau(t^{}_0)\big]^2}{\epsilon_0^2},
	\label{zTzTFapproxET}
\end{equation} 
which is roughly $10^{-8}$ times the $P$-part in amplitude, still a very small correction to (\ref{zTzTapproxET}).
The above behavior is very similar to (\ref{zjzjPFrestAsmp}) for a charged particle at rest at $t-t^{}_0 \gg \epsilon'_0$, with the Minkowski time there generalized to the proper time of the particle here. 

In Figure \ref{FIGcorrUACT} (right), one can see that
\begin{equation}
  \langle \hat{p}_{}^{\sf T}(t), \hat{p}_{}^{\sf T}(t') \rangle \approx 
  \langle \hat{p}_{}^{\sf T}(t), \hat{p}_{}^{\sf T}(t') \rangle^{}_P \approx 
	\langle p_{\sf T}^2\rangle^{}_{\rm I} \left(\frac{\bar{m}'}{\bar{m}}\right)^2, \label{pTpTapproxET}
\end{equation}
which is not distinguishable from the $P$-part $\langle \hat{p}_{}^{\sf T}(t), \hat{p}_{}^{\sf T'}(t') \rangle^{}_P$ there. The $F$-part of the transverse momentum correlator in Figure \ref{FIGcorrUACT} ($\alpha\epsilon_0\ll 1$) is approximately 
\begin{eqnarray}
&&\langle \hat{p}_{}^{\sf T}(t), \hat{p}_{}^{\sf T'}(t') \rangle^{}_F \approx  
  \langle \hat{p}_{}^{\sf T}(t), \hat{p}_{}^{\sf T'}(t') \rangle^{\{0\}}_F \approx
	-\frac{\hbar s\eta_{}^{\sf TT'}}{\bar{m}\pi} \times \nonumber\\ && \hspace{1cm}
	\left\{ 3 \bar{m}'{}^2 \alpha^2 \big[ \tau(t)-\tau(t_0)\big]
	+\frac{\bar{m}'{}^2}{\epsilon_0^2}
	+\left(\bar{m}'{}^2-3 \bar{m}'\bar{m}+\frac{3}{2}\bar{m}^2\right) \frac{1}{\epsilon_1^2}
	\right\},	\label{pTpTFapproxET}
\end{eqnarray} 
which is much smaller than $\langle \hat{p}_{}^{\sf T}(t), \hat{p}_{}^{\sf T'}(t') \rangle^{}_P$.
The $[\epsilon^{}_1(t)]^{-2}= \bar{\gamma}^2(t)/\epsilon'_1{}^2$ term in (\ref{pTpTFapproxET}) is positive ($\bar{m}\approx \bar{m})$ and behaves like $t^2$ in Figure \ref{FIGcorrUACT} (right). It dominates over the other two (negative) terms for $\alpha t > O(1)$.
A term with $[\epsilon^{}_1(t)]^{-4}$ 
in the leading order contribution $\langle \hat{p}_{}^{\sf T}(t), \hat{p}_{}^{\sf T'}(t') \rangle^{(0)}_F$ is not significant yet in the plot, and will become important much later as $\alpha t$ gets very large. 
The amplitude of the next-to-leading order contribution in $s$, 
$\langle \hat{p}_{}^{\sf T}(t), \hat{p}_{}^{\sf T'}(t') \rangle^{\{1\}}_F \approx 
	9 \hbar s^3\alpha^2 \bar{m}/\big[2\pi(\epsilon^{}_1 \bar{\gamma})^2\big]$,
is about $s^2 \alpha^2 \sim 10^{-45}$ times the leading order contribution (\ref{pTpTFapproxET}) in the whole interval $0<\alpha t<10$ in Figure \ref{FIGcorrUACT} (right). 
As $\alpha\to 0$, (\ref{pTpTFapproxET}) reduces to (\ref{pjpjPFrestAsmp}). 

Finally, in Figure \ref{FIGcorrUACT} (middle), one has
\begin{equation}
  \langle \hat{p}_{}^{\sf T}(t), \hat{z}_{}^{\sf T}(t') \rangle \approx 
  \langle \hat{p}_{}^{\sf T}(t), \hat{z}_{}^{\sf T}(t') \rangle^{}_P \approx \frac{\langle p_{\sf T}^2 \rangle^{}_{\rm I}}{\bar{m}}
	  \left(\frac{\bar{m}'}{\bar{m}}\right)\big[\tau(t)-\tau(t^{}_0)\big],  \label{pTzTapproxET}
\end{equation}
which is again not distinguishable from  $\langle \hat{p}_{}^{\sf T}(t), \hat{z}_{}^{\sf T}(t') \rangle^{}_P$ in the plot. 
The $F$-part of the correlator 
for $\alpha\epsilon_0$, $\alpha\epsilon_1 \ll 1$ is approximately
\begin{eqnarray}
  \langle \hat{p}_{}^{\sf T}(t), \hat{z}_{}^{\sf T'}(t') \rangle^{}_F &\approx& 
	\langle \hat{p}_{}^{\sf T}(t), \hat{z}_{}^{\sf T'}(t') \rangle^{\{0\}}_F \nonumber\\ &\approx& 
	\frac{\hbar s}{\pi}\eta_{}^{\sf TT'} \left\{-\frac{\bar{m}'}{\bar{m}}\frac{\big[ \tau(t)-\tau(t^{}_0)\big]}{\epsilon_0^2}
	+\left(\frac{3}{2}-\frac{\bar{m}'}{\bar{m}}\right) \frac{1}{\epsilon_1}\right\}. \label{pTzTFapproxET} 
\end{eqnarray}
The above negative $\epsilon_0^{-2}$ term dominates the $F$-part in the time domain of Figure \ref{FIGcorrUACT} (middle), while its value is very small compared with the $P$-part. 
The positive $\epsilon_1^{-1}$ term in (\ref{pTzTFapproxET}) is not significant until $\bar{\gamma}(t) \sim O\big[ (\tau-\tau^{}_0)\epsilon'_1/\epsilon_0^2\big]$, then it takes over so that $\langle \hat{p}_{}^{\sf T}(t), \hat{z}_{}^{\sf T'}(t') \rangle^{\{0\}}_F$ will be approximately growing linearly since $\epsilon_1^{-1} \approx \bar{\gamma}(t)/\epsilon'_1$ is roughly proportional to $t$ at late times.
The next-to-leading order contribution $\langle \hat{p}_{}^{\sf T}(t), \hat{z}_{}^{\sf T'}(t') \rangle^{\{1\}}_F \approx 3\hbar s^2/(4\pi \epsilon_1^2)$ is about $s\alpha\bar{\gamma}^2(t)/\tau(t)$ times the value of (\ref{pTzTFapproxET}) 
in the whole interval $0<\alpha t<10$ in Figure \ref{FIGcorrUACT} (middle).

\subsection{Quantum coherence of particle state}
\label{SecCoherenceUAC}

In Section \ref{calcUACcorr} we have already obtained all the correlators needed to fully describe the reduced state of the Gaussian wavepacket of the charged particle. 
The uncertainty relation of the particle state is again in the form of (\ref{Uui}), and the purity of the particle can be factorized into ${\bf P}^{}_i$ defined in (\ref{Puri}) for the particle-motion deviation in each direction.

\begin{figure}
\includegraphics[width=7cm]{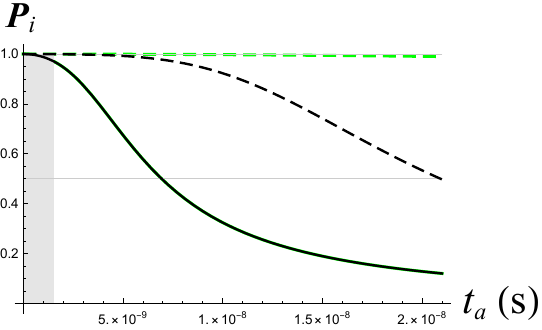} 
\includegraphics[width=7.5cm]{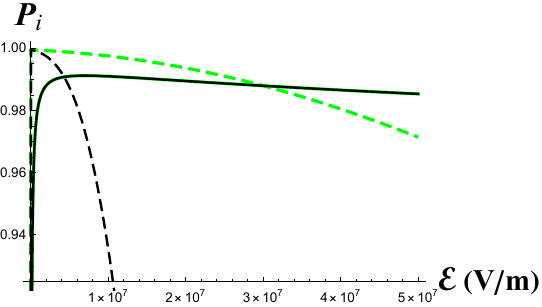} 
\caption{(Left) Purities of the transverse (black solid curve) and longitudinal (black dashed) deviations against the accelerating time $t^{}_a$ of a Gaussian electron wavepacket accelerated in a constant electric field of $5 \times 10^5\, {\rm V/m}$, or $\alpha \approx 2.9 \times 10^8\, {\rm s}^{-1}$. Here $\tau^{}_0=0$, and the initial state is the same as those in Figures \ref{FIGcorrUACT}. The gray region with $t^{}_a = 1.54 \times 10^{-9}$ s at the right edge represents the acceleration stage of an electron (initially at rest) in an acceleration tube of 10 cm. 
(Right) Purities of the transverse (black solid) and longitudinal (black dashed) deviations 
against the electric field ${\cal E}$ (V/m) in the acceleration tube of fixed length 10 cm.
The green solid (dashed) curves in both plots represent the purities obtained from the transverse (longitudinal) deviation correlators in Section \ref{SecQCoherenceRest} for a wavepacket of the same initial dimensions describing inertial single electrons with only the coordinate effects taken into account.} 
\label{FIGpurityUAC}
\end{figure}

In \cite{TE89}, the electrons will be accelerated to the anode at a potential of 50 kV after they first appear around the tip of the FEG. Suppose the acceleration tube is $L=$ 10 cm long for simplicity, and the electric field ${\cal E}$ in the acceleration tube is roughly uniform. Then we have ${\cal E} \approx 5\times 10^5$ V/m, which produces a scaled acceleration $\alpha = q{\cal E}/(\bar{m}c) \approx 2.9\times 10^8 \, {\rm s}^{-1}$. Assume the initial purity of the electron wavepacket is $1$ and its initial speed is zero in the laboratory frame, so $t^{}_0=\tau^{}_0=0$ for the worldline (\ref{z1z1PUAC}). From (\ref{zintauUAC}), an electron initially at rest will spend $t^{}_{a} =\alpha^{-1}\sinh\alpha \tau^{}_a = \sqrt{(L/c)^2+ 2L/(\alpha c)} \approx 1.54\times 10^{-9}$ s 
in this acceleration tube since $L = (c/\alpha)(\cosh\alpha\tau^{}_a - \cosh 0)$ with $\tau^{}_a$ the electron's proper time at the end of the tube. 
With the other parameter values the same as those set in Section \ref{SecQCoherenceRest} and the quantum-state renormalization like (\ref{QSrenorm}), we obtain the evolution of purities ${\bf P}^{}_i$ in Figure \ref{FIGpurityUAC} (left) as a function of accelerating time $t^{}_a$ in the laboratory frame.
At the moment of leaving the tube, the electron will have ${\bf P}^{}_1 \approx 0.9992$ 
for the longitudinal deviation and ${\bf P}^{}_{\sf T}\approx 0.9697$ 
for the transverse deviations ${\sf T}=2,3$, leading to an overall purity ${\bf P}={\bf P}^{}_1{\bf P}^{}_2{\bf P}^{}_3 \approx 0.9395$. 
As expected in Section \ref{SecQCoherenceRest}, the purity of electron wavepacket is not seriously decreased during this acceleration stage.

Applying the correlators in Section \ref{SecQCoherenceRest} for an electron wavepacket at rest 
with the flying time $t^{}_F$ replaced by the acceleration proper time $\tau^{}_{a}$ and the regulator $\epsilon^{}_1$ substituted by (\ref{e0e1UAC}), we find ${\bf P}^{}_1 \approx 0.9994 > 0.9992$, 
and ${\bf P}^{}_{\sf T}\approx 0.9697$ 
slightly greater than the values of ${\bf P}^{}_{\sf T}$ we obtained above with the Unruh effect [the difference is $O(10^{-8})$]. 
This shows that {\it in each direction the purity of an accelerated electron wavepacket with the Unruh effect decays faster than the one obtained with the Unruh effect removed artificially,} and such a tendency is more significant in the longitudinal direction as shown in Figure \ref{FIGpurityUAC} (left).

The above scaled acceleration $\alpha$ is so large that $\langle \overline{\hat{z}_{1}^2}\rangle^{}_{\rm I} \gg \langle \hat{p}_{1}^2\rangle^{}_{\rm I}/(\bar{m}\alpha)^{2} $, 
while $\alpha\epsilon_1$ is small so that $g(\epsilon_1)\approx (\alpha\epsilon_1)^{-2} \gg 1$. From (\ref{z1z1PUAC}), (\ref{p1p1IUAC}), (\ref{z1z1FUAClate}--\ref{p1p1FUAClate}), and other results in Section \ref{LongCorrUAC}, we have
\begin{eqnarray}
  u^{}_1 &\approx& \left[\langle \overline{\hat{z}_{1}{}^2}\rangle^{}_{\rm I}  
	 - \frac{\hbar s}{\pi\bar{m}} g(\epsilon_0)\right] \times\nonumber\\ &&
	\left[ 
	  \frac{\hbar^2}{4\langle\overline{\hat{z}_{1}{}^2}\rangle^{}_{\rm I}}+\frac{3\hbar s\bar{m}}{2\pi \epsilon'_1{}^2}
	+ \frac{3\hbar s}{8\pi}\bar{m}\alpha^2 e^{2\alpha\tau(t)}
	 g(\epsilon_1)\left(2\frac{\bar{m}'}{\bar{m}}-\frac{2 \bar{m}'{}^2}{3\bar{m}^2} \right)\right], 
	\label{u1UACapprox}
\end{eqnarray}
where the contribution by $\langle \hat{z}_{}^{1}(t), \hat{p}_{}^{1}(t') \rangle\,\langle \hat{p}_{}^{1}(t), \hat{z}_{}^{1}(t') \rangle$ is $O\big[\langle \hat{p}_{1}{}^2\rangle^{}_{\rm I}/(\bar{m}\alpha)^{2}\big]$ compared with the leading order, and thus negligible.
Then the decoherence time for the longitudinal deviation in the laboratory frame can be estimated as
\begin{equation}
  \bar{\gamma}(T^{\rm dc}_1) e^{\alpha\tau(T^{\rm dc}_1)} \approx 
	\sqrt{\frac{3\pi \hbar \epsilon'_1{}^2 }{2 s\bar{m}\langle \overline{\hat{z}_{1}^2}\rangle^{}_{\rm I}}}
\end{equation} 
for $\bar{m}'\approx\bar{m}$, such that ${\bf P}^{}_1(T^{\rm dc}_1) \approx 1/2$. Applying (\ref{zintauUAC}) and (\ref{gammaUAC}), and inserting the parameter values in Figure \ref{FIGpurityUAC} (left), we get $T^{\rm dc}_1 \approx 2.1 \times 10^{-8}$ s, which agrees with the result in the plot (the intersection of the black dashed curve and the horizontal gray line of ${\bf P}_i=1/2$) quite well.

In the transverse directions, applying the late-time limits of the correlators from (\ref{zTzTapproxET}) to (\ref{pTzTFapproxET}),
the decoherence time in the laboratory frame is estimated by 
\begin{equation}
   \bar{\gamma}( T^{\rm dc}_{\sf T})\, \tau( T^{\rm dc}_{\sf T}) = 
  \sqrt{\frac{6\pi\bar{m}\epsilon'_1{}^2}{\hbar s} \langle\overline{\hat{z}_{\sf T}^2}\rangle^{}_{\rm I}}
\end{equation}
under the same initial conditions in obtaining (\ref{Tdecoh}) for an electron at rest.
Then we get $T^{\rm dc}_{\sf T}\approx 7.1 \times 10^{-9}$ s in Figure \ref{FIGpurityUAC} (left). 

In Figure \ref{FIGpurityUAC} (right), we show the purities of a uniformly accelerated electron against the accelerating electric field ${\cal E}$ 
with the acceleration tube of fixed length 10 cm. Again one can see that the purity with the Unruh effect (black) in each direction is always lower than the one without the Unruh effect (green), although those in the transverse directions are not distinguishable in the plot (solid curves). Thus we may say that {\it the Unruh effect enhances decoherence, as measured by a decrease of purity.} This is more significant in the longitudinal direction (dashed curves). All the curves go to zero as ${\cal E}\to 0$ because the acceleration time increases as the proper acceleration decreases. For small ${\cal E}$, the transverse purity right after the acceleration stage is less than the longitudinal purity. Figure \ref{FIGpurityUAC} (left) is in this regime.
For ${\cal E}$ greater than about $4.5\times 10^6$ V/m ($\gamma(t^{}_a) > 9.8$), however, the transverse purity overtakes and keeps robust as ${\cal E}$ increases.
Quantum coherence of the particle-motion deviation in the longitudinal direction drops very rapidly as the accelerating electric field and thus the proper acceleration of the electron wavepacket gets larger.

\section{Summary}
\label{SecSumOut}

We have constructed a linear effective 
theory to describe a single relativistic particle moving in the EM vacuum. We consider the quantum state of the deviation of the particle's motion from its classical trajectory (shortened to `particle-motion deviation')  as a Gaussian wavepacket, while the vacuum state of the EM fields is also Gaussian. The Gaussianity of the quantum state for the whole system persists by virtue of the linearity of the theory.  

We quantized the theory in the laboratory frame, and then wrote down the quantum counterpart of the Lorentz-Abraham-Dirac equation for the particle mode functions including the self fields and radiation reactions after mass renormalization. We obtained the analytic solutions of the mode functions, with which we calculated the two-point correlators of the particle-motion deviations and their canonical momenta. Then the reduced state of the particle can be fully determined.

Divergences arise in the mode sums over infinite degrees of freedom of the fields when calculating the two-point correlators of the particle. To deal with the divergences, our regularization is done by introducing a UV cutoff $1/\epsilon$ in the mode sum, and the uncertainties of time tagging $\epsilon_0$ and $\epsilon_1$ for the initial and final moments in the history of the charged particle in the coincidence limit. 
To exclude particle-antiparticle pair production, we set the value of $c\epsilon$ as the electron  Compton wavelength $\lambda^{}_C$ for an electron at rest. 
We chose the values of $\epsilon_0$ and $\epsilon_1$ for single charged particles according to the energy uncertainty of the electrons produced by a field emission electron gun. With these physically meaningful and finite valued regulators
explicitly present in the correlators, we have avoided the absolute coincidence limit and the associated divergences. Moreover, we have $\epsilon \ll \epsilon_0$, which is consistent with the requirement we found in \cite{LH10} to obtain the results relevant to the Unruh effect, and the exact value of $\epsilon$ would be masked by $\epsilon_0$ and $\epsilon_1$ in our results. 

We calculated the purity of the reduced state of the particle, which makes physical sense only after quantum state renormalization. 
We found that the purity of a single electron in the electron interference experiment described in Ref. \cite{TE89} could be close to $1/2$ in the transverse direction when it approaches the screen, while its  purity is not seriously decreased during the early acceleration stage of the flying electron. Our result suggests that vacuum fluctuations may play a major role in blurring the interference pattern in Ref. \cite{TE89}.  

We addressed the Unruh effect on a uniformly accelerated charge. We found that, in order to obtain the Planck factor and identify a regime satisfying the Fermi golden rule in the Unruh effect, the effect of length contraction should be included in the UV cutoff, 
namely, $c\epsilon = \lambda^{}_C/\bar{\gamma}(t)$ with  $\bar{\gamma}(t)\ge 1$ being the Lorentz factor of the particle's classical motion.
Thus the cutoff for a particle in non-inertial motion is apparently time-dependent in the laboratory frame. Furthermore, since the measurement time or energy uncertainty for an electron is determined in the laboratory and is supposed to be a constant in the laboratory time, the resolution $\epsilon^{}_1$ in the proper time of an accelerated particle should also be varying in the laboratory time as the electron's speed changes. It turns out that such an apparently time-dependent regulator is what keeps the momentum correlators positive for large accelerations at late times, and thus this time-dependence is a necessity for the consistency of our effective theory.  
However, these time-dependent regulators imply that if a photoelectron is initially moving at $t^{}_0$ at a highly relativistic speed ($\bar{\gamma}(t^{}_0)\gg 1$) and then decelerated by an applied negative voltage, then in the period with $\bar{\gamma}(t) \ll \bar{\gamma}(t^{}_0)$, the electron may behave very differently from what the Unruh effect would predict.

We also found that some terms in the two-point correlators have a Planck factor corresponding to a fermionic bath at the Unruh temperature, rather than the bosonic bath that the other terms correspond to. Thus, one cannot trivially apply the Unruh effect to a system by simply introducing a bosonic or a fermionic environment. Finally, we demonstrated that the purity of the particle-motion deviation in each direction with the consideration of the Unruh effect decays faster than the results obtained with the Unruh effect removed artificially, and this tendency is more significant in the direction of acceleration.

For future works, using our effective theory, we are currently working on quantum corrections to the radiations by a charge in uniform acceleration \cite{LH06}, circular motion \cite{BL83, AS07, Sc54, Lin03c}, and oscillatory motion  \cite{DLHM13, Lin17}. 
To get more insight, we are comparing our calculations with those using the worldline influence functional method \cite{JH05}. 
For treating quantum foundation issues, this wavepacket-field theory is very adapt for investigating  fundamental problems involving dephasing or decoherence such as in the study of the equivalence principle for quantum systems \cite{QEP}. It also has good potential to be applied to quantum entanglement problems, e.g., for atoms, charges, flavors \cite{Bruschi,Boyan,Blasone} in quantum fields, and after some reformulation,  for mass-gravitational field systems, in the hotly promoted gravitational entanglement experiments \cite{Bose,Vedral,AHGravCat}. \\

\begin{acknowledgments}
SYL thanks Yu-Che Huang for the comment on the switching function in an early version of this work.
SYL is supported by the National Science and Technology Council of Taiwan under grant No. NSTC 
112-2112-M-018-003 and in part by the National Center for Theoretical Sciences, Taiwan.  BLH appreciates the warm hospitality of Prof. Chong-Sun Chu and Prof. Hsiang-nan Li during his visits to the National Tsing Hua University and the Institute of Physics,  Academia Sinica, Taiwan, where a good part of this work was conducted.   
\end{acknowledgments}


\begin{thebibliography}{00}

\bibitem{La47} W. E. Lamb and R. C. Retherford, {\it Fine structure of the hydrogen atom by a Microwave Method}, {\it Phys. Rev.} {\bf 72} (1947) 241.

\bibitem{Be47} H. A. Bethe, {\it The electromagnetic shift of energy levels}, {\it Phys. Rev.} {\bf 72} (1947) 339.

\bibitem{We48} T. A. Welton, {\it Some observable effects of the quantum-mechanical fluctuations of the electromagnetic field},
{\it Phys. Rev.} {\bf 74} (1948) 1157.

\bibitem{Sa31} F. Sauter, {\it \"Uber das Verhalten eines Elektrons im homogenen elektrischen Feld nach der relativistischen Theorie Diracs}, {\it Z. Phys.} {\bf 69} (1931) 742.

\bibitem{HE36} W. Heisenberg and H. H. Euler, {\it Folgerungen aus der Diracschen Theorie des Positrons}, {\it Z. Phys.} {\bf 98} (1936) 714;
English translation by W. Korolevski and H. Kleinert [arXiv:physics/0605038].

\bibitem{Sc51} J. Schwinger, {\it On gauge invariance and vacuum polarization}, 
{\it Phys. Rev.} {\bf 82} (1951) 664.

\bibitem{IZ80} C. Itzykson and J.-B. Zuber, {\it Quantum Field Theory}, McGraw-Hill, New York (1980).

\bibitem{CH88} E. Calzetta and B. L. Hu, {\it Non-equilibrium quantum fields: Closed time-path effective action, Wigner function and Boltzmann equation}, {Phys. Rev. D} {\bf 37} (1988) 2878.

\bibitem{CHR00}  E. Calzetta, B. L. Hu, S. A. Ramsey,{\it Hydrodynamic transport functions from quantum kinetic field theory}, {\it Phys. Rev. D} {\bf 61} (2000) 125013.

\bibitem{RHIC} U. Heinz,  {\it Relativistic hydrodynamics and the transport properties of QCD matter}, in {\it Relativistic Heavy Ion Physics}, Landolt-B\"ornstein New Series, Vol. I/23, R. Stock ed., Springer Verlag, Berlin Heidelberg, (2010). 

\bibitem{qgp} E. Shuryak, {\it Strongly coupled quark-gluon plasma in heavy ion collisions}. {\it Rev. Mod. Phys.} {\bf 89} (2017) 035001.

\bibitem{Jordan} R. D. Jordan, {\it Effective field equations for expectation values}, {\it Phys. Rev. D} {\bf 33} (1986) 444.

\bibitem{CalHu87} E. Calzetta and B. L. Hu, {\it Closed time-path functional formalism in curved spacetime: Application to cosmological backreaction problems}, {\it Phys. Rev. D} {\bf 35} (1987) 495.

\bibitem{Sch61} J. Schwinger, {\it Brownian Motion of a Quantum Oscillator}, {\it J. Math. Phys.} {\bf 2} (1961) 407.

\bibitem{Kel64} L. V. Keldysh, {\it Diagram technique for nonequilibrium processes},
{\it Zh. Eksp. Teor. Fiz.} {\bf 47} (1964) 1515 [{\it Sov. Phys. JETP} {\bf 20}, 1018.]

\bibitem{Chou} K. C. Chou, Z. B. Su, B. L. Hao, and L. Yu, 1985. {\it Equilibrium and nonequilibrium formalisms made unified}, {\it Phys. Rep.} {\bf 118} (1985) 1.


\bibitem{Berges} J. Berges, {\it Introduction to nonequilibrium quantum field theory}, {\it AIP Conf. Proc.} {\bf 739} (2004) 3.

\bibitem{CalHu08} E. A. Calzetta and B. L. Hu, {\it Nonequilibrium quantum field theory}, 
Cambridge University Press, Cambridge U.K. (2008).

\bibitem{MST}  
Y. Mino, M. Sasaki, and T. Tanaka, {\it Gravitational radiation reaction to a particle motion}, {\it Phys. Rev. D} {\bf 55} (1997) 3457.

\bibitem{QuinnWald} T. C. Quinn and R. M. Wald, {\it Axiomatic approach to electromagnetic and gravitational radiation reaction of particles in curved spacetime}, {\it Phys. Rev. D}, {\bf 56} (1997) 3381.

\bibitem{JH02} P. R. Johnson and B. L. Hu, {\it Stochastic theory of relativistic particles moving in a quantum field: Scalar Abraham-Lorentz-Dirac-Langevin equation, radiation reaction and vacuum fluctuations}, {\it Phys. Rev. D} {\bf 65} (2002) 065015.

\bibitem{JH05} P. R. Johnson and B. L. Hu, {\it Uniformly accelerated charge in a quantum field: From radiation reaction to Unruh effect}, {\it Found. Phys.} {\bf 35} (2005) 1117. 

\bibitem{GalHu05} Chad R. Galley and B. L. Hu, {\it Self-Force with a Stochastic Component from Radiation Reaction of a Scalar Charge Moving in Curved Spacetime},  Phys. Rev. D 72, 084023 (2005). 

\bibitem{GHL06} C. R. Galley, B. L. Hu, and S.-Y. Lin, {\it Electromagnetic and gravitational self-force on a relativistic particle from quantum fields in curved space}, {\it Phys. Rev. D} {\bf 74} (2006) 024017. 

\bibitem{HuVer20} B. L. Hu and E. Verdaguer, {\it Semiclassical and Stochastic Gravity: Quantum Field Effects on Curved Spacetime}, Cambridge University Press, Cambridge U.K. (2020).

\bibitem{CamVer94} A. Campos and E. Verdaguer, {\it Semiclassical equations for weakly inhomogeneous cosmologies}, {\it Phys. Rev. D} {\bf 49} (1994) 1861.

\bibitem{CalHu94} E. Calzetta and B. L. Hu, {\it Noise and fluctuations in semiclassical gravity}, {\it Phys. Rev. D} {\bf 49} (1994) 6636.

\bibitem{HuSin95} B. L. Hu and  S. Sinha, {\it Fluctuation-dissipation relation for semiclassical cosmology}, {\it Phys. Rev. D} {\bf 51} (1995) 1587.

\bibitem{CamVer96} A. Campos and E. Verdaguer, {\it Stochastic semiclassical equations for weakly inhomogeneous cosmologies}, {\it Phys. Rev. D} {\bf 53} (1996) 1927.

\bibitem{Parentani} D. Campo and R. Parentani, {\it Decoherence and entropy of primordial fluctuations. II. The entropy budget}, {\it Phys. Rev. D} {\bf 78} (2008) 065045.

\bibitem{LCHent}  
S.-Y. Lin, C.-H. Chou, and B. L. Hu, {\it Quantum entanglement and entropy in particle creation},  {\it Phys. Rev  D} {\bf 81} (2010) 084018. 

\bibitem{HHent} J. T. Hsiang and B. L. Hu, {\it Intrinsic entropy of squeezed quantum fields and nonequilibrium quantum dynamics of cosmological perturbations},  
{\it Entropy} {\bf 23} (2021) 1544. 

\bibitem{EID} J. P. Paz and W. H. Zurek, {\it Environment-induced decoherence and the transition from quantum to classical}, in {\it Fundamentals of quantum information: quantum computation, communication, decoherence and all that}, 
Springer Verlag, Berlin Heidelberg (2002).

\bibitem{cddn} E. Calzetta and B. L. Hu, {\it Correlations, decoherence, dissipation and noise in quantum field theory},   
in {\it Discourses in Mathematics and Its Applications No. 4},
S. A. Fulling ed., Texas A \& M University Press, College Station (1995) 

\bibitem{CalHu99} E. Calzetta and B. L. Hu, {\it Stochastic dynamics of correlations in quantum field theory: From Schwinger-Dyson to Boltzmann-Langevin equations}, {\it Phys. Rev. D} {\bf 61} (1999) 025012.

\bibitem{Ro65} F. Rohrlich, {\it Classical Charged Particles},
Addison-Wesley, Redwood (1965). 

\bibitem{Sp99} H. Spohn, {\it Dynamics of Charged Particles and Their Radiation Field}, Cambridge University Press, Cambridge U.K. (1999).

\bibitem{HHL23} Y.-C. Huang, F.-M. He, and S.-Y. Lin, 
{\it Quantum mechanical wavepackets of single relativistic particles},
{\it Chinese J. Phys.} {\bf 87} (2023) 486. 

\bibitem{ELINP} D. Doria, M. O. Cernaianu, P. Ghenuche, D. Stutman, K. A. Tanaka, C. Ticosa, and C. A. Ur,
{\it Overview of ELI-NP status and laser commissioning experiments with 1PW and 10PW class-lasers},
{\it JINST} {\bf 15} (2020) C09053 

\bibitem{SULF} Z. Gan {\it et al.}, {\it The Shanghai superintense ultrafast laser facility (SULF) project}, in 
{\it Progress in Ultrafast Intense Laser Science XVI, Topics in Applied Physics 141}, 
K Yamanouchi, K Midorikawa and L Roso eds., Springer Verlag, Berlin Heidelberg (2021). 

\bibitem{EFT} W. D. Goldberger and I. Z. Rothstein, {\it 
Effective field theory of gravity for extended objects},
{\it Phys. Rev. D} {\bf 73} (2006) 104029. 

\bibitem{GH09} C. R. Galley and B. L. Hu, {\it Self-force on extreme mass ratio inspirals via curved spacetime effective field theory}, {\it Phys. Rev. D} {\bf 79} (2009) 064002.

\bibitem{TE89} A. Tonomura, J. Endo, T. Matsuda, and T. Kawasaki, and H. Ezawa,
{\it Demonstration of single-electron buildup of an interference pattern},
{\it Am. J. Phys.} {\bf 57} (1989) 117.

\bibitem{Unr76}  W. G. Unruh, 
{\it Notes on black hole evaporation},
{\it Phys. Rev. D} {\bf 14} (1976) 3251.

\bibitem{BL83} J. S. Bell and J. M. Leinaas, 
{\it Electrons as accelerated thermometers},
{\it Nucl. Phys. B} {\bf 212} (1983) 131.

\bibitem{AS07} E. T. Akhmedov and D. Singleton,
{\it On the relation between Unruh and Sokolov–Ternov effects},
{\it Int. J. Mod. Phys. A} {\bf 22} (2007) 4797.

\bibitem{CT99} P. Chen and T. Tajima, 
{\it Testing Unruh radiation with ultraintense lasers},
{\it Phys. Rev. Lett.} {\bf 83} (1999) 256.

\bibitem{SSH08} R. Schutzhold, G. Schaller, and D. Habs, 
{\it Signatures of the Unruh Effect from Electrons Accelerated by Ultrastrong Laser Fields},
{\it Phys. Rev. Lett.} {\bf 100} (2008) 091301.

\bibitem{OYZ16} N. Oshita, K. Yamamoto, and S. Zhang, 
{\it Quantum radiation produced by a uniformly accelerating charged particle in thermal random motion},
{\it Phys. Rev. D} {\bf 93} (2016) 085016.

\bibitem{DeW79} B. S. DeWitt, 
{\it Quantum gravity: the new synthesis}, 
in {\it General Relativity: an Einstein Centenary Survey}, 
S. W. Hawking and W. Israel eds., Cambridge University Press, Cambridge U.K. (1979).

\bibitem{LH06}  S.-Y. Lin and B. L. Hu, 
{\it Accelerated detector - quantum field correlations: From vacuum fluctuations to radiation flux},
{\it Phys. Rev. D} {\bf 73} (2006) 124018. 

\bibitem{LH07} S.-Y. Lin and B. L. Hu, {\it Backreaction and the Unruh effect: new insights from exact solutions of
uniformly accelerated detectors}, {\it Phys. Rev. D} {\bf 76} (2007) 064008 (2007).

\bibitem{Ha92} B. Hatfield, {\it Quantum Field Theory of Point Particles and Strings},
Addison-Wesley, Redwood (1992).

\bibitem{Lo92a} H. A. Lorentz, {\it Arch. Néer. Sci. Exact. Nat. XXV} (1892) 363.

\bibitem{Lo92b} H. A. Lorentz, {\it The Theory of Electrons, 2nd ed.}, Dover, New York (1952).

\bibitem{Ab05} M. Abraham M, {\it Theorie der Elektrizit\"at, Vol. II}, Teubner, Leipzig (1905).

\bibitem{Di38} P. A. M. Dirac, {\it Classical theory of radiating electrons}, {\it Proc. R. Soc. London A} {\bf 167} (1938) 148. 

\bibitem{LH10} S.-Y. Lin and B. L. Hu, {\it Entanglement creation between two causally disconnected objects}, {\it Phys. Rev. D} {\bf 81} (2010) 045019.

\bibitem{Gr17} D. J. Griffiths, {\it Introduction to Electrodynamics, 4th Ed.}, Cambridge University Press, Cambridge U.K. (2017).

\bibitem{BD82} N. D. Birrell and P. C. W. Davies, {\it Quantum Fields in
Curved Space}, Cambridge University Press, Cambridge U.K. (1982).

\bibitem{HT92} M. Henneaux and C. Teitelboim, {\it Quantization of Gauge Systems}, Princeton University Press, Princeton (1992).

\bibitem{Ro01} F. Rohrlich, {\it The correct equation of motion of a classical point charge}, {\it Phys. Lett. A} {\bf 283} (2001) 276-278.

\bibitem{LL75} L. Landau and E. Lifshitz, {\it The classical theory of fields}, Pergamon Press, New York (1975).

\bibitem{JEOL22} Glossary of TEM/SEM Terms, JEOL Ltd. 
{\sf https://www.jeol.co.jp/en/words/emterms/index.html}
(Retrieved 2 October 2022)

\bibitem{SSS95} M. P. Silverman, W. Strange, and J. C. H. Spence,
{\it The brightest beam in science: New directions in electron microscopy and interferometry},
{\it Am. J. Phys.} {\bf 63} (1995) 800.

\bibitem{To98} A. Tonomura {\it The Quantum World Unveiled by Electron Waves}, World Scientific, Singapore (1998).

\bibitem{TP4} X. Llopart et al. 
{\it Timepix4, a large area pixel detector readout chip which can be tiled on 4 sides providing sub-200 ps timestamp binning}, {\it JINST} {\bf 17} (2022) C01044.

\bibitem{Lin23} S.-Y. Lin, {\it Quantum coherence of relativistic wavepackets in electromagnetic vacuum},  
{\it J. Phys.: Conf. Ser.} {\bf 2482} (2023) 012018.

\bibitem{Sc54} J. Schwinger, {\it The quantum correction in the radiation by energetic accelerated electrons}, 
{\it PNAS} {\bf 40} (1954) 132.

\bibitem{Lin03c} S.-Y. Lin, {\it Quantum corrections to synchrotron radiation from wave-packet},
{\it Phys. Lett. A} {\bf 317} (2003) 37.

\bibitem{DLHM13} J. Doukas, S.-Y. Lin, B. L. Hu, and R. B. Mann, {\it Unruh effect under non-equilibrium conditions: oscillatory motion of an Unruh-DeWitt detector}, {\it JHEP} {\bf 11 (2013)} 119.

\bibitem{Lin17} S.-Y. Lin, {\it Quantum radiation by an Unruh-DeWitt detector in oscillatory motion},
{\it JHEP} {\bf 11 (2017)} 102.

\bibitem{QEP}  C. Anastopoulos and Bei-Lok Hu, {\it Equivalence principle for quantum systems: Dephasing and phase shift of free-falling particles},  {\it Class. Quant. Gravity} {\bf 35} (2018) 035011. 

\bibitem{Bruschi} D. E. Bruschi, J. Louko, E. Martín-Martínez, A. Dragan, and I. Fuentes, {\it Unruh effect in quantum information beyond the single-mode approximation}, {\it Phys. Rev. A} {\bf 82} (2010) 042332.

\bibitem{Boyan} L. Lello, D. Boyanovsky, and R. Holman, {\it Entanglement entropy in particle decay}, {\it JHEP} {\bf 11(2013)} 116. 

\bibitem{Blasone} M. Blasone, F. Dell'Anno, S. De Siena, and F. Illuminati, {\it Flavor entanglement in neutrino oscillations in the wave packet description},  {\it Europhys. Lett.} {\bf 112} (2015) 20007.

\bibitem{Bose} S. Bose et al., {\it A spin entanglement witness for quantum gravity}, {\it Phys. Rev. Lett.} {\bf 119} (2017) 240401.

\bibitem{Vedral} C. Marletto and V. Vedral, {\it Gravitationally induced entanglement between two massive particles is sufficient evidence of quantum effects in gravity}, {\it Phys. Rev. Lett.} {\bf 119} (2017) 240402.

\bibitem{AHGravCat} C. Anastopoulos and B. L. Hu, {\it Quantum superposition of two gravitational cat states}, {\it Class. Quantum Grav.} {\bf 37} (2010) 235012. 



\end{thebibliography}
\end{document}